\documentclass[acmtog,screen]{acmart}
\AtBeginDocument{%
  }

\setcopyright{cc}
\setcctype{by}
\acmJournal{TOG}
\acmYear{2026} \acmVolume{45} \acmNumber{4} \acmArticle{142}
\acmMonth{7} \acmDOI{10.1145/3811327}

\author{Xuan Tang}
\affiliation{%
  \institution{University of California San Diego}
  \city{San Diego}
  \state{California}
  \country{USA}
}

\author{Kemeng Huang}
\affiliation{%
  \institution{University of Hong Kong}
  \country{China}
}

\affiliation{%
  \institution{Carnegie Mellon University}
  \country{USA}
}
\author{Gilbert Bernstein}
\affiliation{%
  \institution{University of Washington}
  \country{USA}
}

\author{Minchen Li}
\affiliation{%
  \institution{Carnegie Mellon University}
  \country{USA}
}
\affiliation{%
  \institution{Genesis AI}
  \country{USA}
}
\author{Tzumao Li}
\affiliation{%
  \institution{University of California San Diego}
  \city{San Diego}
  \state{California}
  \country{USA}
}

\usepackage{tabularx} 
\usepackage{makecell} 
\usepackage{amsmath}
\usepackage{subfig}
\usepackage{algorithm}
\usepackage{algpseudocode}
\usepackage{listings}
\usepackage{xcolor}
\usepackage[most]{tcolorbox}
\usepackage{tikz}
\usepackage{wrapfig}
\usepackage{caption} 
\usepackage{placeins} 
\usepackage{enumitem}
\setlist[enumerate]{leftmargin=20pt}
\setlist[itemize]{leftmargin=20pt}

\usetikzlibrary{arrows.meta,positioning,shapes,fit}
\usetikzlibrary{calc}
\selectcolormodel{rgb}
\definecolor{pythonBackgroundColor}{cmyk}{0.00, 0.00, 0.00, 0.00}
\definecolor{pythonBorderColor}{cmyk}{0.79, 0.35, 0.00, 0.57}
\definecolor{pythonBaseColor}{cmyk}{0.00, 0.00, 0.00, 1.00}
\definecolor{pythonKeywordColor}{cmyk}{0.77, 0.34, 0.00, 0.29}
\definecolor{pythonStringColor}{cmyk}{0.34, 0.39, 0.00, 0.31}
\definecolor{pythonParameterColor}{cmyk}{0.72, 0.00, 0.52, 0.40}
\definecolor{pythonFunctionColor}{cmyk}{0.77, 0.34, 0.00, 0.29}
\definecolor{pythonCallColor}{cmyk}{0.00, 0.88, 0.66, 0.11}
\definecolor{pythonAttributeColor}{cmyk}{0.00, 0.63, 0.95, 0.13}
\definecolor{pythonNumberColor}{cmyk}{0.72, 0.00, 0.52, 0.40}
\definecolor{pythonCommentColor}{cmyk}{0.00, 0.00, 0.00, 0.61}
\definecolor{pythonOperatorColor}{cmyk}{0.00, 0.00, 0.00, 1.00}
\definecolor{pythonDecoratorColor}{cmyk}{0.00, 0.00, 0.00, 0.65}

\usepackage{bm} 

\selectcolormodel{rgb}

\lstdefinestyle{pythonStyle}{
backgroundcolor=\color{pythonBackgroundColor}, 
commentstyle=\color{pythonCommentColor}, 
keywordstyle=\color{pythonKeywordColor}, 
stringstyle=\color{pythonStringColor}, 
basicstyle=\linespread{0.8}\color{pythonBaseColor}\ttfamily\footnotesize, 
breakatwhitespace=false, 
breaklines=true, 
captionpos=b, 
keepspaces=true, 
numbers=none, 
numbersep=5pt, 
showspaces=false, 
showstringspaces=false, 
showtabs=false, 
tabsize=4, 
xleftmargin=0pt, 
otherkeywords={}, 
moredelim=[is][\color{pythonParameterColor}]{|@}{@|},  
moredelim=[is][\color{pythonFunctionColor}]{*|}{|*},  
moredelim=[is][\color{pythonCallColor}]{|!}{!|},  
moredelim=[is][\color{pythonAttributeColor}]{|?}{?|},  
moredelim=[is][\color{pythonNumberColor}]{?@}{@?},  
moredelim=[is][\color{pythonOperatorColor}]{@*}{*@},  
moredelim=[is][\color{pythonDecoratorColor}]{?*}{*?},  
moredelim=[is][\color{pythonKeywordColor}]{!@}{@!},  
emph={}, 
emphstyle=\color{pythonKeywordColor}, 
literate=%
{\{level\}}{{{\color{pythonBaseColor}{\{level\}}}}}{7}
}

\newtcblisting{pythonBlock}[1]{
boxrule=0.26pt, 
arc=2.44pt, 
auto outer arc, 
colframe=pythonBorderColor, 
colback=pythonBackgroundColor, 
listing only, 
listing options={language=Python, style=pythonStyle}, 
title=#1, 
fonttitle=\bfseries, 
top=-2pt, 
bottom=-2pt, 
left=1pt, 
right=1pt, 
}

\usepackage{siunitx}
\begin{document}

\title{YASPS: A Symbolic Framework for Extensible, High-Performance IPC Simulation}


\begin{abstract}
Incremental Potential Contact (IPC) has emerged as a robust and unifying formulation for contact-rich physical simulation by casting elasticity and collision handling as a single energy minimization problem. Achieving high performance, however, typically requires heavily specialized implementations that hard-code assumptions about energies, primitive types, and parameterizations, creating a major barrier to extensibility. Adding new energies or alternative parameterizations often requires re-deriving first and second-order derivatives, and implementing new assembly logic for the global Hessian and gradient. This challenge is further exacerbated by collision energies, where the same energy definition is often applied to mixed parameterizations, which can lead to a combinatorial explosion of parameterization-specific derivative and assembly cases.

In this paper we introduce YASPS, a framework for physical simulation that resolves this limitation by making structural relationships explicit in a differentiable representation. YASPS introduces two relational operators, \texttt{JOIN} and \texttt{UNION}, which encode connectivity and heterogeneous parameterizations directly in the symbolic computation graph. Using symbolic differentiation over these operators, YASPS automatically derives local derivatives, and determines the induced sparsity and block structure of global gradients and Hessians from the same description while avoiding any code explosion induced by mixed-parameterizations.

Targeting IPC workloads, YASPS compiles the resulting symbolic graphs into GPU kernels for local energy evaluation, derivative computation, and block-sparse matrix assembly, and solves the resulting Newton systems using a GPU-based iterative solver. This approach achieves performance competitive with state-of-the-art IPC implementations while enabling new energies and parameterizations to be added through localized symbolic definitions, without hand-written derivative or assembly code.
\end{abstract}


\begin{teaserfigure}
  \includegraphics[width=\textwidth]{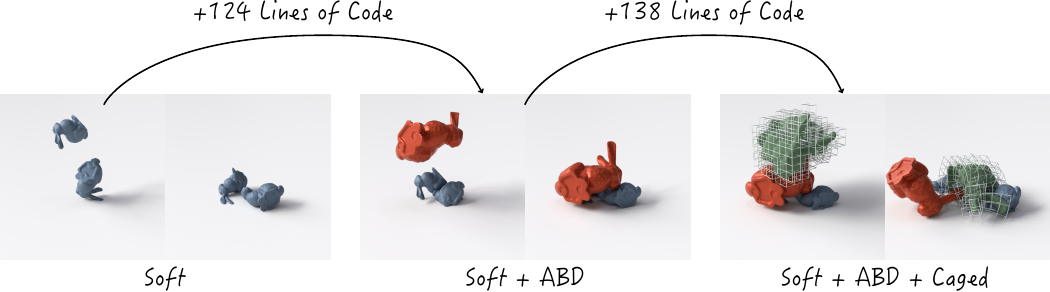}
  \caption{We introduce \textsc{YASPS}, an IPC~\cite{ipc}-based simulation framework that is both extensible and performant. Users define parameterizations, shape primitives, and energies in \textsc{Python}, from which \textsc{YASPS} automatically generates and compiles GPU code for first- and second-order differentiation. YASPS also assembles the global gradient and Hessian and efficiently solves the resulting linear systems on the GPU. Despite its high-level interface, \textsc{YASPS} achieves performance comparable to hand-optimized GPU IPC frameworks such as \textsc{GIPC}~\cite{GIPC}, while significantly reducing implementation complexity: adding a new shape type requires only around 130 lines of \textsc{Python} code.}
  \Description{User can easily define new shapes and energies through YASPS, and get performant code.}
  \label{fig:teaser}
\end{teaserfigure}

\maketitle

\section{Introduction}

\emph{Incremental Potential Contact (IPC)}~\cite{ipc} has become a reliable approach for simulating complex contact because it casts contact and elasticity as a single energy minimization problem. Its robustness comes with substantial computational cost: IPC relies on Newton-type solvers that repeatedly evaluate local energies and second derivatives and solve large, sparse linear systems. Making this pipeline efficient is essential for IPC to be practical.

State-of-the-art IPC implementations such as \textsc{PolyFEM}~\cite{polyfem} and \textsc{GIPC}~\cite{GIPC} achieve high performance through heavily specialized, hand-optimized kernels and assembly logic tailored to a fixed set of energies, primitive types, and parameterizations. This specialization creates an \textbf{extensibility bottleneck}: adding a new energy term, a new primitive type, or a new parameterization (e.g., Affine Body Dynamics~\cite{abdipc}) typically requires deriving and implementing new gradient/Hessian code and new assembly rules for many combinatorial cases. 

The root cause of the extensibility bottleneck is that the simulation pipeline lacks a unified way to represent and propagate \emph{structural information}: 
\begin{enumerate}
    \item Which entities participate in each energy term (connectivity and relations).
    \item How their positions are parameterized (direct coordinates, rigid/affine bodies, etc).
    \item How local derivatives map back to global degrees of freedom (sparsity and block structure).
\end{enumerate}
In hand-written systems, this structure is encoded implicitly across the kernels and case-specific matrix assembly routines, tightly coupling energies to representations and preventing extension.

Automatic and symbolic differentiation~\cite{tinyad, symx, mesh-based-symbolic} reduce the burden of deriving local derivatives. However, heterogeneous parameterizations, commonly found in collision energies, still require either duplicated differentiated code paths or new hand-written projection and assembly logic in those systems, because they lacks a representation of parameterizations and their sparsity pattern. 

To make this concrete, consider a repulsive energy between two points $p_1$ and $p_2$,
\begin{equation}\label{energy-simple-repulsive}
E(p_1,p_2) = \frac{1}{\|p_1 - p_2\|}, \quad p_1,p_2 \in \mathbb{R}^3.
\end{equation}

If points $p_1$ and $p_2$ are directly parameterized by three coordinates each (free vertices), one can implement kernels for $E$, $\nabla E$, and $\nabla^2 E$ directly. In many simulators, however, positions can be derived from different parameterizations. For example, in Affine Body Dynamics~\cite{abdipc} a point is given by

\[
p = A_b r + t_b,
\]
where $A_b \in \mathbb{R}^{3\times 3}$ and $t_b \in \mathbb{R}^3$ describe the transform of body $b$ and $r$ is a rest-pose position. 

With multiple parameterizations available, mixed interactions increase exponentially. Consider a mesh with mixed materials, where each vertex may be realized using one of three parameterizations. A point–triangle collision energy, which involves four vertices, then leads to $3^4 = 81$ distinct parameterization combinations. In a na\"ive implementation, each combination has a different derivative, Hessian block layout, and assembly rule, requiring separate specialized kernels at compile time and explicit case selection at runtime.

Human implementations, such as those in \textsc{PolyFEM} and \textsc{GIPC}, typically avoid the exponentiality by factoring computation in the following steps:
\begin{enumerate}
    \item Differentiate $E$ with respect to $p_1$ and $p_2$.
    \item Differentiate $p_1$ and $p_2$ with respect to their underlying parameters (body transforms, rest positions, or direct coordinates).
    \item Apply the chain rule to obtain the final gradient and Hessian with respect to the global degrees of freedom.
\end{enumerate}
They can exploit such optimization whenever such \textbf{structural information} -- how the energy is constituted from different parameterizations, and what those parameterizations are -- is made explicit to the system.  

In contrast, a straightforward automatic or symbolic differentiation implementation operates on the concrete computation graph produced by a particular choice of parameterizations. When parameterization alternatives are represented via branching,
existing systems differentiate each branch separately produced by different combinations of parameterizations,
and provide no principled way to either factor out shared derivative terms across alternative branches or derive a unified sparsity/assembly strategy from the same description. 
As a result, supporting heterogeneous parameterizations still tends to produce exponentially many sub-kernels for the same energy under different combinations of parameterizations.

Returning to the repulsive energy in Eq.~\ref{energy-simple-repulsive}, we can view its computation as the composition graph below:
\begin{figure}[H]
\centering
\resizebox{\columnwidth}{!}{%
\begin{tikzpicture}[
  font=\LARGE,                                
  node distance=10mm and 10mm,                
  box/.style={draw, rounded corners, align=center, inner sep=4pt, 
              minimum width=24mm, minimum height=9mm}, 
  func/.style={Latex-, thick},                
  inst/.style={Stealth-, dashed},             
  choice/.style={Latex-, dotted}              
]

\node[box] (E) {$E(p_1,p_2)$};

\node[box, right=10mm of E, yshift=6mm] (p1) {$p_1$};
\node[box, right=10mm of E, yshift=-6mm] (p2) {$p_2$};

\node[box, right=10mm of p1, yshift=-6mm] (pos) {Position};

\node[box, right=10mm of pos, yshift=6mm] (freepos) {Free position};
\node[box, right=10mm of pos, yshift=-6mm] (abdpos) {ABD position};

\node[box, right=10mm of abdpos, yshift=6mm] (A) {$A_b$};
\node[box, right=10mm of abdpos, yshift=-6mm] (t) {$t_b$};

\draw[func]  (p1.west) -- (E.east);
\draw[func]  (p2.west) -- (E.east);
\draw[func]  (A.west) -- (abdpos.east);
\draw[func]  (t.west) -- (abdpos.east);

\draw[inst] (pos.west) -- (p1.east) ;
\draw[inst] (pos.west) -- (p2.east);

\draw[choice]  (freepos.west) -- (pos.east);
\draw[choice]  (abdpos.west) -- (pos.east);

\node[draw, rounded corners, inner sep=3pt, align=center,
      above=8mm of freepos, anchor=south east, xshift=18mm] (legend) {
  \begin{tikzpicture}[baseline={(0,0)}]
    \draw[func] (0,0) -- +(8mm,0);
    \node[anchor=west] at (9mm,0) {computed from};

    \draw[inst] (42mm,0) -- +(8mm,0);
    \node[anchor=west] at (51mm,0) {instance of};

    \draw[choice] (77mm,0) -- +(8mm,0);
    \node[anchor=west] at (86mm,0) {choice of};
  \end{tikzpicture}
};

\end{tikzpicture}
}
\vspace{-20pt}
\caption{The energy in Eq.~\eqref{energy-simple-repulsive} can be decomposed into different layers.}
\vspace{-10pt}
\label{figure-energy-composition}
\end{figure}
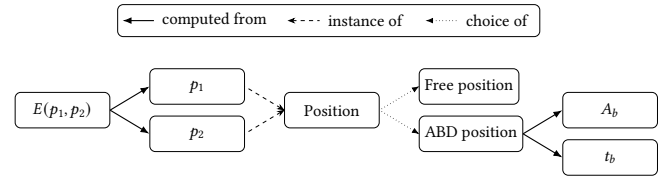

The \textbf{structural information} in Fig.~\ref{figure-energy-composition} that guides the differentiation in a manual implementation can also be used to determine the sparsity pattern of the global Hessian. In hand-written systems, however, this structure is encoded \emph{implicitly} in specialized kernels and parameterization-specific assembly code, rather than represented explicitly in a unified form that a differentiation and assembly pipeline can share and extend.

While prior work~\cite{mesh-based-symbolic} partially addresses the structural problem by asking users to declare which types of geometric primitives (e.g., triangles, tetrahedra, free vertices, rigid-body vertices) participate in each energy term, this approach again hard-codes a fixed set of primitive types and parameterizations into the framework, and still requires hand-written casework when new types are introduced.

\paragraph{Overview.}
In this paper we address these gaps with \textbf{YASPS} (\textbf{Y}et \textbf{A}nother \textbf{Symbolic} framework for \textbf{P}hysical \textbf{S}imulation), a framework that introduces two relational operators that allow users to express computations and parameterizations structurally:
\begin{itemize}
  \item \textbf{\texttt{JOIN}} composes dependent quantities across user-defined relations. For example, one may specify that a tetrahedron depends on four vertices and pull their positions when evaluating a volumetric energy; that each vertex depends on a single affine body transform when evaluating body-space deformation; or that each repulsive energy term retrieves two vertex positions as its inputs (Fig.~\ref{figure-energy-composition}).
  \item \textbf{\texttt{UNION}} represents alternative parameterizations within a relation. In Fig.~\ref{figure-energy-composition}, one may declare that a vertex position is realized either as a directly parameterized variable or as one derived from an affine body (or another model), and that different vertices may choose different alternatives. 
\end{itemize}

Users write energies and parameterizations directly in terms of \texttt{JOIN} and \texttt{UNION}. Because these operators are differentiable in YASPS' symbolic representation, we can obtain the symbolic differentiation, derive the induced sparsity and block structure of global gradient and Hessian, and generate efficient evaluation and assembly kernels.

Because IPC workloads are dominated by embarrassingly parallel evaluation of many local terms, YASPS compiles its symbolic graphs into GPU kernels that compute local energies and derivatives, perform index extraction through the relational layers, and assemble a block-sparse matrix for the Hessian matrix and gradient. A GPU-based conjugate gradient solver then operates directly on this representation to produce the solution of the resulting linear system for Newton iteration. The result is performance competitive with specialized IPC implementations like \textsc{GIPC}~\cite{GIPC} while remaining extensible: adding a new energy or parameterization requires only a high-level description, without hand-written derivative or assembly kernels. In the examples shown in Fig.~\ref{fig:teaser}, adding new parameterizations (Affine Body Dynamics and cage-based deformation) and their corresponding energies—an orthogonality regularizer $E_{\text{affine}}(\mathbf{A})= \frac{1}{2}\left\|\mathbf{A}^{\top}\mathbf{A} - \mathbf{I}\right\|_F^2$, and elastic energies on the cages—requires only around 130 additional lines of \textsc{Python} code. For collision energies, the only required change is to re-declare which attributes are \texttt{UNION}ed. This avoids an exponential blow-up in parameterization-specific kernels (both in user code and generated code), making YASPS scalable on both the frontend and backend with respect to the number of parameterizations.


\paragraph{Contributions.}
In summary, we:
\begin{itemize}
    \item Introduce a relational abstraction based on \texttt{JOIN} and \texttt{UNION} that encodes connectivity, instance relationships, and heterogeneous parameterizations.
    \item Develop symbolic differentiation rules over these operators, including an efficient second-order procedure that reuses intermediate Jacobians and reduces Hessian-projection cost.
    \item Derive global gradient/Hessian sparsity and block layout from the same relational description, enabling structure-aware block-sparse storage and compression.
    \item Implement a GPU-oriented system that uses just-in-time (JIT) compilation to generate evaluation, derivative, and solver kernels directly from the symbolic graph at runtime, enabling fast IPC-style simulation while preserving rapid extensibility.
\end{itemize}

To the best of our knowledge, YASPS is the first IPC-oriented framework that achieves both \emph{extensibility} and GPU-scale \emph{performance} by making relations and parameterization alternatives first-class in a differentiable intermediate representation (IR), and using that same representation to generate derivatives and structure-aware assembly.

\section{Related Work}
\subsection{IPC-based frameworks}
Incremental Potential Contact (IPC) \cite{ipc} has become a widely adopted strategy for preventing interpenetration in contact-rich simulation by combining barrier-style energies with step-size control. 
Recent works have applied IPC to simulate thin solids with codimensional geometries \citep{li2021codimensional}, rigid body systems \citep{ferguson2021intersection,abdipc,chen2022unified}, and the coupling between domains with different discretizations, such as FEM-MPM \citep{li2024dynamic,li2022bfemp}, FEM-SPH \citep{xie2023contact}, and DEM-MPM \citep{jiang2022hybrid}.
IPC has been integrated into or has inspired several modern systems, including \textsc{PolyFEM}~\cite{polyfem}, \textsc{GIPC}~\cite{GIPC} and its successor \textsc{Stiff GIPC}~\cite{libuipc}, etc. Broadly, \textsc{PolyFEM} emphasizes coverage of materials and discretizations, while \textsc{GIPC}/\textsc{Stiff GIPC} emphasize performance and robust contact handling. On the application side, \textsc{Stark}~\cite{stark} targets robotics, providing built-in constraints and priors common in manipulation (e.g., ball joints, hinges).

Despite being broadly applied across different materials, contact models, and application domains, frameworks that employ IPC are largely element-centric: 
they are typically specialized to a fixed set of geometric primitive types and discretizations, most commonly triangle and tetrahedral meshes. Extending such systems with a new shape representation or introducing a new energy generally requires substantial modifications to the core system. In particular, users must manually derive and implement the corresponding local energy, gradient, and Hessian, as well as ensure their correctness and efficient matrix assembly into global sparse structures.

This limited extensibility is further amplified in GPU implementations. To achieve high performance, GPU kernels often rely on fixed memory layouts, compile-time constants, and hand-tuned data structures. As a result, extending an existing framework to support new energies or representations on the GPU typically demands significant additional engineering effort, including redesigning memory layouts and writing specialized kernels.

\subsection{Symbolic and automatic differentiation}
To reduce the burden of manually deriving gradients and Hessians, many simulation systems rely on \emph{automatic differentiation} or \emph{symbolic differentiation with form compilation}~\cite{evaluating-derivatives}. Form-compilation approaches, such as \textsc{FEniCS} and its Unified Form Language (UFL)~\cite{fenics,ufl}, represent variational forms symbolically in a domain-specific intermediate representation and generate specialized low-level code for derivative evaluation and assembly.
\textsc{PolyFEM}~\cite{polyfem} uses automatic differentiation to obtain gradients and Hessians for energies and discretizations. \textsc{TinyAD}~\cite{tinyad} provides a lightweight C++ automatic differentiation layer tailored to geometry-processing energies (e.g., parameterization and smoothing), enabling concise local formulations and rapid prototyping.

Tape- or trace-based implementations of automatic differentiation may incur runtime overhead due to temporary objects, memory traffic, and limited algebraic simplification. To mitigate these costs, several systems adopt differentiation with \emph{code generation}~\cite{symx,mesh-based-symbolic, sparsity-specific}: the energy is expressed in a restricted language, algebraic simplifications are performed ahead of time, and specialized kernels for local gradients, Hessians, and assembly are emitted for the target backend (CPU or GPU). 
Classical IPC implementations~\cite{ipc} also rely on symbolic differentiation for the derivatives of barrier energies.
However, existing symbolic systems are typically specialized to fixed element types and interaction patterns (what pairs of element types can participate in the same computation), making it difficult to extend them to new geometric representations or contact formulations without modifying the underlying compiler or code generator.

To simplify global matrix assembly, some frameworks~\cite{mesh-based-symbolic} further expose interfaces that bind energies to specific element types (e.g., triangles, edges, tetrahedra), enabling automatic sparsity inference and consistent insertion of local contributions into the global system. However, this again limits those frameworks to the known primitives defined within the system. 

Other frameworks~\cite{symx, tinyad} instead require the user to explicitly provide the assembly structure, such as the relevant connectivity and how local quantities map into global degrees of freedom. While this offers more flexibility than hard-coding assembly around fixed primitive types, it still places the burden of structural specification on the user. This becomes especially problematic for energies that admit multiple parameterizations (e.g. collision between a soft body and affine body would normally admit $2\times2 = 4$ different parameterizations for each collision energy). Each parameterization is typically realized as a separate kernel with its own connectivity, requiring the user to manually partition cases, invoke different kernels, and assemble results across multiple connectivity structures. Thus, structural variation is handled through duplicated user effort rather than a unified representation.

Beyond mesh-centric toolchains, general-purpose differentiable programming systems such as \textsc{PyTorch}~\cite{pytorch}, \textsc{DiffTaichi}~\cite{difftaichi}, and \textsc{Warp}~\cite{warp} have been applied to energy-based simulation. These systems excel in dense or grid-structured settings and support sparse computations to varying degrees; however, assembling large, irregular mesh operators typically requires additional domain-specific infrastructure—such as custom kernels, data layouts, and explicit sparsity management—which dedicated geometry and simulation frameworks often provide.

\subsection{Relational mesh representations and DSLs}
Representing meshes as relational data is common in prior work on mesh-based systems, although this structure is often implicit. For example, relationships like triangle to vertices are often encoded through array-based data layouts and indexing schemes rather than explicitly modeled as relations.

Domain-specific languages (DSL) such as \textsc{Ebb}~\cite{ebb} and \textsc{Simit}~\cite{simit} go further by defining explicit relational data models.
In \textsc{Simit}, meshes are modeled using hypergraphs.  (For instance, a tetrahedron is a hyper-edge containing 4 vertices.)
In \textsc{Ebb}, meshes are modeled as relational tables, interconnected by columns of references to other tables.
Both systems support the specification of imperative, local, stencil computations patterned over the data structure.
YASPS follows these prior systems in adopting a relational data model, but differs by specifying simulations in terms of declarative energies, rather than imperative computations.
In order to compute the simulation, YASPS must differentiate these energies.  Doing so requires addressing how differentiation interacts with relational operations like \texttt{JOIN} and \texttt{UNION}.

\section{System Overview}

\begin{figure*}
  \centering
  \includegraphics[width=\linewidth]{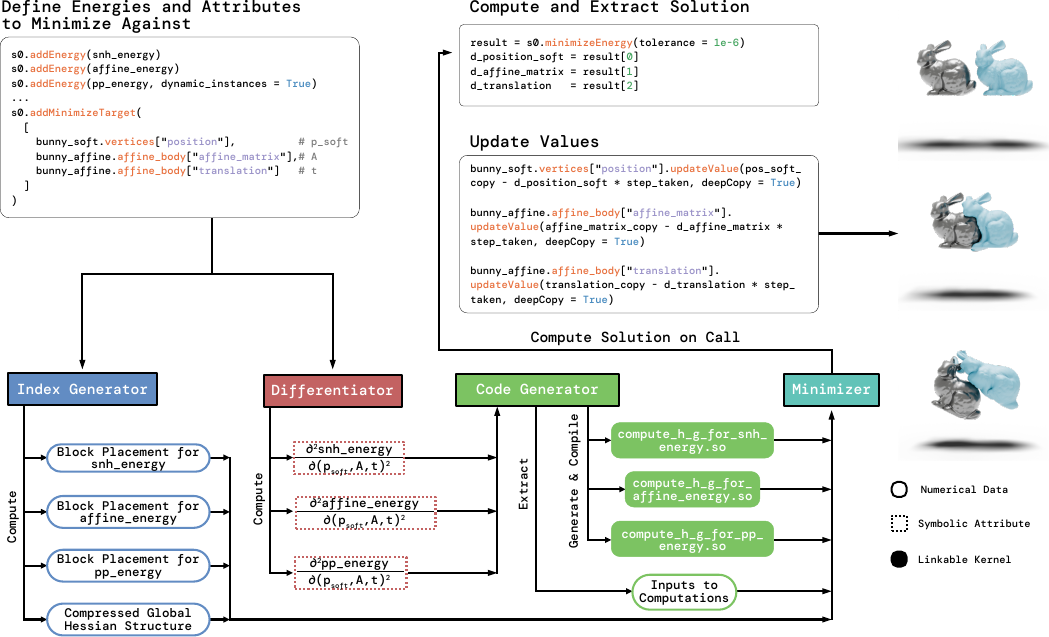}
  \vspace{-15pt}
  \caption{An overview of how the YASPS' backend works once user has specified the energies in the scene and the attributes to minimize against. The symbolic attributes of the energy computation and the minimization target will be passed to YASPS' index generator and differentiator. The differentiator will compute the symbolic Hessian and gradient while the index generator computes the global Hessian and gradient structure, as well as how to place the local Hessian and gradient to the global ones.
  The code generator will be invoked once the symbolic Hessian and gradient is computed, generating and compiling kernels responsible for the computation. Code generator will also extract the pointers to the data so that when the generated kernels are invoked, they know exactly where the input data lies. When user wants to extract the solution, YASPS' minimizer will then use the generated kernels and data to compute the solution to the system $Hx = g$. Here we omit the construction of the scene and energies, which are mostly on the frontend. The detailed example regarding the frontend is shown in Sec.~\ref{section-frontend}.}
  \label{figure-system-overview}
\end{figure*}

YASPS is designed to make it easy to introduce new parameterizations and energy formulations while retaining high performance. Given a user-defined energy, YASPS automatically computes the corresponding local gradient $g_\text{local}$ and local Hessian matrix $H_\text{local}$ (which is also projected to positive semi-definite automatically by perturbing the Eigenvalues of each local Hessian matrix), assembles the global gradient $g$ and global Hessian matrix $H$, and solves the resulting linear system $Hx = g$ as part of a Newton-type optimization loop. 

YASPS provides a frontend implemented in \textsc{Python} that exposes the following core functionality to users:
\begin{itemize}
    \item \textbf{Scene and mesh structure specification.}
    Users define the structure of a scene by adding meshes, creating primitives associated with each mesh (e.g., vertices, triangles, tetrahedra), and declaring the topological relationships between primitives (e.g., triangle--vertex adjacency). See Sec.~\ref{section-system-setup}.

    \item \textbf{Symbolic attribute definition and computation.}
    Attributes can be defined symbolically at the scene, mesh, or primitive level. Users may then express computations directly over these symbolic attributes, including the use of two YASPS-introduced operators, \texttt{JOIN} (Sec.~\ref{section-join-intro}) and \texttt{UNION} (Sec.~\ref{section-union-intro}), which enable structured aggregation across topological relationships. Any computation over YASPS' symbolic attributes produces a new symbolic attribute, allowing complex expressions to be constructed compositionally.

    \item \textbf{Energy definition.}
    Users designate the symbolic attributes as energy terms that guide simulation, and specify the attributes with respect to which the energy is minimized. See Sec.~\ref{section-energy}.

    \item \textbf{Solution extraction.}
    Once all energy terms in the scene are defined, users may compute and extract the solution $x$ of the linear system $Hx = g$. This solution can then be used as the update direction in a Newton-type optimization method.
\end{itemize}

On the backend, YASPS consists of several systems that operate directly on symbolic attributes to construct the linear system and perform the numerical solve:
\begin{itemize}
    \item \textbf{Symbolic differentiator.}
    Once users specify the energy terms and the attributes with respect to which differentiation is performed, the symbolic differentiator computes the corresponding first- and second-order derivatives. The resulting gradient and Hessian are represented as symbolic attributes. See Sec.~\ref{section-differentiation}.

    \item \textbf{Index generator.}
    Given the symbolic gradient and Hessian, the index generator determines how local numerical contributions are assembled into the global gradient and Hessian (Sec.~\ref{section-index}). It is also responsible for generating the compressed global Hessian structure (Appendix~\ref{section-hessian-compression}) to minimize storage and computational cost.

    \item \textbf{Code generator and compiler.}
    For any symbolic attribute, the code generator translates the corresponding computation graph into \textsc{CUDA} kernels and links them into the main program (Sec.~\ref{section-code-generation}). For symbolic attributes computing gradients and Hessians, specialized kernels are generated for projecting the local Hessian matrix to positive semi-definite through eigendecomposition (Sec.~\ref{section-hessian-code}). Those specialized kernels will also compress and scatter the locally computed Hessian and gradient to the global Hessian and gradient (Appendix~\ref{section-hessian-compression}). 
    Additionally, the code generator will support the kernel with the correct numerical data source for the actual execution.

    \item \textbf{Minimizer.}
    Given the symbolic Hessian and gradient, together with the index mappings produced by the index generator and the compiled kernel to perform computation, the minimizer will first invoke the compiled code to compute and store the Hessian and gradient, then a solver kernel will be compiled and invoked to compute the solution to $Hx = g$ by running a conjugate gradient solver. See Sec.~\ref{section-solver}.
\end{itemize}

To remain agnostic to specific parameterizations and energy formulations, YASPS deliberately excludes several components that are typically tightly coupled to particular representations:
\begin{itemize}
    \item \textbf{Collision detection and continuous collision detection.}
    Since the trajectory and geometry of a primitive are determined by its parameterization, supporting arbitrary user-defined parameterizations makes it impractical for YASPS to provide a built-in collision detection or continuous collision detection module. Instead, YASPS is designed to interoperate with external collision handling systems that supply contact constraints or energies compatible with the chosen parameterization. In the examples presented in Sec.~\ref{section-examples}, we use a CCD implementation specialized for linear parameterizations.

    \item \textbf{Newton stepping.}
    YASPS does not prescribe a specific Newton stepping or line search strategy. Instead, it provides the solution of the linear system, which users may integrate into customized stepping, damping, or termination schemes.
\end{itemize}

Figure \ref{figure-system-overview} shows how the frontend call activates our backend. A running example of how the scene and mesh are constructed, how the two new operators \texttt{JOIN} and \texttt{UNION} are represented, and how attributes and energies are specified on the frontend will be shown in the next section.
\section{Frontend Overview}\label{section-frontend}
We now detail the frontend of YASPS by showing a running example of two bunnies, one modeled with soft material, the other rigid with affine body parameterization, colliding (see Fig.~\ref{figure-system-overview} for the rendered scene). This section will also define the two new operators \texttt{JOIN} and \texttt{UNION} and how they are used in our system. We also show the core syntax grammar covering Secs.~\ref{section-system-setup}-\ref{section-union-intro} in Appendix~\ref{section-core-syntax}.

\subsection{Setup}\label{section-system-setup}

At a high level, YASPS organizes a simulation into three layers of abstraction: scenes, meshes, and primitive types.

A scene represents the global scope of a simulation and contains one or more meshes.

In YASPS, the construction of the scene and meshes is performed using the following code:

\begin{pythonBlock}{}
from yasps import scene
s0@* =*@ |!scene!|("scene0")
bunny_affine@* =*@ s0.|!addMesh!|("bunny_affine")
bunny_soft@* =*@ s0.|!addMesh!|("bunny_soft")
\end{pythonBlock}

Each mesh contains a collection of user-defined primitive types, such as vertices, tetrahedra, or affine bodies. Below, we define these primitive types for the affine bunny mesh:

\begin{pythonBlock}{}
bunny_affine.|!addPrimitive!|("vertices", numInstances = NUM_BUNNY_VERTEX)
bunny_affine.|!addPrimitive!|("affine_body", numInstances = ?@1@?)
bunny_affine.|!addPrimitive!|("tets", numInstances = NUM_TETS)
\end{pythonBlock}

These primitive types can later be accessed in the following way:

\begin{pythonBlock}{}
bunny_affine.|?vertices?|
bunny_affine.|?affine_body?|
bunny_affine.|?tets?|
\end{pythonBlock}

\begin{figure}
    \centering
    \includegraphics[width=1.0\columnwidth]{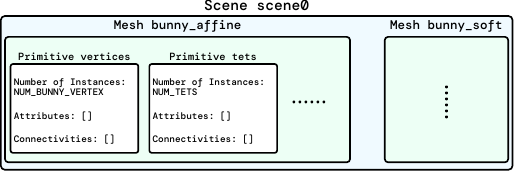}
    \vspace{-20pt}
    \caption{In YASPS, scenes, meshes and primitive types themselves only contain the next level entities and additionally some metadata. The primitive types will also contain a list of attributes (Sec.~\ref{section-attribute-intro}) and connectivities (Sec.~\ref{section-connectivity-intro}), which are empty when the primitive is initialized.}
    \label{figure-4-1-scene-mesh-primitive}
\end{figure}

When a primitive type is first created, it only needs to specify the number of instances it contains. For example, in our current setting, as the affine bunny is only controlled by one affine body, the \texttt{numInstances} is set to 1. The scene structure from the previous codes is illustrated in Fig.~\ref{figure-4-1-scene-mesh-primitive}.

Importantly, scenes, meshes and primitive types in YASPS do not carry intrinsic geometric or physical semantics. In particular, no inherent properties, such as positions, are associated with meshes or primitive types by default. Instead, such properties are expressed through attributes defined on top of them.

\subsection{Attributes}\label{section-attribute-intro}
In YASPS, scene, mesh and primitive types may bind multiple attributes. Formally, let's first define the notions of primitive type and attribute. 

A primitive type is simply a collection of instances. Let \(X\) denote a primitive type with \(n\) instances, and let \([n] := \{1,\dots,n_X\}\) be its index set, an attribute on \(X\) is then a function
\[
\alpha_X : [n] \to \mathbb{R}^{r \times c},
\]
where $r$ and $c$ represents the attribute's per-instance number of rows and columns.

We also use the stacked tensor notation
\[
\boldsymbol{\alpha}_X \in \mathbb{R}^{n \times r \times c}, \qquad
[\boldsymbol{\alpha}_X]_i := \alpha_X(i).
\]

On the frontend, we can define attributes on the affine body and supply its numerical values in the following way:

\begin{pythonBlock}{}
bunny_affine.|?affine_body?|.|!addAttribute!|("affine_matrix", rows = ?@3@?, cols = ?@3@?)
bunny_affine.|?affine_body?|["affine_matrix"].|!updateValue!|(np.|!eye!|(?@3@?, dtype=np.|?float64?|))
bunny_affine.|?affine_body?|.|!addAttribute!|("translation", rows = ?@3@?, cols = ?@1@?)
bunny_affine.|?affine_body?|["translation"].|!updateValue!|(np.|!array!|([?@0.0@?, ?@0.0@?, ?@0.0@?], dtype=np.|?float64?|))
\end{pythonBlock}

As there is only one instance of affine body on the mesh, we only need to supply a data array with total size $1\times3\times3$ for the numerical value of the affine matrix. 

Similarly, we can define rest position of all the vertices and update the value in the following way:
\begin{pythonBlock}{}
bunny_affine.|?vertices?|.|!addConstant!|("rest_position", rows = ?@3@?, cols = ?@1@?)
bunny_affine.|?vertices?|["rest_position"].|!updateValue!|(BUNNY_VERTEX_POSITIONS)
\end{pythonBlock}

\begin{figure}
    \centering
    \includegraphics[width=1.0\columnwidth]{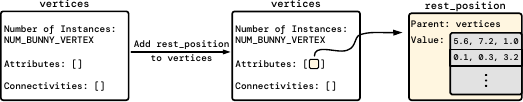}
    \vspace{-20pt}
    \caption{\textsc{Attributes.} When attribute \texttt{rest\_position} is added to the primitive type \texttt{vertices}, YASPS initializes an array of size \texttt{NUM\_BUNNY\_VERTEX} $\times 3 \times 1$, and saves it under the \texttt{rest\_position} attribute, whose reference pointer is then saved under the \texttt{vertices} primitive type. When the user updates the values of this attribute, the new values will be copied into this array.}
    \label{figure-4-2-attributes}
\end{figure}

Here, \texttt{BUNNY\_VERTEX\_POSITIONS} is a flattened array with size $\texttt{NUM\_BUNNY\_VERTEX} \times 3 \times 1$ as each \texttt{rest\_position} is an attribute of size $3\times1$. 

When calling \texttt{addConstant} or \texttt{addAttribute} on a primitive type with number of instances $N$ with the signature (\texttt{name}, \texttt{number of rows} $r$, \texttt{number of columns} $c$), YASPS will initialize a 1-D array of size $N\times r\times c$ on GPU, illustrated in Fig.~\ref{figure-4-2-attributes}. Subsequent value updates to those attributes will directly write to those arrays.

Notably we used two functions \texttt{addAttribute} and \texttt{addConstant} in the previous two code blocks. The naming does not suggest that attributes created through \texttt{addConstant} need to remain unchanged numerically throughout the simulation. It only means those attributes do not participate in differentiation (i.e., differentiation w.r.t. those attributes will always result in 0).


\subsection{Computation of Attributes}
\begin{figure}
    \centering
    \includegraphics[width=1.0\columnwidth]{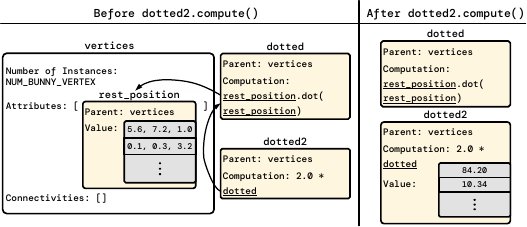}
    \vspace{-20pt}
    \caption{\textsc{Attributes computation.} Any computation of attributes will directly result in another attribute. Attributes not initialized using the \texttt{addAttribute} function will not be added under the corresponding primitive types, but their parents can still be determined immediately based on the lineage of the attributes participating in the computation, hence the dimension of the numeric values can be immediately determined. Notice that after computing \texttt{dotted2}, there is no value attached to \texttt{dotted} even though it is an attribute used in computation as YASPS is designed to not materialize any intermediate computation.}
    \label{figure-4-3-attribute-compute}
\end{figure}

In YASPS, attribute expressions are represented symbolically. For example, the following code:

\begin{pythonBlock}{}
rp@* =*@ bunny_affine.|?vertices?|["rest_position"] # per-vertex, shape 3x1
dotted@* =*@ rp.|!dot!|(rp) # per-vertex, shape 1x1
dotted2@* =*@ ?@2.0@?@* **@ dotted # per-vertex, shape 1x1
\end{pythonBlock}
\noindent will store the computation graph under a new attribute. This new attribute is not stored under the \texttt{vertices} primitive type, but has its parent pointed at \texttt{vertices}, as shown in the middle table in Fig.~\ref{figure-4-3-attribute-compute}.

To obtain the numerical value of an attribute, we can invoke the following function:
\begin{pythonBlock}{}
dotted2.|!compute!|().|?value?| # flattened array of size  NUM_BUNNY_VERTEX * 1 * 1
\end{pythonBlock}

This operation will invoke our code generator that traces the computation graph stored under \texttt{dotted2} all the way to the root variable \texttt{rest\_position}, and compile and execute a piece of kernel corresponding to the computation trace to obtain the numerical result. YASPS also avoids materializing any intermediate variable as the additional writes and reads are more expensive than the computation itself, hence the intermediate variable \texttt{dotted} is not materialized during the computation, as illustrated in Fig.~\ref{figure-4-3-attribute-compute}.

Although an attribute, when evaluated to numerical results,  corresponds to a stacked tensor \(\boldsymbol{\alpha}_X \in \mathbb{R}^{n_X\times r\times c}\),
the symbolic attribute itself only represents the \emph{per-instance} expression of shape \(r\times c\).
When an operator \(f\) is applied to a symbolic attribute on primitive type \(X\), it is implicitly broadcast over all instances:
the generated code for the computation tree stored on the attribute will evaluate \(f(\alpha_X(i))\) for every \(i\in \{1,\ldots,n_X\}\).



However, this also means that for any attribute $\alpha_X$, YASPS needs to know exactly what the primitive type $X$ is. Thus, to preserve this simplicity, YASPS enforces a key rule: all attributes participating in the same computation must share a common \emph{lineage}. The only exceptions are the two new operators \texttt{JOIN} and \texttt{UNION}, which we will introduce soon.

\begin{figure}
    \centering
    \includegraphics[width=1.0\columnwidth]{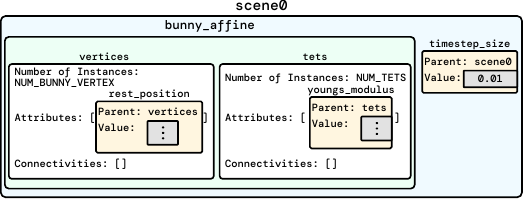}
    \vspace{-20pt}
    \caption{\textsc{Lineage of attributes.} A simple scene hierarchy illustrating an affine-body mesh. Attributes that share the same lineage (within the same box) can participate in the same computation. By contrast, attributes from different hierarchies cannot be directly combined until their relationship is made explicit through a relational operator.}
    \label{figure-4-3-attribute-hierachy}
\end{figure}

To understand this rule, consider our current setup, as shown in Fig.~\ref{figure-4-3-attribute-hierachy}. Any attribute defined on a higher level in the hierarchy, such as the scene attribute \texttt{timestep\_size} or mesh-level attributes, may directly participate in computations involving attributes defined on the primitive types beneath them. Intuitively, a scene-level attribute influences everything within the scene, and a mesh-level attribute influences all primitive types belonging to that mesh. 

Likewise, attributes defined on the same primitive type can freely interact with one another. When a computation involves multiple attributes, YASPS will then determine the deepest entity that lies on the shared lineage chain, and attaches the resulting attribute to that entity, ensuring consistent scoping and lineage.

In contrast, attributes belonging to different primitive types, such as \texttt{youngs\_modulus} and \texttt{rest\_position}, cannot appear in the same computation. Attributes defined on one mesh should also not affect attributes defined on another mesh.

Yet in practice, many quantities do conceptually depend on attributes from different primitive types. For instance, a vertex’s current position depends on the affine body’s affine matrix, and a tetrahedron’s current position depends on the current positions of its four vertices. 

Such dependencies exist because there is a topological relationship between different types of primitives. To make such a relationship explicit, YASPS uses connectivity.

\subsection{Connectivity}\label{section-connectivity-intro}
\begin{figure}
    \centering
    \includegraphics[width=1.0\columnwidth]{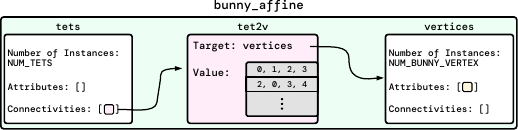}
    \vspace{-20pt}
    \caption{\textsc{Connectivities.} When connectivity is added to a primitive type, we create a \texttt{connectivity} object and put it in the list of connectivities under the corresponding primitive type. The connectivity object simply contains a list, which indicates the relation between two primitive types, and the target, which points to the other primitive type.}
    \label{figure-4-4-connectivity}
\end{figure}
To express the relationships between different primitive types, we need to introduce a mapping from the index set of one primitive type to tuples of indices from another primitive type. In YASPS we refer to such a mapping as a \texttt{connectivity}.

Formally, let \(A\) and \(B\) be primitive types with \(n_A\) and \(n_B\) instances, respectively. Define the index sets \(I_A := \{1,\ldots,n_A\}\) and \(I_B := \{1,\ldots,n_B\}\).

A connectivity of arity \(k \ge 1\) from \(A\) to \(B\) is a function
\[
C:\ I_A \;\to\; I_B^k, \quad
C(i) = \big(j_1(i),\ldots,j_k(i)\big)
\]
where each \(j_\ell(i) \in I_B\) is the index of the \(\ell\)-th instance of \(B\) referenced by instance \(i \in I_A\), and $I_B^k$ denotes the Cartesian product \(I_B \times \cdots \times I_B\) ($k$ copies), i.e.\ the set of ordered $k$-tuples of indices from \(I_B\).

As an example, consider \texttt{tetrahedra} and \texttt{vertices}. 
Since each tetrahedron should connect to 4 vertices, the connectivity 
\(C_{\mathrm{tet2v}}\) maps every index \(i \in I_{\texttt{tets}}\) to a 4-tuple of 
vertex indices in \(I_{\texttt{vertices}}\), i.e.,
\[
C_{\mathrm{tet2v}}(i) = \big(j_1(i), j_2(i), j_3(i), j_4(i)\big), 
\qquad j_\ell(i) \in I_{\texttt{vertices}}
\]

In YASPS, connectivity is created through the following code:
\begin{pythonBlock}{}
# create connectivity between vertices and affine body
bv2abd@* =*@ bunny_affine.|?vertices?|.|!addConnectivity!|("bunny_vertex_to_affine_body", bunny_affine.|?affine_body?|, [?@0@?]@* **@ (NUM_BUNNY_VERTEX), ?@1@?)
# create connectivity between tetrahedron and vertices
affine_tet2v@* =*@ bunny_affine.|?tets?|.|!addConnectivity!|("tet2v", bunny_affine.|?vertices?|, TET_INDICES, ?@4@?)
\end{pythonBlock}

The four arguments of the \texttt{addConnectivity} function are the name, the target primitive type, the connectivity list and the arity. 

The first connectivity \texttt{bv2abd} is defined between the vertices primitive type and the affine body primitive type. Since each vertex corresponds to only one affine body, and there is only one affine body under the mesh, the third argument to the function is an array of zeros.

The second connectivity \texttt{affine\_tet2v} is defined between the tetrahedra of the bunny and the vertices of the bunny. The variable \texttt{TET\_INDICES} is a list of size $\texttt{NUM\_TETS}\times 4$ that contains the actual index mapping from tetrahedra to vertices. The illustration of how this connectivity is stored is shown in Fig.~\ref{figure-4-4-connectivity}.

While \texttt{connectivity} provides the relationship between two primitive types, the attributes defined on those primitives still have different parents and cannot participate in the same computation. As such, we need to introduce a method that is able to transfer an attribute from one primitive type to another.

\subsection{JOIN}\label{section-join-intro}

\begin{figure*}[t]
  \centering
  \includegraphics[width=\textwidth]{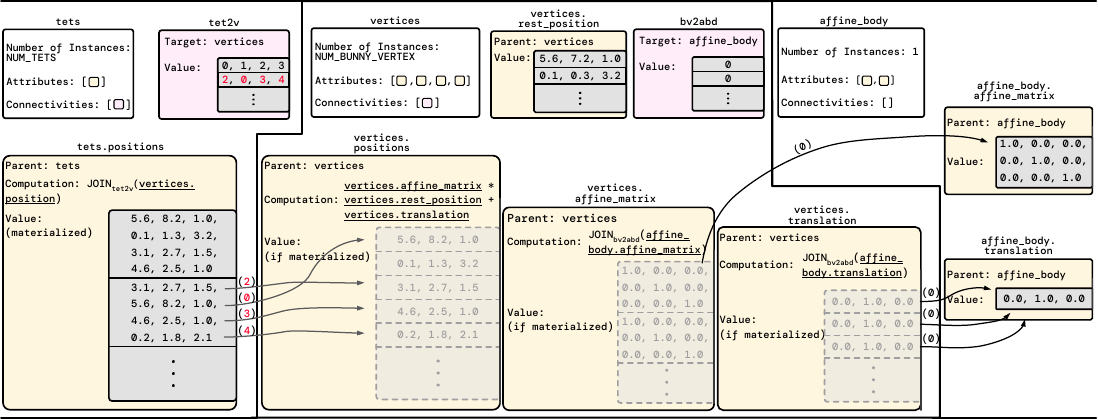}
  \vspace{-20pt}
  \caption{\textsc{Joins.} Illustration of how \texttt{tets.position} is computed using \texttt{JOIN} operators. The \texttt{JOIN} operator uses the connectivity (e.g., \texttt{tet2v}) to gather the corresponding entries from the target attribute \texttt{vertices.position}. Intermediate attributes, such as \texttt{vertices.position}, \texttt{vertices.affine\_matrix}, and \texttt{vertices.translation}, are not materialized during this computation. Instead, when a required attribute is itself defined by a computation, YASPS evaluates the specific entries referenced by the \texttt{JOIN} on the fly, temporarily materializing only those rows in memory before assembling the final values for \texttt{tets.position}.}
  \label{figure-4-5-join}
\end{figure*}
\subsubsection{Motivation}

Many simulation quantities are defined in terms of attributes that originate on different primitive types. 
To handle those cases systematically, we require a mechanism that makes cross-primitive access to attributes explicit. 

Given a \texttt{connectivity} that relates two primitive types, we introduce an operator that uses this
relationship to gather an attribute defined on one primitive type and make it available on another.
We call this operator \texttt{JOIN}.

\subsubsection{Definition}
A \texttt{JOIN} uses a \texttt{connectivity} to gather attributes from one primitive type and make them available on another.

Formally, let \(C : I_A \to I_B^k\) be a \texttt{connectivity} from primitive type \(A\) to primitive type \(B\), 
and let \(\alpha_B : I_B \to \mathbb{R}^{r \times c}\) be an attribute on \(B\). 
The result of the \texttt{JOIN} of \(\alpha_B\) through \(C\) is an attribute on \(A\):
\begin{equation}
\mathrm{JOIN}_{C}(\alpha_B)(i)
  = \big(\,\alpha_B(j_1(i)), \ldots, \alpha_B(j_k(i))\,\big)
  \;\in\; \mathbb{R}^{k \times (rc)}
\label{equation:join-definition}
\end{equation}
where \((j_1(i),\ldots,j_k(i)) = C(i)\).
In other words, each instance of \(A\) collects the attributes of the \(k\) referenced instances of \(B\), stacked along a new leading dimension of size \(k\). To avoid tensor representation, we also flatten the last two dimensions so that each instance $\mathrm{JOIN}_{C}(\alpha_B)(i)$ is a 2D attribute.

By construction, \(\mathrm{JOIN}_{C}(\alpha_B)\) is owned by \(A\), so it can participate in expressions
with other attributes on \(A\) under YASPS' lineage constraints. The \texttt{JOIN} operation also does not abide by the lineage rule as it is designed specifically to operate on two different primitive types. However, the resulting attribute still does.

\subsubsection{Code}
On the frontend, a \texttt{JOIN} is performed by creating a new attribute on the source primitive type
and specifying (i) the connectivity (\texttt{through}) and (ii) the attribute to gather (\texttt{source}):

\begin{pythonBlock}{}
# JOIN affine_body.affine_matrix onto vertices through bv2abd (arity 1)
bva@* =*@ bunny_affine.|?vertices?|.|!addAttribute!|("affine_matrix", through = bv2abd, source = bunny_affine.|?affine_body?|["affine_matrix"]) # bva has dimension 1x9
# JOIN vertices.rest_position onto tets through tet2v (arity 4)
btrp@* =*@ bunny_affine.|?tets?|.|!addAttribute!|("rest_positions", through = affine_tet2v, source = bunny_affine.|?vertices?|["rest_position"]) # btrp has dimension 4x3
\end{pythonBlock}

The first line performs:
\[
\mathrm{JOIN}_{\texttt{bv2abd}}\!\left(\texttt{bunny\_affine.affine\_body["affine\_matrix"]}\right)
\]
and attaches the result to the \texttt{vertices} primitive type. Since \(\texttt{bv2abd}\) has arity \(k=1\)
and the source attribute has shape \(3\times 3\), the per-instance shape of \texttt{bva} is \(1\times 9\)
(i.e.\ \(1\times (3\cdot 3)\)).

Similarly, \texttt{btrp} gathers the \(3\times 1\) rest position of each of the four vertices referenced by a tet,
so its per-instance shape is \(4\times 3\) (i.e.\ \(4\times (3\cdot 1)\)).

Because \texttt{JOIN} flattens the trailing \(r\times c\) block, we may reshape the result back to a matrix form
when needed:

\begin{pythonBlock}{}
bva@* =*@ bva.|!resize!|(?@3@?, ?@3@?) # 1x9 -> 3x3
\end{pythonBlock}

Whenever a user performs a join operation, the newly generated attribute is also a symbolic attribute and can participate in computations the same way any other attributes do. For example, we can now compute the current position of the vertices as follows:

\begin{pythonBlock}{}
bvt@* =*@ ... # get the translation to vertices through join and resize it to 3x1

# perform p = Ap_rest + t
current_position@* =*@ bva@* **@ bunny_affine.|?vertices?|["rest_position"]@* +*@ bvt

# optionally give this attribute a name
bvp@* =*@bunny_affine.|?vertices?|.|!addAttribute!|("position", computed_attribute = current_position)
\end{pythonBlock}

Since \texttt{bvp} is an attribute on \texttt{vertices}, it can be gathered again onto \texttt{tets}:

\begin{pythonBlock}{}
btp@* =*@ bunny_affine.|?tets?|.|!addAttribute!|("positions", through = affine_tet2v, source = bunny_affine.|?vertices?|["position"])
\end{pythonBlock}

This produces, for each \texttt{tet} instance, a \(4\times 3\) matrix whose rows are the positions of its four vertices. We illustrate how the \texttt{JOIN} operator constructs the cross primitive relation and how \texttt{tets.positions} can be materialized in Fig.~\ref{figure-4-5-join}.

With the \texttt{JOIN} operator, we can already express most of the standard relationships in mesh data structures, such as edge–vertex, edge–triangle, or triangle–vertex associations. However, in settings that involve collision, the \texttt{JOIN} operator alone is insufficient. 
\subsection{UNION}\label{section-union-intro}
\begin{figure}
    \centering
    \includegraphics[width=1.0\columnwidth]{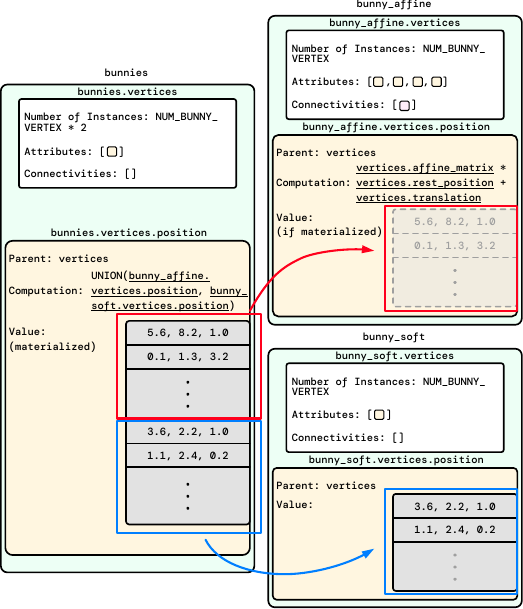}
    \vspace{-10pt}
    \caption{\textsc{Unions.} The \texttt{UNION} attribute, if materialized, stacks the materialized values of the children being unioned. Just like \texttt{JOIN}, this operator can be performed on attributes across different primitive types, or even different meshes, making this operator, in combination of \texttt{JOIN}, particularly suitable for describing computations involving collisions.}
    \label{figure-4-6-union}
\end{figure}

\subsubsection{Motivation}

In simulations that involve contact, or alternative parameterizations, computations may depend on quantities originating from different parameterizations. Returning to the collision energy example, both soft-body and affine-body vertices contribute world-space positions of identical dimension (\(3 \times 1\)), yet their constructions differ fundamentally: a soft-body vertex stores its position directly as a free variable, while an affine-body vertex computes its position from the body’s transform and translation.

If we write separate code paths for each parameterization, the computation graph fragments into multiple versions (free–free, free–ABD, ABD–ABD, ABD-free), each requiring its own derivative kernel and assembly logic. If instead we precompute world‑space positions numerically and then forget where they came from, we lose the ability to propagate derivatives back to their parameters.

We therefore need a way to merge those heterogeneous but shape‑compatible attributes (e.g., soft and ABD positions) into a single differentiable attribute that preserves each element’s dependency trace. This attribute should ensure that computation and differentiation still produces a single attribute instead of fragmenting into multiple independent attributes, which leads to even more fragmentation with subsequent computation or differentiation.

YASPS introduces this mechanism through the \texttt{UNION} operator, which combines attributes from heterogeneous primitive types into a single differentiable attribute.

\subsubsection{Definition}

Suppose we have primitive types \(X_1,\dots,X_m\) with attributes \(\alpha_{X_j} : [n_j] \to \mathbb{R}^{r \times c}\). The \texttt{UNION} of these 
attributes is the function
\[
\mathrm{UNION}\big(\alpha_{X_1},\dots,\alpha_{X_m}\big) 
:\ \bigsqcup_{j=1}^m [n_j] \;\to\; \mathbb{R}^{r \times c}
\]
where $\bigsqcup$ denotes a disjoint union of index sets.  An element of the
domain is a tagged pair $(j,i)$ with $i \in [n_j]$, and the mapping is defined
by
\[
\mathrm{UNION}(\alpha_{X_1},\dots,\alpha_{X_m})(j,i)
   = \alpha_{X_j}(i)
\]

Although in the implementation the pair $(j,i)$ is compressed into a single integer index, the mapping is reversible and can be recovered as $(j,i)$ during runtime evaluation.

Unlike \texttt{JOIN}, where the destination primitive type of the resulting attribute is unambiguous, the result of a \texttt{UNION} naturally spans multiple primitive types. This multi-origin correspondence appears to violate the rule introduced earlier, that all attributes participating in a computation must share a common lineage.

YASPS resolves this by introducing a dedicated construct, the \texttt{primitiveUnion}. A \texttt{primitiveUnion} is a new primitive type that explicitly declares which \texttt{primitive}s and \texttt{primitiveUnion}s it unifies, and exists one level below \texttt{mesh}. Attributes defined on this entity are gathered from the constituent primitives through the \texttt{UNION} operator and attached to the new \texttt{primitiveUnion} rather than to the original primitive types. This design preserves the single-lineage invariant while allowing computations to operate over unified attributes spanning heterogeneous representations.

The \texttt{UNION} operator itself does not abide by the lineage rule by design. However, as the resulting attribute now resides on the \texttt{primitiveUnion}, it can only perform computation with other attributes defined on the same \texttt{primitiveUnion} or the attributes of the ancestors.

\subsubsection{Code}
Consider the example in Fig.~\ref{figure-energy-composition} where we want to formulate a point-point collision energy. As there are two types of vertices in the scene (soft bunny vertices and affine bunny vertices), we therefore want to create a unioned representation of vertices. To do this, we first create the \texttt{primitiveUnion} under a new mesh:

\begin{pythonBlock}{}
bunnies@* =*@ s0.|!addMesh!|("bunnies")
bunnies.|!addPrimitiveUnion!|("vertices", [bunny_affine.|?vertices?|, bunny_soft.|?vertices?|])
\end{pythonBlock}

We then add attributes that are unioned:
\begin{pythonBlock}{}
bunnies.|?vertices?|.|!addAttribute!|("rest_position")
bunnies.|?vertices?|.|!addAttribute!|("position")
\end{pythonBlock}

Note that this \texttt{addAttribute} function accepts a string input. This means attributes with those names must be available in both primitive types \texttt{bunny\_affine.vertices} and \texttt{bunny\_soft.vertices}. Additionally, those attributes must have the same shape, as enforced by the definition of the \texttt{UNION} operation. As demonstrated in Fig.~\ref{figure-4-6-union}, once we add the attribute \texttt{position} to \texttt{bunnies.vertices}, YASPS will create a new attribute under the primitive union. This new attribute's materialized value will be the stacked materialized values of the children \texttt{bunny\_affine.vertices.position} and \texttt{bunny\_soft.vertices.position}. And just like before, materializing this \texttt{UNION} attribute will not result in the immediate materialization of the children attributes. 

Since the collision energy depends on exactly two position vectors, we introduce a dedicated primitive type to represent point-point collision pairs:

\begin{pythonBlock}{}
bunnies.|!addPrimitive!|("pp", numInstances = ?@0@?, isDynamic = !@True@!) # for point point collision
pp2v@* =*@ bunnies.|?pp?|.|!addConnectivity!|("pp2v", bunnies.|?vertices?|, [], ?@2@?)
\end{pythonBlock}

Unlike the primitives defined earlier, the number of point-point pairs is not known at construction time, as it depends on the runtime collision detection results. We therefore initialize \texttt{pp} with \texttt{numInstances=0} and mark the primitive type as dynamic. For the same reason, we create the connectivity \texttt{pp2v} with an empty index list, specifying only its arity \(2\); its contents will be populated at runtime once the set of collision pairs is determined.

We can then apply \texttt{JOIN} to gather the two positions for each pair from the unioned vertex representation:

\begin{pythonBlock}{}
pp_positions@* =*@ bunnies.|?pp?|.|!addAttribute!|("positions", through = pp2v, source = bunnies.|?vertices?|["position"])
\end{pythonBlock}

The resulting attribute \texttt{pp\_positions} has per-instance shape \(2\times 3\) (two 3D points per pair) and can be used directly in downstream computations. As an example, we define a point-point barrier energy, and add the attribute back to \texttt{bunnies.pp} with a given name:

\begin{pythonBlock}{}
def*| point_point|*(|@position@|,|@ dHat@|,|@ kappa@|):
  p0@* =*@ position.|!row!|(?@0@?)
  p1@* =*@ position.|!row!|(?@1@?)
  d@* =*@ (p1@* -*@ p0).|!dot!|(p1@* -*@ p0)
  I5@* =*@ d@* /*@ dHat
  lenE@* =*@ d@* -*@ dHat
  I5log@* =*@ I5.|!log!|()
  return kappa@* **@ lenE@* **@ lenE@* **@ I5log@* **@ I5log

pp@* =*@ |!point_point!|(pp_positions, DHAT, KAPPA)
pp_energy@* =*@ bunnies.|?pp?|.|!addAttribute!|("point_point", computed_attribute = pp)
\end{pythonBlock}
\subsection{Energies}\label{section-energy}

So far we have shown how to construct symbolic computations using attributes. To run a simulation step, we must additionally 
\begin{enumerate}
    \item Define and register system energies using computed attributes.
    \item Specify the attributes that are treated as optimization variables.
\end{enumerate}

In the previous subsection we defined an attribute \texttt{pp\_energy} that evaluates the point-point barrier energy per collision pair. We now register it as an energy term in the scene:

\begin{pythonBlock}{}
s0.|!addEnergy!|(pp_energy, dynamic_instances = !@True@!)
......|? # add other energies
\end{pythonBlock}

Here \texttt{dynamic\_instances=True} indicates that the primitive instances (e.g., the set of point-point pairs) may change at runtime, so the number of energy contributions is not fixed at construction.

Since the soft bunny is directly controlled by the position attribute of each vertex, and the affine bunny is controlled by the affine matrix and translation vector of the affine body, we therefore want to minimize the energies in the scene w.r.t. those attributes:

\begin{pythonBlock}{}
s0.|!addMinimizeTarget!|(
  [bunny_soft.|?vertices?|["position"],
   bunny_affine.|?affine_body?|["affine_matrix"],
   bunny_affine.|?affine_body?|["translation"]]
)
\end{pythonBlock}

Once energy terms and minimization targets are declared, YASPS can follow the workflow in Fig.~\ref{figure-system-overview} to (1) symbolically differentiate the registered energies with respect to the targets, and (2) determine the sparsity structure of the global Hessian.

\subsection{Solution Extraction}\label{section-solution-extraction}
At each minimization step, YASPS assembles the global gradient vector \(g\) and the projected Hessian matrix \(H\), and solves a linear system to obtain an update direction. Concretely, the update \(\Delta x\) is obtained by solving $H\Delta x = g$.
We did not negate the right-hand side $g$, but users can simply negate $\Delta x$ to get the true update direction.

On the frontend, the (negative of) update directions can be obtained by calling:

\begin{pythonBlock}{}
result@* =*@ s0.|!minimizeEnergy!|(tolerance = ?@1e-6@?)
\end{pythonBlock}

Since we previously registered three minimization targets, \texttt{result} is a list of length three, where each entry contains the update direction corresponding to one target attribute in the same order as passed to \texttt{addMinimizeTarget}.

The first time \texttt{minimizeEnergy} is invoked on a scene, YASPS triggers code generation to produce specialized kernels that numerically evaluate the (symbolically derived) derivatives and assemble their contributions into the global \(H\) and \(g\). Subsequent calls reuse these generated kernels.

After assembling \(H\) and \(g\), the minimizer computes the solution using a conjugate gradient solver and returns the resulting update directions in \texttt{result}.
\section{Symbolic Differentiation}\label{section-differentiation}
To compute gradient and Hessian for the Newton iterations, YASPS performs symbolic differentiation and then generates GPU kernels to numerically evaluate the resulting expressions.

The design of YASPS' differentiation system must satisfy two requirements and two performance needs.

The two requirements are:
\begin{itemize}
    \item Differentiating a computation graph in YASPS should yield another graph expressed entirely in terms of operators already defined in YASPS (including \texttt{JOIN} and \texttt{UNION}).
    \item Differentiating a computation involving \texttt{UNION} (i.e., alternative parameterizations) should produce a single unified computation graph rather than multiple branched versions, which leads to exponential growth of such branches with more differentiation and computation.
\end{itemize}

The two performance needs are:
\begin{itemize}
\item Maximize reuse of symbolic intermediates to avoid redundant computation.
\item Produce Hessians with structures that are well-suited for eigendecomposition, 
enabling efficient projection of the local Hessian matrix to positive semi-definite form (required by conjugate gradient solvers).
\end{itemize}

To fulfill the two requirements, we introduce explicit differentiation rules for the \texttt{JOIN} and \texttt{UNION} operators.

To meet the two performance needs, YASPS adopts a two-pass symbolic differentiation scheme to compute both gradients and Hessians efficiently.
During the first pass, YASPS traverses the computation graph to derive local Jacobians, gradients and Hessians.
In the subsequent pass, these symbolic expressions are propagated to assemble the final gradient and Hessian of the energy following the second-order chain rule:
\begin{equation}
\nabla_x^2 f(g(x)) =
J_g(x)^\top \, \nabla^2_g f(g(x)) \, J_g(x)
+ \sum_{j=1}^m \frac{\partial f}{\partial u_j}(g(x)) \,\nabla_x^2 g_j(x),
\label{equation:chain-rule}
\end{equation}
where \(u_j\) denotes the \(j\)-th component of \(g(x)\) and \(J_g(x)\) is the Jacobian of \(g\).

In the remainder of this section, we will first establish the differentiation rules for the two operators \texttt{JOIN} and \texttt{UNION}, followed by an explanation of how the two-pass differentiation scheme satisfies the performance needs.

\subsection{Differentiation of Joined Attributes}\label{section-diff-join}

Because the \texttt{JOIN} operator concatenates attributes without modification, its derivative with respect to the joined attributes is the identity. Specifically, for an attribute 
\(\mathrm{JOIN}_{C}(\alpha_B)(i)\) that gathers elements 
\(\big(\alpha_B(j_1(i)), \ldots, \alpha_B(j_k(i))\big)\), we have
\begin{equation}
\frac{\partial \,\mathrm{JOIN}_{C}(\alpha_B)(i)}
     {\partial \big(\alpha_B(j_1(i)), \ldots, \alpha_B(j_k(i))\big)}
  = I \in 
  \mathbb{R}^{(krc)\times(krc)},
\end{equation}
where \(k\) is the number of joined instances and each instance has shape \(r \times c\).

If each instance of \(\alpha_B\) depends on another set of parameters \(\beta\)
(e.g., the position of each vertex depends on an affine matrix and a translation vector), 
then the derivative of the \texttt{JOIN} operator with respect to \(\beta\)
becomes block-diagonal, with each block corresponding to the derivative of one joined instance:
\begin{equation}
\frac{\partial \,\mathrm{JOIN}_{C}(\alpha_B)(i)}{\partial \beta}
  =
  \begin{bmatrix}
  \tfrac{\partial \alpha_B(j_1(i))}{\partial \beta} & 0 & \cdots & 0 \\
  0 & \tfrac{\partial \alpha_B(j_2(i))}{\partial \beta} & \cdots & 0 \\
  \vdots & \vdots & \ddots & \vdots \\
  0 & 0 & \cdots & \tfrac{\partial \alpha_B(j_k(i))}{\partial \beta}
  \end{bmatrix}.
\end{equation}

A simple performance optimization at this stage is to ignore the zero entries in the block-diagonal structure. 
After eliminating these zeros, the differentiation of \texttt{JOIN} becomes equivalent to applying \texttt{JOIN}
to the differentiations of its child attributes:
\begin{equation}
\frac{\partial\,\mathrm{JOIN}_{C}(\alpha_B)}{\partial \beta}
\;\equiv\;
\mathrm{JOIN}_{C}\!\left(
  \frac{\partial \alpha_B}{\partial \beta}
\right).
\end{equation}

By Eq.~\eqref{equation:chain-rule}, the second derivative of \texttt{JOIN} with respect to \(\beta\) has the same structure: 
since the operator is linear, the first term of the Hessian chain rule vanishes, leaving another block-diagonal matrix:
\begin{equation}
\frac{\partial^2 \,\mathrm{JOIN}_{C}(\alpha_B)(i)}{\partial \beta^2}
  =
  \begin{bmatrix}
  \tfrac{\partial^2 \alpha_B(j_1(i))}{\partial \beta^2} & 0 & \cdots & 0 \\
  0 & \tfrac{\partial^2 \alpha_B(j_2(i))}{\partial \beta^2} & \cdots & 0 \\
  \vdots & \vdots & \ddots & \vdots \\
  0 & 0 & \cdots & \tfrac{\partial^2 \alpha_B(j_k(i))}{\partial \beta^2}
  \end{bmatrix}.
\label{equation:join-hessian}
\end{equation}

Thus, the Hessian of \texttt{JOIN} with respect to any parameter (or other node in the graph) 
is computationally equivalent to the \texttt{JOIN} of the Hessians of its children with respect to those parameters.


\subsection{Differentiation of Unioned Attributes}\label{section-diff-union}
We now establish the rule to differentiate a unioned attribute.

Conceptually, what \texttt{UNION} does is take an index and return the corresponding attribute by checking the stacked attributes. Hence, when differentiating a unioned attribute w.r.t. its direct children, it becomes a union of identity matrices:

\[
\frac{\partial \,\mathrm{UNION}(\alpha_{X_1}, \ldots, \alpha_{X_m})}{\partial (\alpha_{X_1}, \ldots, \alpha_{X_m})}
  \;=\;
 \,\mathrm{UNION}(I_{X_1}, \ldots, I_{X_m}),
\]

where $I_{X_n}$ is an identity matrix bound to the primitive type $X_n$.

Similarly, when computing the Jacobian matrix of the unioned attribute w.r.t. any other parameter $\beta$, it becomes the union of Jacobians:

\[
\frac{\partial \,\mathrm{UNION}(\alpha_{X_1}, \ldots, \alpha_{X_m})}{\partial \beta}
  \;=\;
 \,\mathrm{UNION}\left(\frac{\partial\alpha_{X_1}}{\partial \beta}, \ldots,\frac{\partial\alpha_{X_m}}{\partial \beta}\right)
\]

Hessian computation also follows the same fashion: the Hessian of \texttt{UNION} w.r.t. any parameter becomes the \texttt{UNION} of each child's Hessian w.r.t. the parameters.

However, a subtle issue arises because \texttt{UNION} is designed to combine attributes originating from different parameterizations. As a result, not all unioned attributes necessarily depend on the same set of parameters.

To illustrate this, suppose that \(\beta\) is a collection of parameter sets \(\{\beta_1, \beta_2, \ldots, \beta_m\}\),where each \(\beta_n\) corresponds to the parameters governing the computation of \(\alpha_{X_n}\)and may have a different dimensionality. 
In this case, while the partial derivatives 
\(\tfrac{\partial \alpha_{X_1}}{\partial \beta}, \ldots, \tfrac{\partial \alpha_{X_m}}{\partial \beta}\)
share the same shape, the derivatives with respect to their true parameter sets,\(\tfrac{\partial \alpha_{X_1}}{\partial \beta_1}, \ldots, \tfrac{\partial \alpha_{X_m}}{\partial \beta_m}\), generally have different shapes.

From a performance standpoint, it is inefficient to compute the differentiation of each \(\alpha_{X_n}\) with respect to the full parameter set \(\beta\), since the resulting Jacobian or Hessian matrices would be largely sparse. To address this, whenever such differentiation is required, we first identify the parameter set \(\beta_l\) with the largest dimension 
\(q_l = r_l \times c_l\). 
Each derivative \(\tfrac{\partial \alpha_{X_n}}{\partial \beta_n}\) is then padded so that its shape becomes \(p \times q_l\), where \(p\) denotes the flattened dimension of \(\alpha_{X_n}\).

This padding ensures consistent dimensions across all derivatives, allowing them to be safely combined within a single \texttt{UNION}. The result of the differentiation is then a new \texttt{UNION} attribute on the \texttt{primitiveUnion} in which $\mathrm{UNION}(\alpha_{X_1}, \ldots, \alpha_{X_m})$ resides in.

By treating the result of differentiation as a \texttt{UNION} attribute, the derivative paths are merged symbolically into a single node in the computation graph. Although the \texttt{UNION} operator still performs path selection at runtime, it no longer induces code fragmentation at compile time. As a result, all parameterization-specific derivatives are represented within one unified symbolic structure, 
ensuring both correctness and compile-time efficiency.

\subsection{Reusing Intermediate Symbolic Differentiation}\label{section-reuse}


As YASPS explicitly assembles the global Hessian matrix and gradient vector, it must also explicitly compute the corresponding \emph{local} Hessian and gradient for each energy term. Therefore, YASPS first performs symbolic differentiation on the computation graph of an energy to obtain a per-instance computation trace for the local gradient and Hessian. From this symbolic trace, YASPS later generates code; in this subsection we focus only on the symbolic stage.

Just like any differentiation system, we would like to reuse intermediate results whenever possible. Here we focus on \emph{symbolic} reuse: when multiple energies (or intermediate computations) depend on the same intermediate attribute, we want the differentiation information for that attribute to be computed once and reused.

To illustrate, consider simulation on a mixed-material bunny mesh with elasticity and collision (a similar example is shown in Section \ref{section-mixed-material}). A subset of the bunny's vertices' positions, marked as \texttt{bunny}.\texttt{abd\_vertices}.\texttt{position}, is controlled by affine transformation, while the other vertices, \texttt{bunny.soft\_vertices.position}, have independent $3$-DoF parameters. 

In this example, we first \texttt{UNION} these two position fields to obtain \texttt{bunny.vertices.position}. We then \texttt{JOIN} this unified position field using different connectivities to form three attributes:
\begin{itemize}
  \item \texttt{bunny.tets.position} for the elastic energy,
  \item \texttt{pt-pairs.positions} for point--triangle collisions and point--triangle frictions,
  \item \texttt{pe-pairs.positions} for point--edge collisions.
\end{itemize}
Each of these attributes receives \texttt{bunny.vertices.position} through its own \texttt{JOIN} operator and evaluates its corresponding energy term on the resulting attributes.

When computing first- and second-order derivatives, it is desirable to reuse the derivatives of the unified position attribute \texttt{bunnies.vertices.position}. Without reuse, every computation tree that passes through this attribute would redundantly recompute its symbolic Jacobian and Hessian. The generated code will also contain duplicate information which may not be detected through common-subexpression elimination.

The desire for such reuse motivates a differentiation method that preserves the structure during differentiation, which is why we choose to explicitly follow the second-order chain rule (Equation~\eqref{equation:chain-rule}). The use of chain rule allows us to factor a computation into smaller parts $f \circ g$. Through this decomposition, any differentiation through $g$ will reuse symbolic computation generated for the first- and second-order differentiation on $g$ regardless of what $f$ is.

However, we still need to choose the decomposition $f \circ g$. 
In reality, there are many ways to cut the graph, and globally optimizing the choice of $f$ and $g$ to minimize the number of operations in differentiation can be hard. Fortunately, we do not need to pick a single decomposition because the second-order chain rule can be composed recursively: we can further decompose both $f$ and $g$ recursively into smaller function combinations. 

YASPS takes advantage of this by using the structural information provided by the \texttt{JOIN} and \texttt{UNION} operators to drive the decomposition of the computation graph. Concretely, we restrict attention to a small set of nodes, which we call \emph{boundary nodes}:
\begin{itemize}
  \item the energy attribute
  \item \texttt{JOIN} attributes
  \item \texttt{UNION} attributes
  \item data attributes, whose values are directly supplied by users
\end{itemize}

The boundary nodes are not themselves the cut points for $f$ and $g$, but they tell us where separations should occur. Algorithm~\ref{alg:boundary-dfs} in the appendix traverses the computation graph and, for each boundary node, extracts its neighboring boundary-node descendants. This induces a coarsened “boundary graph” whose edges indicate where local Jacobians and Hessians must be computed.

Once we have identified all neighboring boundary nodes, we compute the local derivatives, namely $\nabla_g f$ and $\nabla^2_g f$ for each pair. 

However, as the differentiation rules for \texttt{JOIN} and \texttt{UNION} in Secs.~\ref{section-diff-join} and~\ref{section-diff-union} show, differentiating these operators is computationally equivalent to applying the operators to the differentiation. In other words, the Jacobian and Hessian of a \texttt{JOIN} or \texttt{UNION} node can be obtained from the Jacobians and Hessians of its children via another \texttt{JOIN} or \texttt{UNION}.

Consequently, once we obtain the local boundary-node pairs $(v, u)$ (where $u$ is a tuple of boundary nodes, which we treat here as a merged attribute), we do not compute derivatives directly on $v$ when $v$ is a \texttt{JOIN} or \texttt{UNION} attribute. Instead, we differentiate the direct children of $v$ (the attributes being joined or unioned) with respect to $u$, as detailed in Algorithm~\ref{alg:neighbor-differentiation} in the appendix.

The final step at this stage is to assemble the symbolic gradient and Hessian expressions from the decomposed graph. This is done by recursively applying the second-order chain rule along the boundary graph, using the local boundary-node pairs identified by Algorithm~\ref{alg:boundary-dfs}. Although we only explicitly compute local derivatives for the direct children of boundary \texttt{JOIN} and \texttt{UNION} nodes, the derivatives of their parent nodes can be symbolically inferred using the differentiation rules in Secs.~\ref{section-diff-join} and~\ref{section-diff-union}. As a result, the symbolic computation tree for the Hessian and gradient for the energy $E = f_1 \circ f_2 \circ \cdots f_n$ is fully determined by the local derivatives on the boundary-node pairs (namely $(f_1, f_2), (f_2, f_3)\dots, (f_{n-1}, f_n)$). This is detailed in Algorithm~\ref{alg:final-hessian} in the appendix.




Eq.~\eqref{equation:chain-rule} is written for a scalar-valued function $f$, whereas boundary nodes in YASPS often are not. 
In practice, we apply the chain rule component-wise and then stack the results to avoid tensor representations. 
For simplicity, in the algorithms we also assume that all data nodes are the target nodes of differentiation, while in reality we simply discard the boundary data nodes (and the branch that only contains this node) that are not the differentiation targets.

By recursively applying the chain rule over the boundary graph, YASPS reuses derivatives at two levels. Local Jacobians and Hessians on boundary edges are computed once and shared by all energies that touch those edges, while intermediate gradients and Hessians (w.r.t. the target attributes) stored on boundary nodes are reused across the assembly phase for the final gradient and Hessian.

\subsection{Eigendecomposition (EVD) and PSD Projection} \label{section-evd}
Newton-type solvers for energy minimization typically require a (locally) positive semi-definite (PSD) Hessian to produce a descent direction and to keep the linear solve well behaved. In IPC-style simulations, this is commonly enforced by projecting each \emph{local} Hessian block onto the PSD cone, usually by clamping negative eigenvalues after an eigendecomposition.

In practice, the eigendecomposition itself often becomes a bottleneck during Hessian assembly. Existing systems reduce this cost using three broad strategies:
\begin{enumerate}
    \item \textbf{Reduce the dimension} of the matrix being projected.
    \item \textbf{Derive closed-form} eigenvalues/eigenvectors for specific energy Hessians.
    \item \textbf{Reformulate energies} so that their Hessians are PSD by construction.
\end{enumerate}
Because the latter two strategies require a priori knowledge of the energy’s algebraic form, they are difficult to apply in a general system such as YASPS. We therefore focus on the first strategy, which is universally applicable.

\paragraph{Projecting in reduced coordinates.}
Recall the second-order chain rule  for a composed energy \(E(x)=f(g(x))\) in Eq.~\eqref{equation:chain-rule}.

During symbolic differentiation, YASPS determines whether the second term \(\sum_{j=1}^m \frac{\partial f}{\partial u_j}(g(x)) \,\nabla_x^2 g_j(x)\) vanishes. This occurs, for example, when each \(g_j\) is a linear function of \(x\), so that \(\nabla_x^2 g_j(x)=0\). In this case, the Hessian reduces to
\[
\nabla_x^2 E(x)=J_g(x)^\top \,\nabla_g^2 f(g(x))\, J_g(x).
\]
To enforce PSD in this setting, it suffices to project the \emph{reduced} Hessian \(\nabla_g^2 f(g(x))\), since
\[
A \succeq 0 \;\Rightarrow\; J^\top A J \succeq 0.
\]
Therefore, if we replace \(\nabla_g^2 f(g(x))\) by its PSD projection, the resulting Hessian is guaranteed PSD, while the eigendecomposition is performed in a (typically) smaller space.

\paragraph{Example.}
Consider the repulsive energy in Equation~\eqref{energy-simple-repulsive}, where \(f\) is the scalar energy as a function of two world-space positions \((p_1,p_2)\in\mathbb{R}^6\), and \(g\) gathers these positions via \texttt{JOIN}. If both points are controlled by affine bodies, then \(x\in\mathbb{R}^{24}\) (two affine bodies, \(12\) DoF each), while \(g(x)\in\mathbb{R}^6\). Consequently, \(J_g(x)\in\mathbb{R}^{6\times 24}\) and \(\nabla_g^2 f(g(x))\in\mathbb{R}^{6\times 6}\). Moreover, because the map from affine parameters to positions is linear, the curvature term \(\sum_j \frac{\partial f}{\partial u_j}\nabla_x^2 g_j\) is identically zero.

\paragraph{Symbolic rewrite.}
When the curvature term can be determined to be zero at compile time, YASPS rewrites the local block
\[
J_g(x)^\top \,\nabla_g^2 f(g(x))\, J_g(x)
\]
as
\[
J_g(x)^\top \,\mathsf{Project}\!\big(\nabla_g^2 f(g(x))\big)\, J_g(x)
\]
before lowering to generated code. Here \(\mathsf{Project}(\cdot)\) is a symbolic operator that is compiled into an eigendecomposition-and-clamping kernel. This reduces the projection dimension from the expanded DoF space to the smaller \(g\)-space, substantially lowering the cost of EVD in common IPC energy terms.

\section{Index Generator}\label{section-index}

While the symbolic differentiation gives us the ability to obtain the computation graph for Hessian and gradient locally (for each instance of the energy), to assemble the global linear system $Hx = g$, we still need to determine where each per-instance Hessian and gradient contribution belongs in the global system. 

In YASPS, such information can be gathered by traversing the computation graph. By lowering \texttt{JOIN} and \texttt{UNION} to the connectivity list instead of the actual numerical data, YASPS can easily extract the placement index for per-instance Hessian and gradient. 

Furthermore, such operation can be parallelized on GPU as the computation graph is fixed at compile time. This allows YASPS to easily handle such index computation in scenes with collision, where connectivities constantly change.

We defer the detailed implementation of how the index is computed to Appendix \ref{section-appendex-index}. Specifically, such implementation asserts static memory usage per device kernel. We also show the performance of our index computation kernel in Sec.~\ref{section-evaluation-index}.



\section{Code Generator}\label{section-code-generation}

YASPS uses a just-in-time (JIT) compiler to translate a symbolic computation graph to GPU-optimized code. In this section, we first introduce how the JIT generates code for any symbolic attribute, then discuss special code emitted only for attributes that compute Hessian and its projection.


\subsection{Modular Code Generation}\label{section-modular-code-generation}

Just like differentiation, YASPS associates each symbolic operator with a code generation rule that translates its computation into low-level GPU code. Code generation proceeds by traversing the symbolic computation tree in a depth-first manner, eliminating common sub-expressions, and composing the results into a GPU kernel. The generated kernel is then compiled to a linkable module and bound to the attribute for subsequent execution.

Because the traversal also records which data attributes (leaf nodes) are accessed, users do not need to manually specify kernel arguments. Instead, generated kernels read the required data attributes directly.

However, rather than emitting a single monolithic kernel for an entire computation, YASPS exploits the layered composition of attributes to generate \emph{modular} kernels that can be linked into other computations. Going back to the mixed-material example we briefly mentioned in Sec.~\ref{section-reuse}, multiple energy terms depend on the same unioned position attribute. Re-generating identical code for this shared subcomputation would be wasteful. Instead, YASPS compiles a dedicated object file (e.g., \texttt{.o}) for the shared subgraph and allows downstream kernels to link against it.

The key question is which attributes should be compiled as standalone modules. Analogous to the boundary-node notion in Sec.~\ref{section-reuse}, YASPS treats a small set of nodes as \emph{semantically important} and compiles them separately:
\begin{itemize}
    \item the requested output attribute (the attribute whose value is being generated),
    \item \texttt{JOIN} and \texttt{UNION} attributes, and
    \item named attributes (attributes explicitly given a name via \\ \texttt{addAttribute}).
\end{itemize}

During traversal, when YASPS encounters such a node, it triggers JIT compilation for that node and caches the resulting object file on the attribute. When generating code for a larger kernel, these cached object files are reused and linked rather than regenerated.

The code generation process is then separated into three stages. 

In the first stage, YASPS performs a DFS from the requested output attribute and collects the set of semantically important nodes \(N\). When a node in \(N\) is discovered, YASPS schedules it for separate code generation and compilation, and does not traverse its descendants in the current generation process. All other attributes are pushed in a duplicate-free stack \(S\). (See Algorithm~\ref{alg:gen-code-order} in the appendix.)

In the second stage,  YASPS loops over $S$ and, using per-operator code generation rules, emits code strings while replacing repeated sub-computations with intermediate values. (Algorithm~\ref{alg:codegen-dfs} in the appendix.)

In the final stage, YASPS links all previously generated object files for the semantically important nodes to the current kernel, and compiles the result into a final linkable kernel. (Algorithm~\ref{alg:gen-code-and-compile} in the appendix.)

When generating an object file for a node, the emitted code does not need to inline all of its descendants. For example, if \(\beta = 2.0 \cdot \alpha\) and \(\alpha\) is a \texttt{JOIN} node compiled as a standalone module, then the kernel for \(\beta\) only needs a declaration (a symbol reference) for \(\alpha\). The implementation of \(\alpha\) (stored in the object file compiled for \(\alpha\)) is linked later when \textsc{CompileFinalKernel} is invoked.
\subsection{Hessian Code Generation}\label{section-hessian-code}
\begin{figure*}[t]
  \centering
  \includegraphics[width=\textwidth]{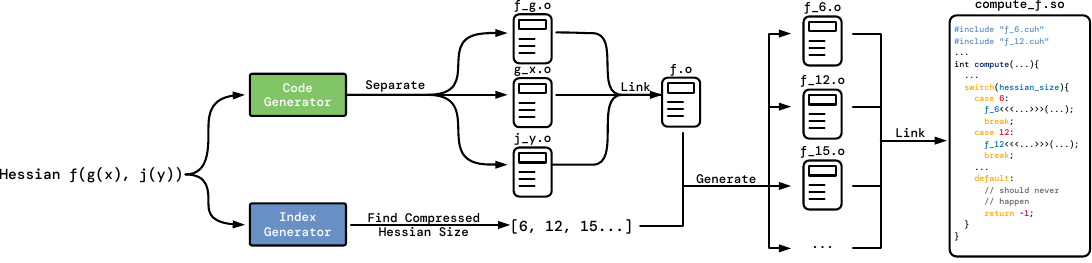}
  \vspace{-20pt}
  \caption{Illustration of the compilation pipeline for a Hessian computation into a final linkable dynamic library. When the Hessian computation has a structure of the form $f(g(x), j(y))$, YASPS first analyzes the computation graph and separates it into multiple kernels, such as $g(x)$, $j(y)$, and $f(g,j)$. This separation process is described in Algorithm~\ref{alg:gen-code-order}. YASPS then generates code for each kernel and compiles each one into a separate object (\texttt{.o}) file. These object files are subsequently linked to produce the final computation kernel for $f(g(x), j(y))$. At runtime, after the index of each Hessian block has been computed, YASPS determines the compressed size of each local Hessian block. Whenever a new compressed Hessian size $k$ is encountered, YASPS triggers an additional code generation and compilation step to produce a new \texttt{f\_k.o} file. This kernel is responsible for compressing the local Hessian to size $k \times k$ before accumulating it into the global Hessian matrix. Finally, these size-specific compression kernels are linked with the overall coordination kernel to produce the final shared library, which orchestrates the separated Hessian evaluation and compression pipeline. Notably, YASPS does not generate separate kernels on different combinations of parameters, which may lead to combinatorial explosion, but on the final compressed Hessian size, which almost always leads to smaller kernel counts in practice.}
  \label{figure-hessian-code-generation}
\end{figure*}

The code generation process for local gradient/Hessian evaluation uses the same modular infrastructure as ordinary symbolic attribute evaluation, but it additionally offers two optimizations tailored to second-order derivatives and PSD projection.

When the condition in Sec.~\ref{section-evd} holds, the local Hessian takes the form \(H_\text{final} = J^\top\, \mathsf{Project}(H_\text{inner})\, J\).
In this case, instead of explicitly materializing \(H_\text{final}\), YASPS offers the option to materialize only \(J\) and the projected inner Hessian \(H_\text{inner}\), and compute the contributions of \(J^\top H_\text{inner} J\) on the fly as it scatters blocks into the global matrix.

If \(H_\text{final}\) would be larger than the combined storage of \(J\) and \(H_\text{inner}\), this strategy can reduce peak memory usage. The trade-off is additional arithmetic during assembly (partial evaluation of \(J^\top H_\text{inner} J\)). In practice this option is rarely needed because typical local blocks remain small. But in cases with significantly rank-deficient local Hessian matrices (e.g. Sec.~\ref{section-cage}), generating the Jacobian and inner Hessian matrices can lead to a 6$\times$ size reduction compared to the final local Hessian. We detail this in Sec.~\ref{section-separation-evaluation}.

When the condition in Sec.~\ref{section-evd} does not hold, YASPS falls back to projecting the full local Hessian,
i.e., applying \(\mathsf{Project}(H_\text{final})\). From a performance perspective, however, the uncompressed local block can still be larger than necessary.

\begin{figure*}[t]
  \centering
  \includegraphics[width=\textwidth]{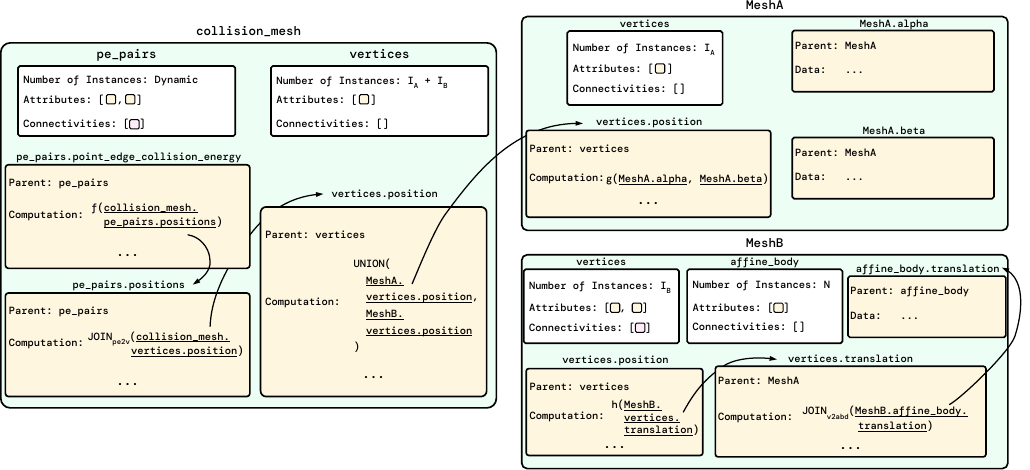}
  \vspace{-20pt}
  \caption{A small example of collision energy involving two types of vertices. \texttt{MeshA}'s vertex positions are controlled by two parameters $\alpha$ and $\beta$, which are attached to \texttt{MeshA} itself. \texttt{MeshB}'s vertex positions are controlled by the translations of multiple affine bodies. When a collision occurs, it is possible that two vertices participating in the collision correspond to the same underlying attributes. The resulting Hessian is then compressible.}
  \label{figure-small-example-collision}
\end{figure*}

\begin{figure}
  \centering
  \includegraphics[width=\columnwidth]{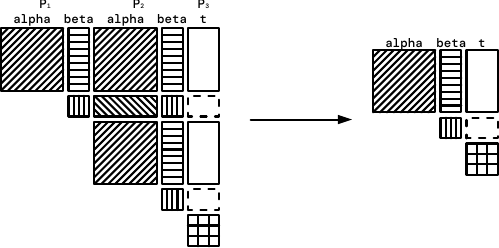}
  \vspace{-10pt}
  \caption{The local Hessian generated in the example from Fig.~\ref{figure-small-example-collision}, where $P_1$ and $P_2$ are both from \texttt{MeshA} and $P_3$ is from \texttt{MeshB}. Blocks with the same pattern correspond to the same global block in the global Hessian and can be merged before projection is applied. Note that the block representing $\frac{\partial^2 E}{\partial \beta \partial \alpha}$ vanished in the compressed Hessian. This is because it is compressed into the lower triangle of the matrix, which is simply the transpose of $\frac{\partial^2 E}{\partial \alpha \partial \beta}$ in the compressed Hessian.}
  \label{figure-small-example-collision-hessian}
\end{figure}

Consider the point-edge collision example in Fig.~\ref{figure-small-example-collision}. The energy depends on three position vectors, where each position is obtained from a \texttt{UNION} of multiple parameterizations. If two of the participating vertices are controlled by the same underlying parameters (e.g., both come from \texttt{MeshA} and share \(\alpha,\beta\)), then multiple entries in the local Hessian may map to identical coordinates in the global Hessian. As shown in Fig.~\ref{figure-small-example-collision-hessian}, several local blocks correspond to the same global block and can be aggregated prior to projection, reducing the dimension of the matrix on which eigendecomposition is performed.

In YASPS, after index generation is complete, we compute an additional per-instance \emph{permutation/aggregation} pattern. This pattern is derived by inspecting the index array for the current instance and detecting repeated global coordinates. When repetitions are found, YASPS records a permutation and aggregation rule so that the generated Hessian kernel first compresses the local Hessian into a smaller matrix by merging duplicate rows/columns.

In addition, because \texttt{UNION} reserves a maximum-sized representation, the resulting local blocks may contain structurally zero rows and columns corresponding to inactive branches. At this stage, YASPS also removes such zero rows/columns, further shrinking the matrix before projection.

This local compression is performed even when the reduced-space projection optimization from Sec.~\ref{section-evd} applies (in that case, it is applied to the assembled contributions without an additional projection step). Besides reducing the dimension for eigendecomposition, compression also minimizes the number of atomic additions required when scattering into the global Hessian, which is itself stored in a duplicate-free compressed format. We detail the global Hessian storage in Appendix~\ref{section-hessian-compression}.

In Fig.~\ref{figure-hessian-code-generation} we illustrate the flow of how a Hessian computation $f$ is compiled to the final kernel.
\section{Solver}\label{section-solver}

Since all other computations in YASPS are performed on the GPU, it is natural to run the linear solver, which solves the system $Hx = g$, on the GPU as well. YASPS therefore uses the modified preconditioned conjugate gradient (PCG) method \cite{baraff-witkin}, where most of the work consists of sparse matrix–vector multiplications (SpMV).

The optimization of this step has two main aspects:
\begin{itemize}
    \item Reducing the cost of SpMV
    \item Reducing the number of PCG iterations
\end{itemize}

To reduce the cost of SpMV, YASPS further compresses the global Hessian matrix by eliminating repeated blocks. At the same time, we implement a special kernel for SpMV that is designed for our storage layout. We detail the compression in Appendix \ref{section-hessian-compression} and the SpMV algorithm in Appendix \ref{section-spmv}.

To reduce the number of PCG iterations, YASPS employs a block Jacobi preconditioner, which is detailed in Appendix \ref{section-preconditioner}.


\section{Examples}\label{section-examples}



We include all the code of the examples, and YASPS itself, in the supplementary material.

\subsection{Cloths on Bunny on Cloth}\label{section-soft}

\begin{figure}[h]
    \centering
    \includegraphics[width=0.33\columnwidth]{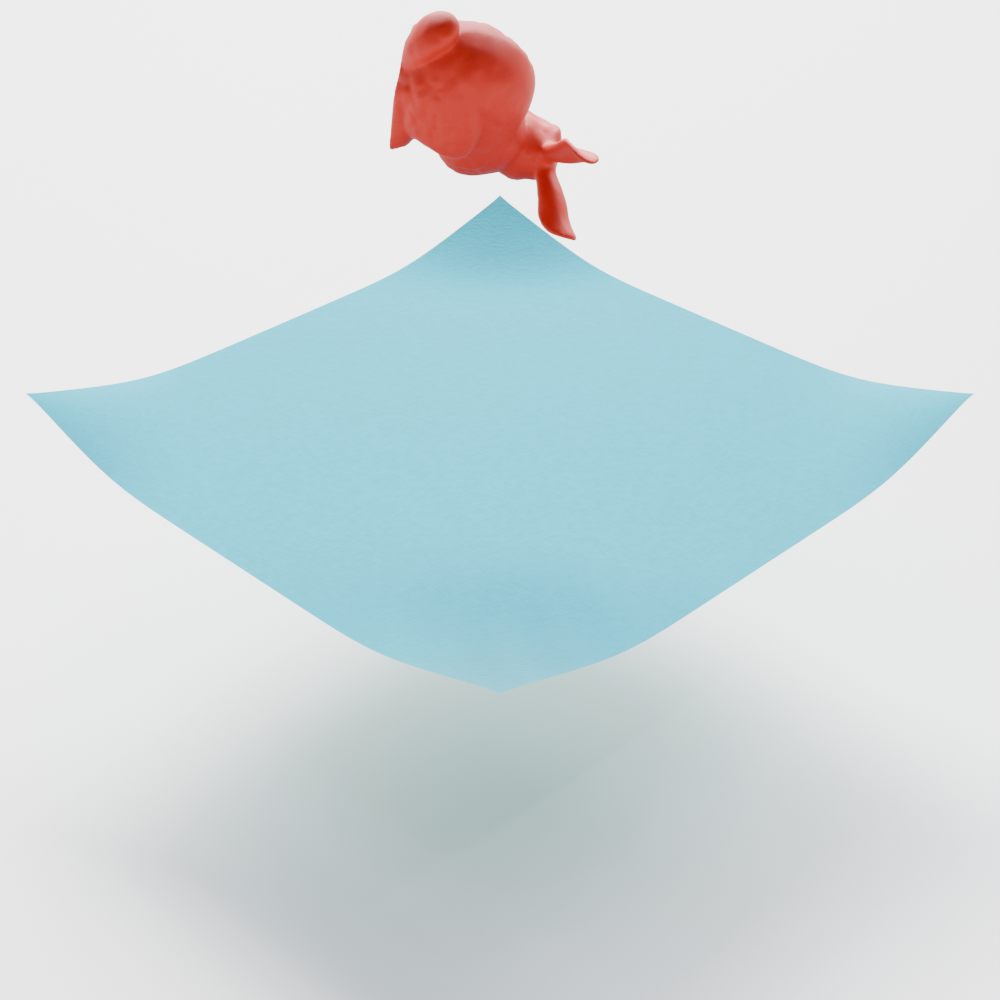}%
    \includegraphics[width=0.33\columnwidth]{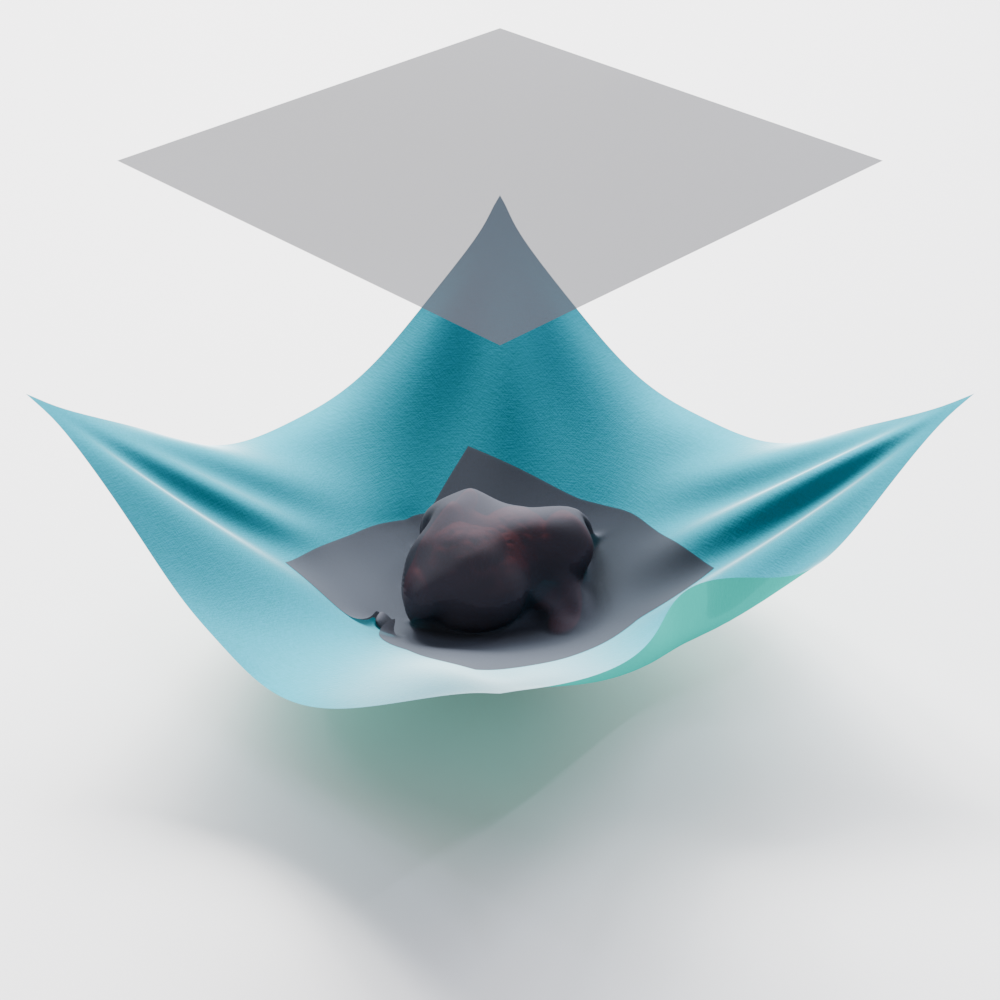}%
    \includegraphics[width=0.33\columnwidth]{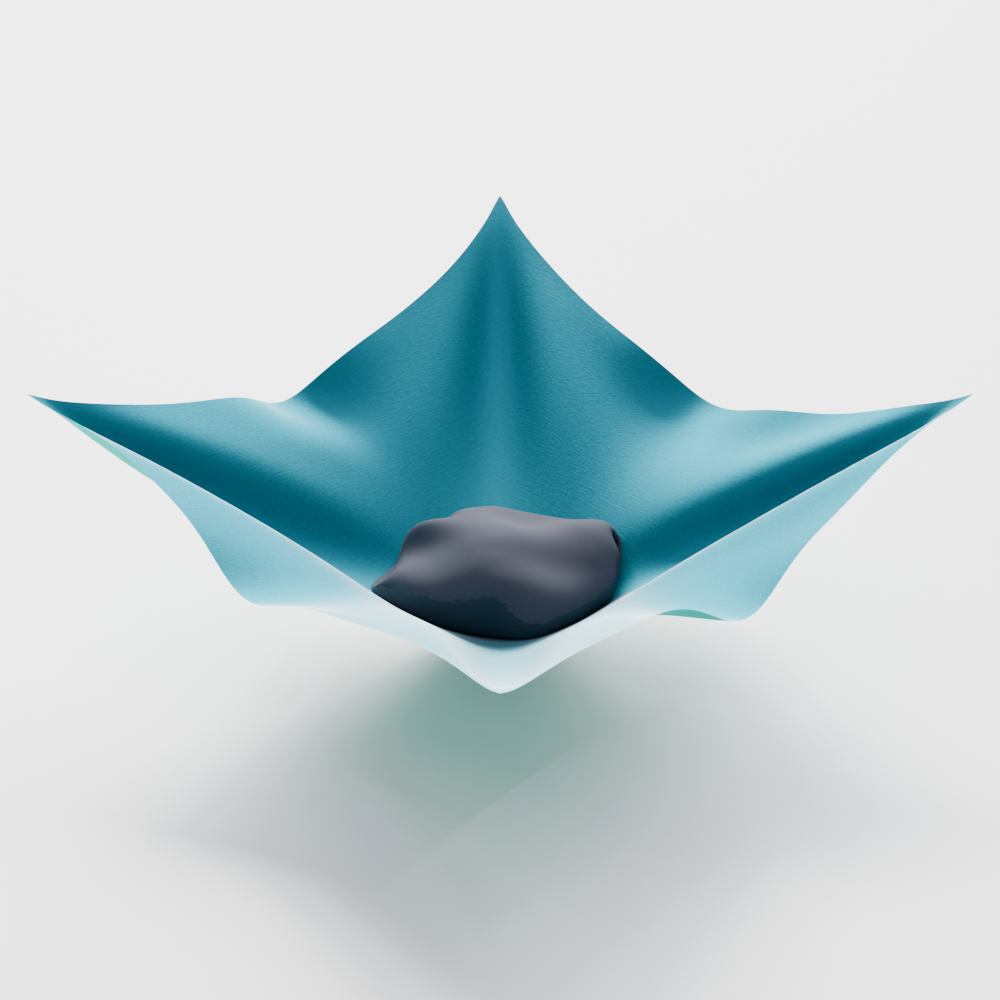}
    \vspace{-10pt}
    \caption{A soft bunny is dropped onto a cloth sheet whose four corner vertices are fixed. Extra cloth layers are then dropped sequentially onto the bunny.}
    \label{figure-one-bunny-many-cloth}
\end{figure}

For our first example, we simulate a soft bunny falling onto a cloth sheet, followed by additional layers of cloth dropping onto the bunny (Fig.~\ref{figure-one-bunny-many-cloth}). The bunny’s deformation is governed by a stable Neo-Hookean material with Young’s modulus $10259\ \mathrm{Pa}$ and Poisson’s ratio $0.205$. All cloth meshes share identical parameters and are modeled using the Baraff-Witkin energy, with stretch stiffness $33570\ \mathrm{Pa}$, shear stiffness $100607\ \mathrm{Pa}$, and a bending term with stiffness $0.055$. The CG solver uses a relative tolerance of $10^{-4}$, and each Newton iteration terminates when the maximum vertex displacement, normalized by the timestep $dt = 0.01\ \mathrm{s}$, falls below $10^{-2} \ \mathrm{m/s}$.
We selected this scene because an analogous setup can be conveniently constructed in both \textsc{GIPC}~\cite{GIPC} and \textsc{Stark} \cite{stark}. For \textsc{GIPC}, we match the material and scene parameters exactly, while for \textsc{Stark} we modify the material model to approximate the same behavior as closely as possible. The time comparison is reported in Sec.~\ref{section-total-performance}.
\subsection{Many Bunnies on Cloth}\label{section-many-bunnies-on-cloth}

\begin{figure}
    \centering
    \includegraphics[width=0.33\columnwidth]{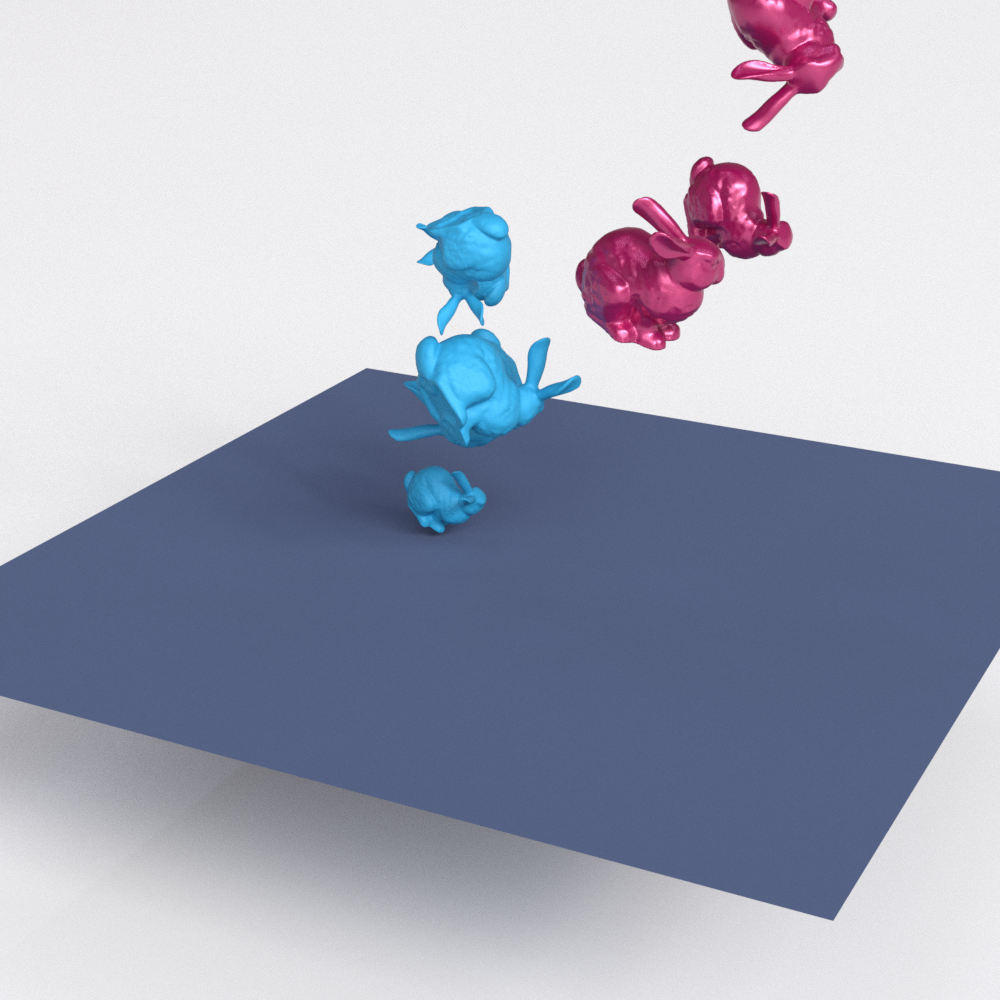}%
    \includegraphics[width=0.33\columnwidth]{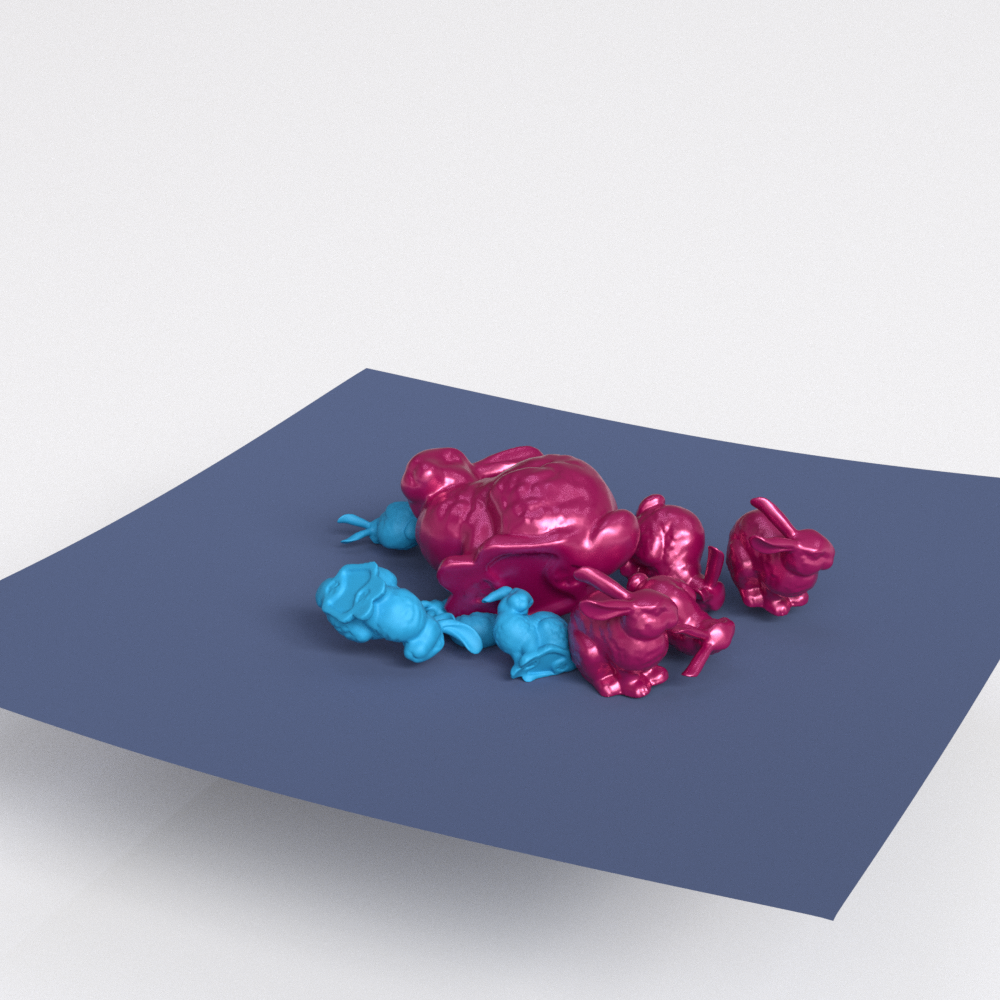}%
    \includegraphics[width=0.33\columnwidth]{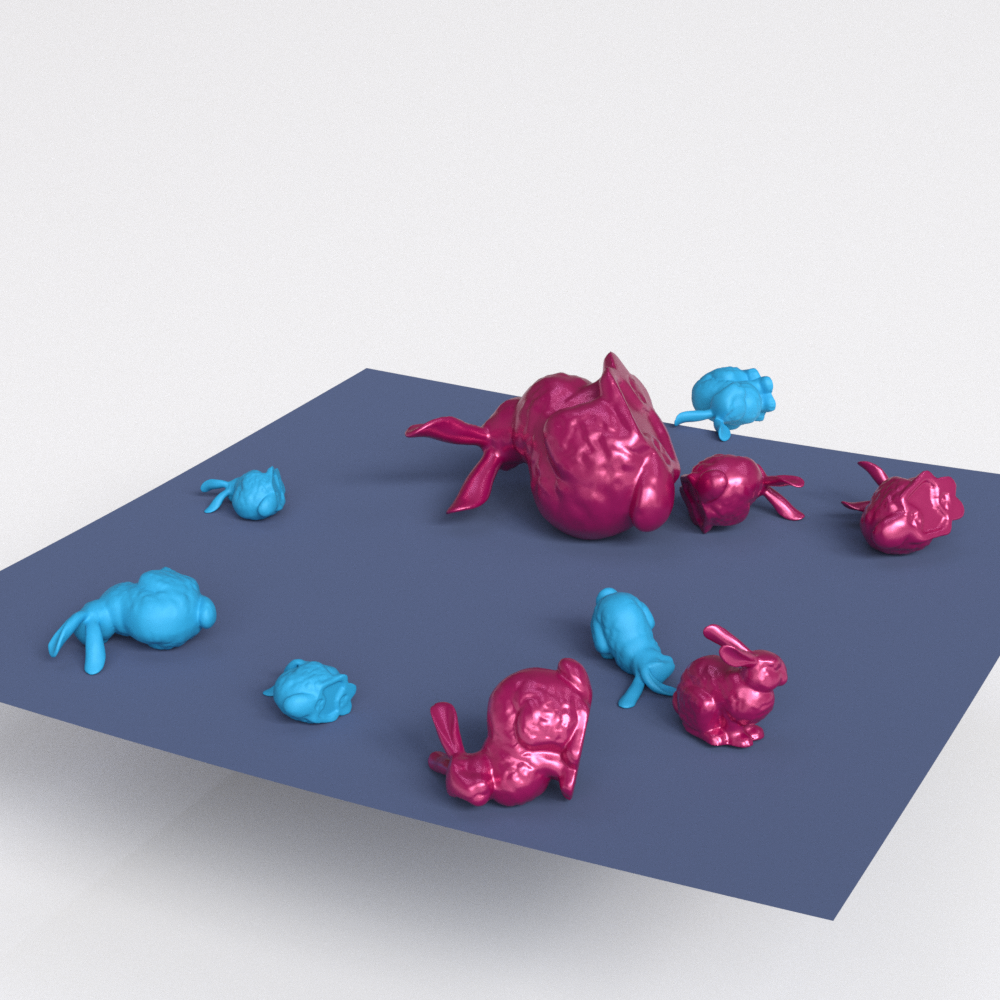}
    \vspace{-10pt}
    \caption{A set of soft (light blue) and rigid (dark pink) bunnies falling onto a cloth sheet whose four corner vertices are fixed. The rigid bunnies are controlled by their own affine transformation matrices.}
    \label{figure-many-bunny-one-cloth}
\end{figure}

For our second simulation example, we drop multiple bunnies - some deformable and some rigid - onto a cloth whose corners are fixed (Fig.~\ref{figure-many-bunny-one-cloth}). This example, in combination with the evaluations shown in Sec.~\ref{section-line-count}, demonstrates how users can use YASPS to rapidly add new materials and energies.

\subsection{Caged Bunny}\label{section-cage}
\begin{figure}
    \centering
    \includegraphics[width=1.0\columnwidth]{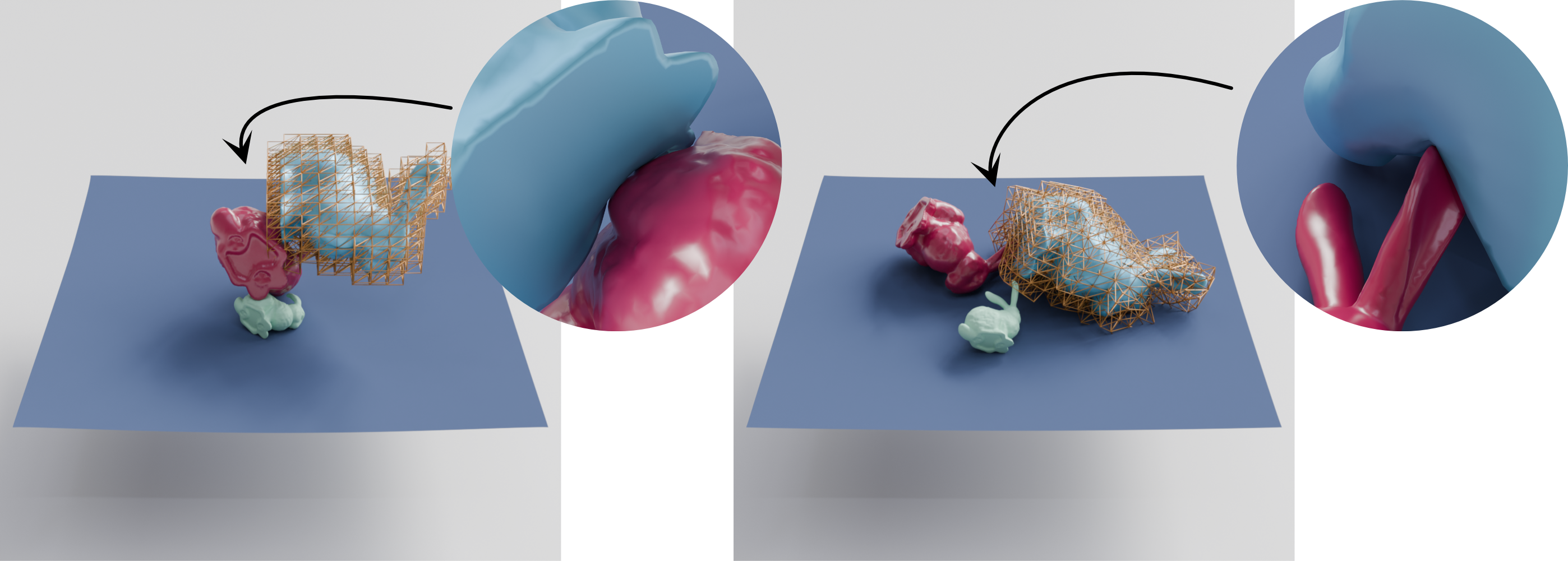}
    \vspace{-10pt}
    \caption{We additionally drop a bunny controlled by a cage deformation. The cage points do not participate in collision detection but the bunny controlled by the cage points does. During the simulation, the cage nicely deforms to avoid collision.}
    \label{figure-cage}
\end{figure}

In Fig.~\ref{figure-cage}, in addition to the soft body mesh and the affine body mesh, we implement a soft bunny controlled by a set of cages forming a uniform grid pre-computed for this mesh.

Each surface point of the bunny is controlled by exactly 8 vertices of the cage (a cube) enclosing that point. Each cage is then divided into 6 tetrahedra so that we can directly apply stable Neo-Hookean energy to the entire cage structure. The bunny's surface is affected by collision, which in turn deforms the cage points. 

Since the third bunny is a surface mesh, volume preservation is instead enforced on the cages. As shown in Fig.~\ref{figure-cage}, for the \texttt{Tet4} discretization we decompose each cage into 6 tetrahedra by connecting its vertices and apply the stable Neo-Hookean energy on those tetrahedra. In Appendix~\ref{section-hex8} we compare this against a \texttt{Hex8} discretization, where each cage is treated as a hexahedral element with eight Gaussian quadrature points for evaluating the same energy.

This example adds an additional 140 lines of code to the previous example to support the caged bunny. No modification to YASPS itself is needed.

\subsection{Mixed Materials}\label{section-mixed-material}

\begin{figure}
    \centering
    \includegraphics[width=1\columnwidth]{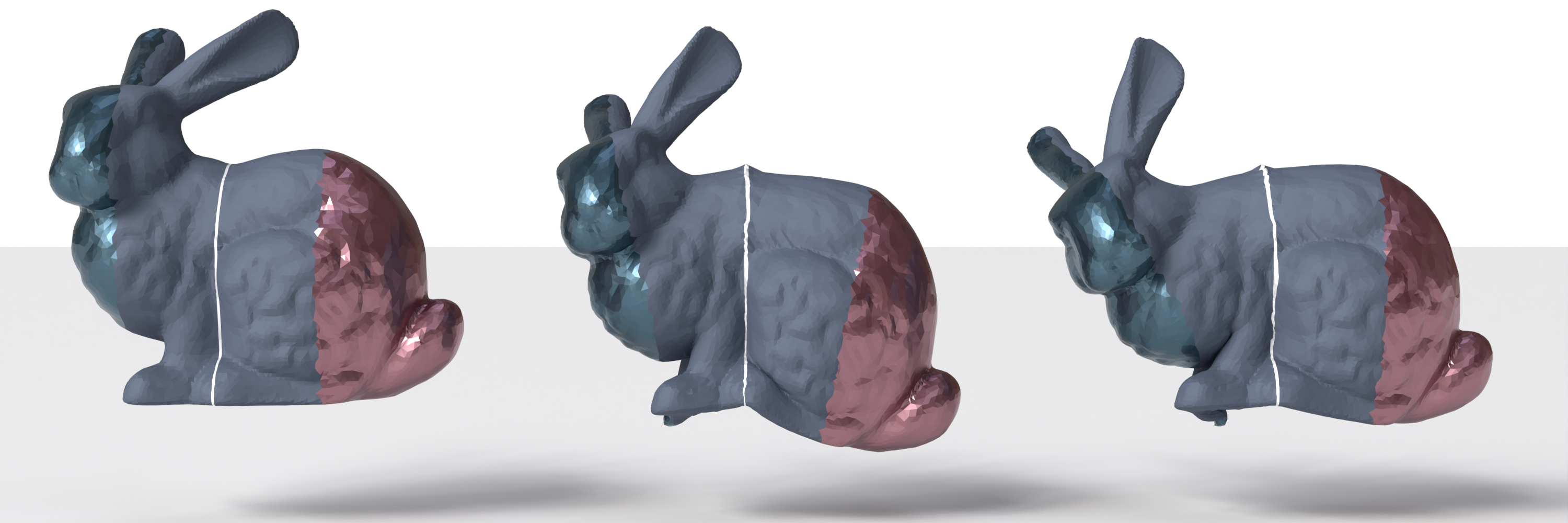}%
    \vspace{-10pt}
    \caption{A bunny model with mixed materials. The middle of the bunny is fixed in space (white strip) while the two ends of the bunny are modeled using two different affine bodies. The rest of the bunny can be deformed freely under elasticity constraints.}
    \vspace{-10pt}
    \label{figure-mixed-material}
\end{figure}

Mixed materials can be easily represented in YASPS. By creating separate primitive types for each material region and then forming a union over them, we can get a unified representation of vertices. Those vertices can then be used to form energies like elasticity and collision. Fig.~\ref{figure-mixed-material} shows a bunny which is modeled with two affine bodies, soft materials, and fixed vertices.

\subsection{Repulsive Curve On Bunny}
\begin{figure}
    \centering
        \includegraphics[width=1\columnwidth]{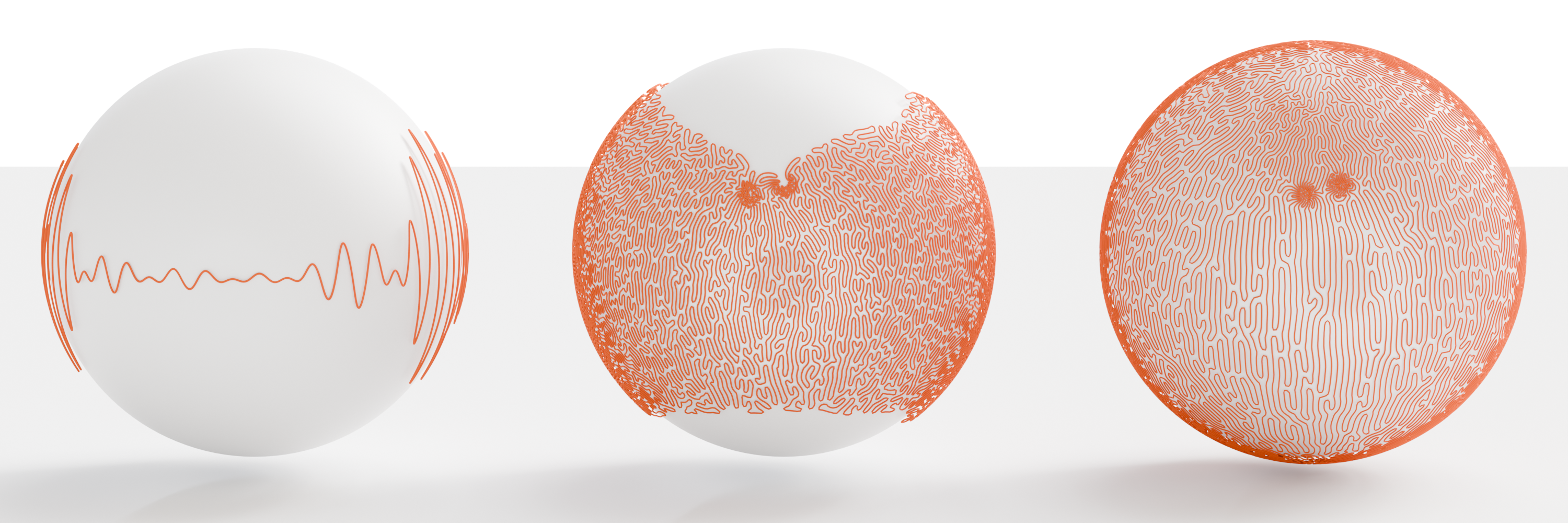}\\
        \includegraphics[width=1\columnwidth]{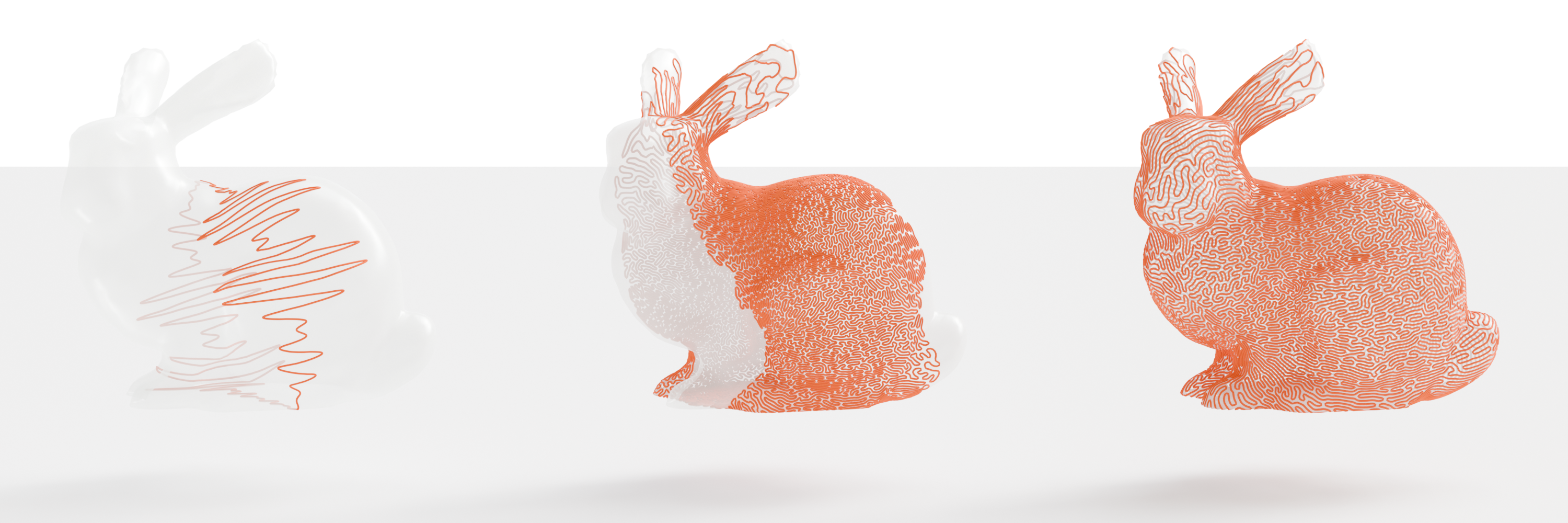}%
    \vspace{-10pt}
    \caption{We compute a repulsive curve on a bunny. To do this we first continuously smooth the bunny until we obtain a sphere as a parameterization. We then repulse a curve by adding barrier energy as well as inverse length penalty. Each point also gets its mass from the spherical parameterization. In the end we map the points on the curve back to the original bunny via the parameterization.}
    \label{figure-repulsive}
\end{figure}

In this example, we implement a repulsive curve on the surface of a bunny. To fully use what YASPS already offers, instead of the technique used in the recent paper ~\cite{repulsive}, we simulate the curve on a sphere and then map it to the surface of the bunny.

To do this, we first need to obtain a mapping from the input bunny mesh to the sphere. This is done by deforming the bunny mesh to a sphere via energy minimization, also using YASPS. A bending energy is added to each edge with its two incident triangles:
\[
  E_\text{bending} = l_\text{init} \|N_1 - N_2\|,
\]
where $l_\text{init}$ is the length of the edge at rest pose, and $N_1, N_2$ are the current normals of the two incident triangles.
To maintain the relative triangle shape and size for the mapping, we add the Baraff-Witkin energy to the mesh. Additionally, we also add an energy to push the points on the bunny surface to the surface of a sphere centered at the bunny's geometry center (average of all bunny's vertices' rest position). Barrier energies are also added to avoid penetration during the process.

Once the sphere is obtained, we want to use the density of the surface as a damping force. The density is computed as
$
  \rho = \frac{A_\text{bunny}}{A_\text{surface}}
$,
where $A_\text{bunny}$ is the area of the triangle on the original bunny surface, and $A_\text{surface}$ is the area of the same triangle on the sphere surface.
The density is then distributed to the 3 points of the triangle. A final UV map is then computed and stored through interpolation.

When simulating the repulsive curve on the sphere, we use the density information read from the UV map as the mass of the vertex. A higher mass means it's harder to move the points in that region. As shown in Fig.~\ref{figure-repulsive}, the part of the sphere surface that corresponds to the bunny ear has high density due to how the mesh was constructed, making it harder to move the loop around the region.
Once the curve is obtained on the sphere, we map it back to the bunny surface through the mapping we obtained earlier.
\section{Evaluation}
In this section, we quantify the implementation effort needed to support new mesh types by reporting the lines of code required for the first three examples in Sec.~\ref{section-examples}. This analysis demonstrates how extensibility in YASPS translates to concrete development cost. We then examine how YASPS’ optimizations in differentiation, code generation, and assembly improve overall performance.

\subsection{Lines of Code}\label{section-line-count}
We first demonstrate the amount of code required to implement a complete simulation pipeline using YASPS. Since YASPS is implemented as a Python package, all reported line counts refer exclusively to Python source code.
When using YASPS, the overall workflow naturally decomposes into six conceptual stages:
Setup, YASPS Integration, Energy Definition, Collision Detection, Initialization, Visualization (optional), and Newton Iteration. 

\begin{figure}
    \centering
    \includegraphics[width=1.0\columnwidth]{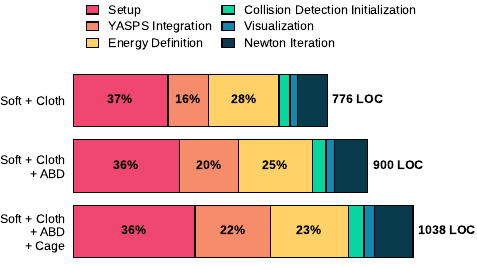}
    \vspace{-15pt}
    \caption{The total lines of code (LOC) for each part of the first three simulations discussed in Sec.~\ref{section-examples}. Adding a new material does not need a major addition to YASPS itself, but translates to an additional 130 lines of Python code.}
    \vspace{-5pt}
    \label{figure-code-count}
\end{figure}

We show the total lines of code (LOC) for each part for the first 3 examples (Secs.\ref{section-soft}, \ref{section-many-bunnies-on-cloth}, \ref{section-cage}) in the previous section in Fig.~\ref{figure-code-count}. While the number is not exact due to white space, comments and formatting, it gives a good sense of how many lines are added for any additional material. 
As each new material is added, the part of code that YASPS handles only increase by a small fraction in terms of lines of code. This allows users to quickly prototype new materials and energy using YASPS.


For comparison, we implemented the example in Sec.~\ref{section-soft} in \textsc{GIPC} and the example in Sec.~\ref{section-many-bunnies-on-cloth} in \textsc{StiffGIPC} \cite{libuipc}. Similarly we also implement the two examples in \textsc{Stark}. While \textsc{Stark}'s formulation for the elastic material and contact forces are different from that of \textsc{GIPC}, and the support for affine body is replaced with rigid body, we mainly focus on achieving similar simulation settings instead of a one-to-one replication. The increased line count on the frontend for \textsc{GIPC} and \textsc{Stark} are 53 and 45 respectively.

While the increase of LOC is lower compared to YASPS, it is only because those frameworks already included routines for loading triangle and tetrahedron meshes, as well as for assigning soft and stiff materials (ABD for \textsc{StiffGIPC} and rigid body for \textsc{Stark}). 
For example, to support affine bodies, \textsc{GIPC} added at least $1,000$ LOC just for the energy, gradient and Hessian computation due to the new parameterization and new block sizes induced by the new parameterization.
Similarly, \textsc{Stark}'s contact and friction module already contains $1,400$ lines of code, which exhaustively lists out all possible contact scenarios for each energy (soft-soft, soft-rigid, rigid-soft, rigid-rigid). Adding a new parameterization like the caged deformation in Sec.~\ref{section-cage} will only increase the workload significantly.

At the same time, YASPS does not need to modify anything in the backend. All changes are made at the frontend. For example, in terms of energy definition, YASPS only added 8 lines to define the rigidity energy
$
E_{\text{affine}}(\mathbf{A})
= \frac{1}{2}
\left\|
\mathbf{A}^{\top}\mathbf{A} - \mathbf{I}
\right\|_F^2
$
which penalizes deviations of the affine matrix $\mathbf{A}$ from an orthogonal matrix.

\subsection{Simulation Time Comparison}\label{section-total-performance}

\begin{table*}[htbp] 
\small
\centering 
\noindent
\caption{Performance comparison against \textsc{GIPC} and \textsc{Stark}.
We benchmark a scenario in which 1–3 cloth layers are dropped onto the Stanford bunny with $dt = 0.01$s, and the bunny itself is dropped onto another piece of cloth whose 4 corners are fixed in space. Each cloth contains 10{,}201 vertices and 20{,}000 triangles, and the bunny mesh contains 19{,}193 vertices and 79{,}935 tetrahedra.
The ``Diff Total'' column reports the total time spent computing all per-element Hessians (including the projection) and gradients and assembling them into the global system as the assembly is often performed right after the numerical value is computed. ``Diff Average'' reports the average time spent per Newton iteration. Similarly, the ``CG Total'' column reports the total time to perform the CG solve, while ``CG Average'' reports the average time spent per CG iteration.
For \textsc{Stark}, hard constraints are not supported, so the four corner vertices of the bottom cloth are fixed using penalty forces. In \textsc{YASPS}, these corners are instead treated as a special primitive whose degrees of freedom are excluded from differentiation, resulting in a slightly smaller linear system ($4 \times 3$ fewer DoF).
Although \textsc{Stark} is unable to complete the full 200-frame simulation due to their Newton iterations failed to converge on collision, the partial timing still provides a useful indication of its relative performance.}

\begin{tabularx}{\textwidth}{|l|l|X|X|X|X|X|r|r|r|}
    \hline
    \thead{System} & \thead{\# Vertices} & \thead{Diff\\Total (s)} & \thead{Diff\\Average (ms)} & \thead{CG\\Total (s)} & \thead{CG\\Average (ms)} & \thead{Total (s)} & \thead{\# Newton} & \thead{\# CG} & \thead{\# Frames} \\ \hline \hline
    \textsc{YASPS}  & 39,595 & 53.25 & 13.53 & 44.91 & 0.075 & 139.31 & 3,934 & 597,428 & 200 \\ \hline
    Optimized  & 39,595 & 43.55 & 11.06 & 44.82 & 0.075 & 130.20 & 3,939 & 600,303 & 200 \\ \hline
    \textsc{GIPC} & 39,595 & 25.46 & 8.81 & 156.32 & 0.97 & 199.06 & 2889 & 160,751 & 200\\ \hline\hline
    
    \textsc{YASPS}  & 49,796 & 81.23 & 15.83 & 68.91 & 0.083 & 218.29 & 5,132 & 825,519 & 200 \\ \hline
    Optimized & 49,796 & 65.17 & 12.89 & 67.35 & 0.083 & 201.26 & 5,055 & 811,759 & 200 \\ \hline
    \textsc{GIPC} & 49,796 & 49.02 & 14.80 & 264.09 & 1.31 & 342.76 & 3313 & 200,886 & 200\\ \hline\hline
    
    \textsc{YASPS}  & 59,997 & 124.64 & 18.99 & 95.36 & 0.092 & 328.51 & 6,564 & 1,036,455 & 200 \\ \hline
    Optimized  & 59,997 & 98.09 & 15.21 & 93.83 & 0.092 & 300.62 & 6,448 & 1,022,530 & 200 \\ \hline
    \textsc{GIPC} & 59,997 & 92.13 & 11.12 & 439.22 & 0.67 & 586.46 & 8,287 & 655,328 & 200\\ \hline\hline
    \textsc{Stark} & 59,997 & 29.1 & 124.89 & 110.9 & 0.91 & 141.90 & 233 & 122,357 & 64\\ \hline
    
\end{tabularx}
\label{table-performance-gipc}
\end{table*}

Table ~\ref{table-performance-gipc} shows the overall runtime for the examples in Sec.~\ref{section-soft}. While YASPS and \textsc{GIPC} are able to finish the simulation, \textsc{Stark} failed when collision happens as its line search algorithm wasn't able to find a valid step size that progresses the simulation without intersection. Hence, for 
\textsc{Stark} we only report the time steps before failure. All experiments are conducted on RTX 4090 and i9-13900F.

This benchmark demonstrates the performance of our system without any manual optimization as well as with manual optimization (the ``Optimized'' rows), which boosts the performance of the Hessian computation and projection (the ``Diff Average'' column). We will explain in detail how the manual optimization is implemented in Sec.~\ref{section-evaluation-evd-manual}.

Notably, \textsc{GIPC} already uses analytic Hessians for both the collision and stable Neo-Hookean energies to accelerate the PSD projection step.
This gives it advantage in differentiation shown in the ``Diff Average'' column compared to YASPS' un-optimized version. 
The version of \textsc{GIPC} we compare to also implemented the MAS preconditioner, which significantly reduces the average CG iterations per Newton solve by $2-3\times$ compared to YASPS. 
However, \textsc{GIPC} does not optimize for the storage compression like YASPS (\autoref{section-hessian-compression}). As a result, even with the reduced CG count, the larger average CG iteration time, which is dominated by sparse matrix-vector multiplication (SpMV), makes the overall solver performance worse. 
In comparison, the compression technique implemented by YASPS gives a close to  $10\times$ performance gain when performing CG iterations. As a comparison, while \textsc{Stark} is evaluated on CPU, the effective storage compression they perform on the Hessian matrix makes its per CG iteration time similar to that of \textsc{GIPC}.

However, even with this close performance, the total runtime of YASPS can still be theoretically optimized. To illustrate, we break down the time distribution of each important routine during the simulation loop for the un-optimized version. 
\begin{figure}[h]
\centering
    \vspace{-10pt}
    \includegraphics[width=\columnwidth]{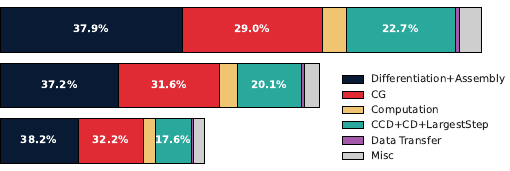}
    \vspace{-15pt}
    \caption{Runtime distribution for the cloths-on-bunny simulation (Sec.~\ref{section-soft}) using 3 (top), 2 (middle), and 1 (bottom) cloth layers. Each bar visualizes the contribution of the major computational components to the total simulation time. 
    \textbf{\emph{Differentiation+Assembly}} is the time to compute the gradient, Hessian, and putting them back to the global gradient vector and Hessian matrix.
    \textbf{\emph{CG}} marks the time spent on CG solver.
    \textbf{\emph{Computation}} is the time to evaluate some required attributes, like current positions.
    \textbf{\emph{CCD+CD+LargestStep}} is the time to invoke our CCD.
    \textbf{\emph{Data Transfer}} is the time to transfer any data from host to device or device to host.
    \textbf{\emph{Misc}} is the accumulation of any other parts whose individual contribution is negligible.
    The total runtime is shown in Table \ref{table-performance-gipc}.}
    \vspace{-10pt}
\label{figure-time-distribution}
\end{figure}

Shown in Fig.~\ref{figure-time-distribution}, since collision detection (CD) and continuous collision detection (CCD) are not the primary focus of \textsc{YASPS}, we directly integrated the corresponding components from \textsc{GIPC}. As a result, those routines, which are not optimized for our usage, take a significant portion of the execution. In comparison, \textsc{GIPC}'s native integration of the collision module makes these same parts more than $2\times$ faster than ours, nearly $10\%$ of our total execution time.
The performance of Hessian computation and projection (the ``Differentiation+Assembly'' bar in Fig.~\ref{figure-time-distribution}) can also be optimized further by manually changing how the energy is formulated in YASPS, which we detail in Sec.~\ref{section-evaluation-evd-manual}.

\begin{figure}
    \centering
    \includegraphics[width=1.0\columnwidth]{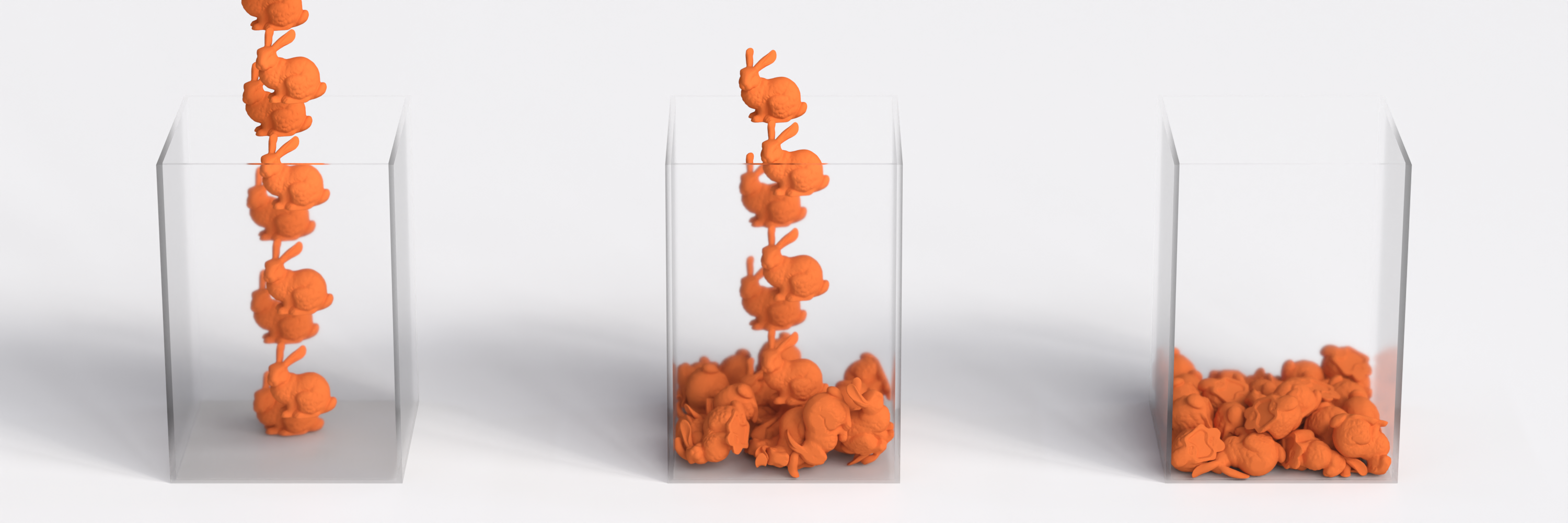}
    \caption{We show scalability by dropping 1-25 bunnies inside a glass container. All bunnies have the same material property. Each bunny contributes $57,579$ DoF with $79,935$ tetrahedra.}
    \label{figure-dropping-in-container}
\end{figure}

Additionally, we show the scalability of the system by dropping 1-25 soft bunnies inside a container shown in Fig.~\ref{figure-dropping-in-container}.
\begin{figure}
\centering
    \includegraphics[width=\columnwidth]{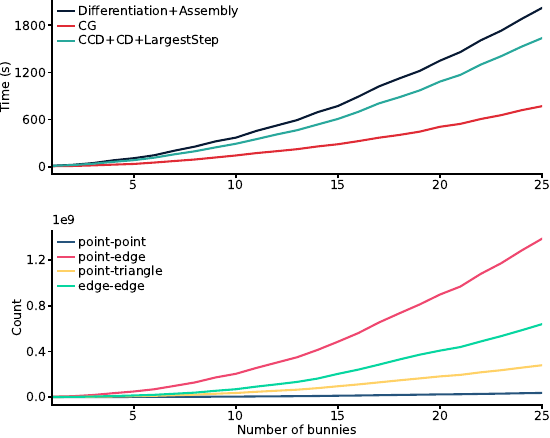}
    \vspace{-20pt}
    \caption{
    Scalability with respect to the number of bunnies for the simulation shown in Fig.~\ref{figure-dropping-in-container}.
    \textbf{Top:} Runtime breakdown of the major simulation components.
    \textbf{Bottom:} Total number of detected collision pairs by type (point--point, point--edge, point--triangle, and edge--edge).}
\label{figure-scalability}
\end{figure}
As shown in Fig.~\ref{figure-scalability}, as the number of collision pairs increases due to the increased number of bunnies in a confined space, the runtime for each major component of the simulation increases in a similar trend.
\subsection{Eigendecomposition: Automatic Optimization}\label{section-evaluation-evd-automatic}

The usage of the Conjugate Gradient (CG) solver requires the entire matrix to be positive semi-definite. YASPS makes this guarantee by projecting each local Hessian matrix to positive semi-definite by setting the negative eigenvalues of the matrix to $0$.
YASPS relies on the \textsc{C++} \textsc{Eigen} library to perform explicit eigendecomposition (EVD), which runs on GPU due to its templated implementations. 
As this operation is quite expensive, i.e.  $O(n^3)$ where $n$ is the matrix size, it can become a bottleneck during the Hessian computation if the energy is simple but the resulting matrix is large in size.

For this reason, YASPS places strong emphasis on minimizing the size of the matrix that must be projected, rather than merely accelerating numerical differentiation. The core optimizations that enable this reduction are described in Secs~\ref{section-evd} and~\ref{section-hessian-code}.
To illustrate the impact of these optimizations, we consider the point–point barrier energy commonly used in IPC-based frameworks. This energy has a very simple formulation, which makes EVD the major bottleneck in the Hessian computation:
\begin{equation}
E(\mathbf{p}_0, \mathbf{p}_1)
=
\kappa \,
(d - \hat d)^2
\left(
\log\left(\frac{d}{\hat d}\right)
\right)^2,
\end{equation}
with
\begin{equation}
d = \|\mathbf{p}_1 - \mathbf{p}_0\|^2,
\qquad
\mathbf{p}_0, \mathbf{p}_1 \in \mathbb{R}^3,
\end{equation}
where $\mathbf{p}_0, \mathbf{p}_1$ are the two position vectors, $\kappa$ is the barrier stiffness and $\hat d$ is the activation distance.
In a scene where all vertices are free (each has its own 3 DoF), no optimization can be done for the projection step from YASPS' perspective as every matrix simply has size $6\times6$. However, if the collision may occur between free vertices (3 DoF each) and vertices controlled by an affine body (12 DoF each), a naïve differentiation strategy must account for four possible cases: free–free (6 DoF), free–ABD (15 DoF), ABD–free (15 DoF), and ABD–ABD (24 DoF). This is where YASPS' optimization can help.

\begin{figure}[h]
    \centering
    \includegraphics[width=1.0\columnwidth]{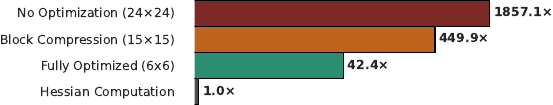}
    \vspace{-15pt}
    \caption{Time comparison for PSD projection with and without YASPS optimizations for the point-point barrier energy in a scene with free and ABD vertices, normalized by the cost of Hessian computation. The results are shown on a log scale. Coincidentally, the 3 unique matrix sizes also correspond to the 4 separate cases, with $24\times24$ corresponds to ABD-ABD, $15\times15$ corresponds to both free-ABD and ABD-free, and $6\times6$ corresponds to free-free.}
    \vspace{-10pt}
    \label{figure-projection-comparison}
\end{figure}

YASPS conservatively represents all possible collision configurations using the \texttt{UNION} operator. As a result, the symbolic Hessian is initially constructed at the maximum theoretical size of $24\times24$. Directly projecting this matrix—even though many rows and columns may be structurally zero—incurs a substantial cost, as shown by the “No Optimization” baseline in Fig.~\ref{figure-projection-comparison}.

A first optimization becomes possible once the collision type is known at runtime. For example, when a free vertex collides with an affine-body vertex, the effective degrees of freedom reduce to 15. In this case, rows and columns introduced solely by symbolic \texttt{UNION} operations can be safely eliminated. This block compression strategy (Sec.~\ref{section-hessian-code}) reduces the matrix size to $15\times15$ and yields the performance shown in the “Block Compression” row of Fig.~\ref{figure-projection-comparison}.

However, since both free-vertex and affine-body parameterizations are linear, the effective matrix that must be projected is in fact only $6\times6$, as shown in Sec.~\ref{section-evd}. Performing PSD projection on this reduced matrix results in a $44\times$ speedup compared to the unoptimized $24\times24$ projection, clearly demonstrating that preserving and exploiting structural information during symbolic differentiation is critical for performance.

In contrast, differentiation systems that manually enumerate and separate cases must both branch at runtime and perform PSD projection on larger matrices, even when the underlying degrees of freedom are much smaller. YASPS avoids this complexity entirely: the \texttt{UNION} operator enables automatic handling of all cases within a single symbolic representation, while still allowing aggressive matrix-size reduction at projection time.

\subsection{Eigendecomposition: Manual Optimization}\label{section-evaluation-evd-manual}

\begin{figure}[t]
    \centering
    \includegraphics[width=1.0\columnwidth]{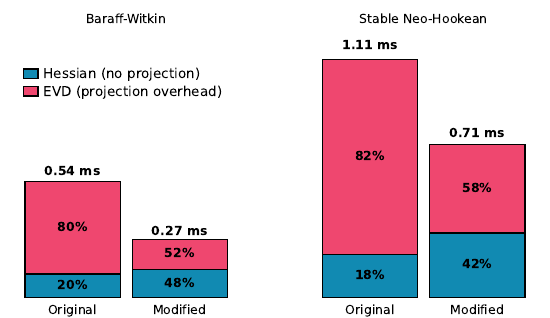}
    \vspace{-15pt}
    \caption{We compute the Hessian and its projection of the Baraff-Witkin energy and stable Neo-Hookean energy on 20k instances. The original method always uses the current position as the direct input to the energy computation, while the modified method uses the deformation gradient $\textbf{F}$ as the input to the energy computation.}
    \vspace{-10pt}
    \label{figure-projection-comparison-optimized}
\end{figure}

On the other hand, for energies that are complex, like the stable Neo-Hookean and the Baraff-Witkin, the ratio of EVD projection to base Hessian computation is significantly lower (5 to 1, shown in Fig.~\ref{figure-projection-comparison-optimized}). However, with YASPS' reuse strategy of differentiation (Sec.~\ref{section-reuse}), it is still possible to optimize this performance from the user's perspective without any modification to the system itself.

Take the formulation of the stable Neo-Hookean energy:
\[
\mathbf{F}(\mathbf{x})
=
\begin{bmatrix}
\mathbf{x}_1 - \mathbf{x}_0 &
\mathbf{x}_2 - \mathbf{x}_0 &
\mathbf{x}_3 - \mathbf{x}_0
\end{bmatrix}\in \mathbb{R}^{3\times3}
\]
\[
\mathbf{F}_I = \mathbf{F}^T\mathbf{B}^{-1}, \quad
J = \det(\mathbf{F}_I), \quad
I_C = \mathrm{tr}(\mathbf{F}_I),
\]
\begin{equation}
\label{equation-snh}
\Psi(\mathbf{F}) =
V \left[
\frac{\mu}{2}(I_C - 3)
- \frac{\mu}{2}\log(I_C + 1)
+ \frac{\lambda}{2}
\left(J - \left(1 + \frac{3\mu}{4\lambda}\right)\right)^2
\right]
\end{equation}
where $\mathbf{x}\in\mathbb{R}^{4\times3}$ is the position vector of a tetrahedron and $\mathbf{B}$ is formulated by the rest position of the tetrahedron.
While it is trivial that $\Psi$ is essentially a function of $\mathbf{x}$ (which is what we did for comparison in Table \ref{table-performance-gipc}), if we explicitly follow the formulation of $\Psi(\mathbf{F}(\mathbf{x}))$, the Hessian can then be written as:
\[
\frac{\partial^2 \Psi}{\partial x^2}
=
J_\mathbf{F}(\mathbf{x})^T
\frac{\partial^2 \Psi}{\partial \mathbf{F}^2}
J_\mathbf{F}(\mathbf{x})
\]
This means instead of projecting a $12\times12$ matrix, we can instead project a $9\times9$ matrix since $\mathbf{F}$ only produces 9 outputs. Similarly, for the Baraff-Witkin elasticity energy, the effective matrix we need to project can be reduced from $9\times9$ to $6\times6$. 

In YASPS, users can do this by first creating the attribute $\mathbf{F}$ under the tetrahedron, then creating a new primitive type, which has a one-to-one relationship to the tetrahedra. By performing a \texttt{JOIN} operation to pull $\mathbf{F}$ from tetrahedron to the new primitive type, we are effectively concretizing the formulation of $\Psi(\mathbf{F}(\mathbf{x}))$.
As shown in Fig.~\ref{figure-projection-comparison-optimized}, although the addition of a new primitive type does make the base Hessian computation slightly worse due to additional Jacobian matrix multiplication, the reduction in the projection cost makes the entire computation significantly faster. ($2\times$ performance gain for Baraff-Witkin and $1.56\times$ for stable Neo-Hookean).
In Table \ref{table-performance-gipc}, for the ``Optimized'' rows, we apply this trick to the stable Neo-Hookean energy ($12\times12 \rightarrow 9\times9$), Baraff-Witkin energy ($9\times9 \rightarrow 6\times6$) and the bending energy ($12\times12 \rightarrow 9\times9$). Those size reductions give a significant boost to the Hessian computation and projection.
\subsection{Hessian Computation}

Next, we evaluate the performance of our base Hessian computation. As shown in Fig.~\ref{figure-projection-comparison-optimized}, even though in all cases the projection takes more time than the base Hessian computation, the latter is still not negligible.

To demonstrate our base Hessian computation performance, we replicate a single element $N$ times and evaluate the gradient and Hessian individually on different energies in parallel using both \textsc{PyTorch} and \textsc{JAX}. Since all replicated elements are independent, the resulting Hessian for an energy such as stable Neo-Hookean elasticity has dimensions $N \times 12 \times 12$, rather than a single $12 \times 12$ block.
In addition, we evaluate a fully symbolic approach using \textsc{SymPy}. The gradient and Hessian are derived symbolically, followed by \textsc{SymPy}’s built-in common sub-expression elimination (CSE), and the resulting expression trees are translated directly into custom GPU kernels.
\begin{figure}[h]
    \centering
    \includegraphics[width=1.0\columnwidth]{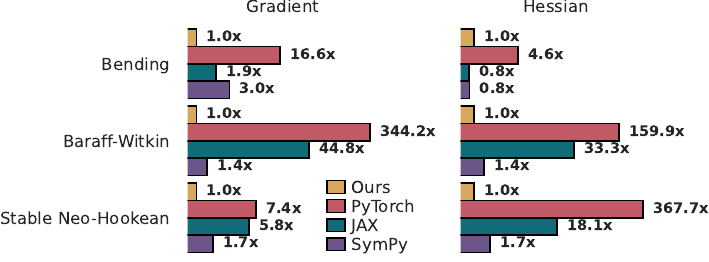}
    \caption{Average time required by \textsc{PyTorch}, \textsc{JAX}, and \textsc{SymPy} to compute the gradient and Hessian of a single instance, normalized by the runtime of YASPS. YASPS exhibits increasing performance advantages as the energy formulation becomes more complex. The x-axis is on a log scale.}
    \label{figure-hessian-comparison}
\end{figure}
Despite the fact that YASPS only performs very basic CSE, we match and in some cases outperform all baseline methods, as shown in Fig.~\ref{figure-hessian-comparison}. Notably, the performance advantage becomes more pronounced as the complexity of the energy formulation increases.

The speedup can partially be attributed to the fact that differentiation in YASPS is carried out at the matrix level rather than at the scalar level. This is true both at the symbolic differentiation phase and in our generated code. For example, consider the stable Neo-Hookean energy in Eq.~\eqref{equation-snh}. 
In this energy, the variable $J$ is the determinant of $\mathbf{F}_I\in \mathbb{R}^{3\times3}$. If this formulation is expanded explicitly, the determinant becomes a sequence of scalar multiplications and additions, which in turn leads to a combinatorial growth of operations when computing first- and second-order derivatives. 

In contrast, the derivative of the determinant admits a compact matrix-level expression:
\[
\frac{\partial \det(\mathbf{A}(x))}{\partial x}
=
\det(\mathbf{A}(x))\,
\mathrm{tr}\!\left(
\mathbf{A}(x)^{-1}\,
\frac{\partial \mathbf{A}(x)}{\partial x}
\right)
\]
Crucially, both the first- and second-order derivatives of $\det(\mathbf{A}(x))$ are expressed by $\det(\mathbf{A}(x))$, $\mathbf{A}(x)^{-1}$ and trace operator. In particular, the second-order derivative repeatedly involves the same matrix quantities like $\det(\mathbf{A}(x))$ and $\mathbf{A}(x)^{-1}$ across multiple terms. This repetition exposes substantial common sub-expressions, enabling CSE to effectively remove redundant matrix operations.

Additionally, as YASPS chooses to differentiate on a matrix level, we can then fully use the \textsc{Eigen} library for those matrix operations.
As an experiment, we can replace the computation of $J$ from a determinant operation to a sum operation (summation over all elements of $\mathbf{F}_I$) and perform the differentiation again.
\begin{figure}[h]
    \centering
    \includegraphics[width=1.0\columnwidth]{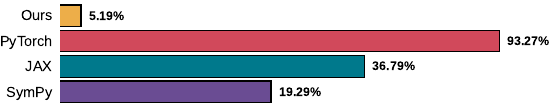}
    \caption{Overhead of computing the Hessian of the determinant operator for YASPS, \textsc{PyTorch}, \textsc{JAX}, and \textsc{SymPy}, normalized by their time of computing the Hessian of the stable Neo-Hookean energy (with determinant) respectively. The x-axis is on a log scale.}
    \label{figure-hessian-comparison-replace-det}
\end{figure}
As shown in Fig.~\ref{figure-hessian-comparison-replace-det}, YASPS is minimally impacted by the replacement of the determinant operator (only $5\%$ time difference) while other methods get a major performance boost when computing the Hessian without the determinant operator.

YASPS' symbolic differentiation is also significantly faster than \textsc{SymPy} at compile time. For example, computing the symbolic Hessian and gradient and performing CSE for the stable Neo-Hookean energy takes about 500ms for YASPS while \textsc{SymPy} takes 11 seconds.

Note that we did not compare to recent work \textsc{SymX} \cite{symx}, as its target platform is CPU.
Its CPU benchmark also shows that the performance of \textsc{SymPy} is close on the Hessian computation of stable Neo-Hookean energy (\textsc{SymPy} is $27\%$ slower than \textsc{SymX}). Similarly, we did not compare to \textsc{TinyAD} \cite{tinyad}, as \textsc{SymX} is $40\times$ faster than \textsc{TinyAD} on the stable Neo-Hookean example on CPU.

\subsection{Index Computation}\label{section-evaluation-index}
As described in Sec.~\ref{section-index}, Appendix \ref{section-index-computation} and Appendix \ref{section-hessian-compression}, YASPS performs a global compression step to eliminate repeated blocks in the assembled global Hessian.
In scenes with contact, this procedure must be invoked whenever the set of collision pairs changes. However, the connectivity of most of the scene can be naturally decomposed into two parts: \emph{static} and \emph{dynamic}. Static connectivity arises from topological relationships that do not change over time (e.g., elasticity), since the tetrahedralization connectivity of a mesh remains fixed even in the presence of collisions. In contrast, dynamic connectivity arises from interactions whose incidence changes over time (e.g., collision pairs), and can be treated as frequently varying.

Motivated by this observation, YASPS separates index computation and global Hessian construction into two components. Rather than assembling a single matrix $H_{\text{global}}$, we construct
\[
H_{\text{global}} = H_{\text{static}} + H_{\text{dynamic}}
\]
where $H_{\text{static}}$ corresponds to contributions with fixed connectivity, and $H_{\text{dynamic}}$ corresponds to contributions whose connectivity depends on the current collision set. This decomposition is also exposed in the frontend (Sec.~\ref{section-many-bunnies-on-cloth}): the construction of primitive types and energies can explicitly specify whether they are dynamic.

\begin{figure}[h]
    \centering
    \vspace{-5pt}
    \includegraphics[width=1.0\columnwidth]{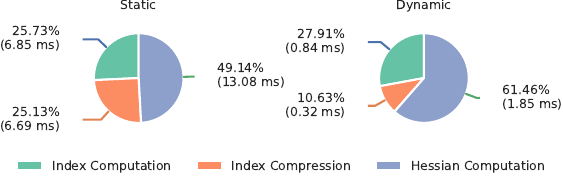}
    \vspace{-10pt}
    \caption{Percentage and the wall-clock time spent on index computation, index compression, and Hessian computation for the static (often elasticity) and dynamic (often collision) parts, measured for a single iteration in the example from Sec.~\ref{section-soft} with 1 piece of cloth falling on top of the bunny.}
    \vspace{-5pt}
    \label{figure-index}
\end{figure}

As a concrete example, consider the scenario in Sec.~\ref{section-soft}. We extract a representative frame containing 840 collision pairs,
and report the time breakdown for index computation, compression, and Hessian evaluation for both $H_{\text{static}}$ and $H_{\text{dynamic}}$ and show it in Fig.~\ref{figure-index}.

While index computation and compression account for a substantial fraction of the runtime in both cases, the absolute time for index computation for the dynamic part is only around 4\% of the total time (index computation plus Hessian computation for both the static and dynamic parts), while the index computation for the static part takes around 46\% of the total time. As such, even though the compressed matrix from $H_\text{dynamic}$ may still overlap with $H_\text{static}$ in structure, by not performing the index computation for the largely static part, we are saving more time than what a total compression can save us.



\subsection{Compile Time: Modular Code Generation}

While compilation time for frameworks that deploy a just-in-time compiler is often discarded as amortized time, since the code generated for the same computation can be reused, it is still important as we want to reduce the wait time for whenever users want to try something new.
In Sec.~\ref{section-modular-code-generation}, we described how YASPS generates separate object files (\texttt{.o}) for semantically meaningful nodes in the computation graph. This strategy allows the computation graph to be naturally segmented and compiled in parallel. Such parallelization is essential, as NVCC’s compilation time scales poorly with code size. Moreover, because YASPS currently employs only a basic form of common subexpression elimination (CSE), the generated kernels can become quite large in practice.

We report the compilation time of both computation kernels (which compute a specific attribute) and Hessian kernels (which compute the gradient, Hessian, and perform projection) with and without our optimizations, using the example from Sec.~\ref{section-many-bunnies-on-cloth}.

\begin{figure}[h]
    \centering
    \vspace{-5pt}
    \includegraphics[width=1.0\columnwidth]{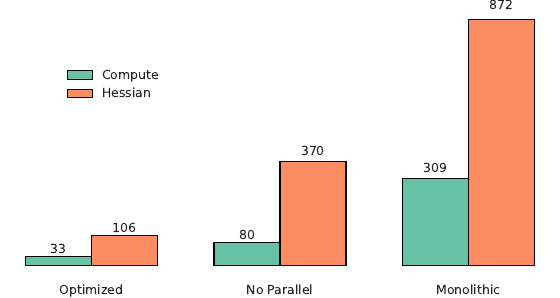}
    \vspace{-10pt}
    \caption{
    Compilation time under different compilation strategies (in seconds).
    ``Optimized'' uses our fully modularized pipeline with parallel compilation.
    ``No Parallel'' compiles the same modularized object files sequentially.
    ``Monolithic'' corresponds to generating and compiling a single monolithic kernel for each attribute, which includes the code for all of its children attributes. The y-axis is on a log scale.
    }
    \vspace{-5pt}
    \label{figure-compilation}
\end{figure}

As shown in Fig.~\ref{figure-compilation}, our modularized code generation and parallel compilation strategy is significantly faster than generating and compiling a monolithic kernel for each attribute (and Hessian).
While the overall compilation time remains substantial, this cost is largely attributable to the simplicity of our current CSE implementation, which can result in occasionally large generated kernels.

Nevertheless, this modularization strategy remains highly effective in certain cases even when the code size is relatively small.

\begin{figure}[t]
    \centering
    \includegraphics[width=0.245\columnwidth]{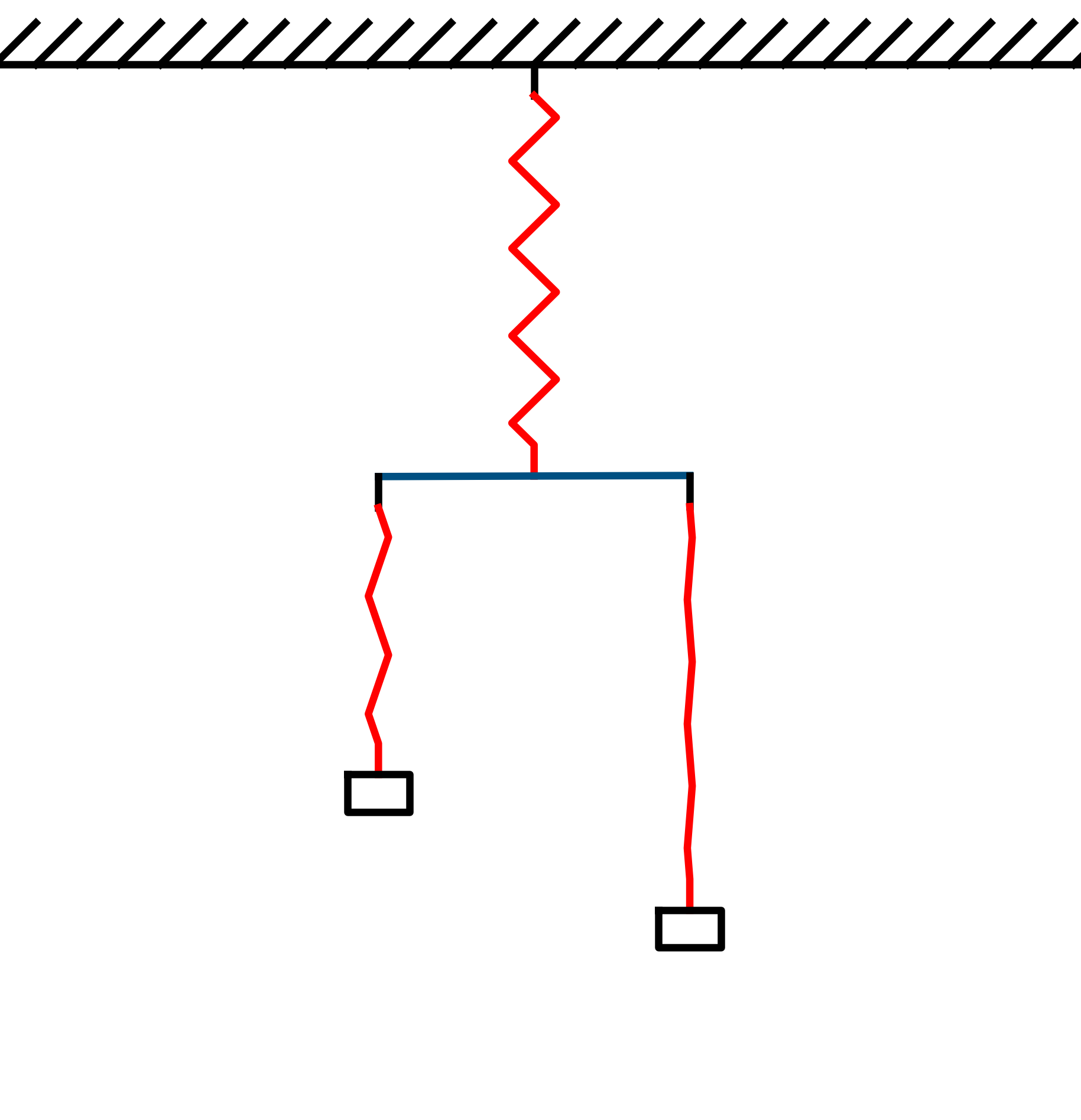}%
    \includegraphics[width=0.245\columnwidth]{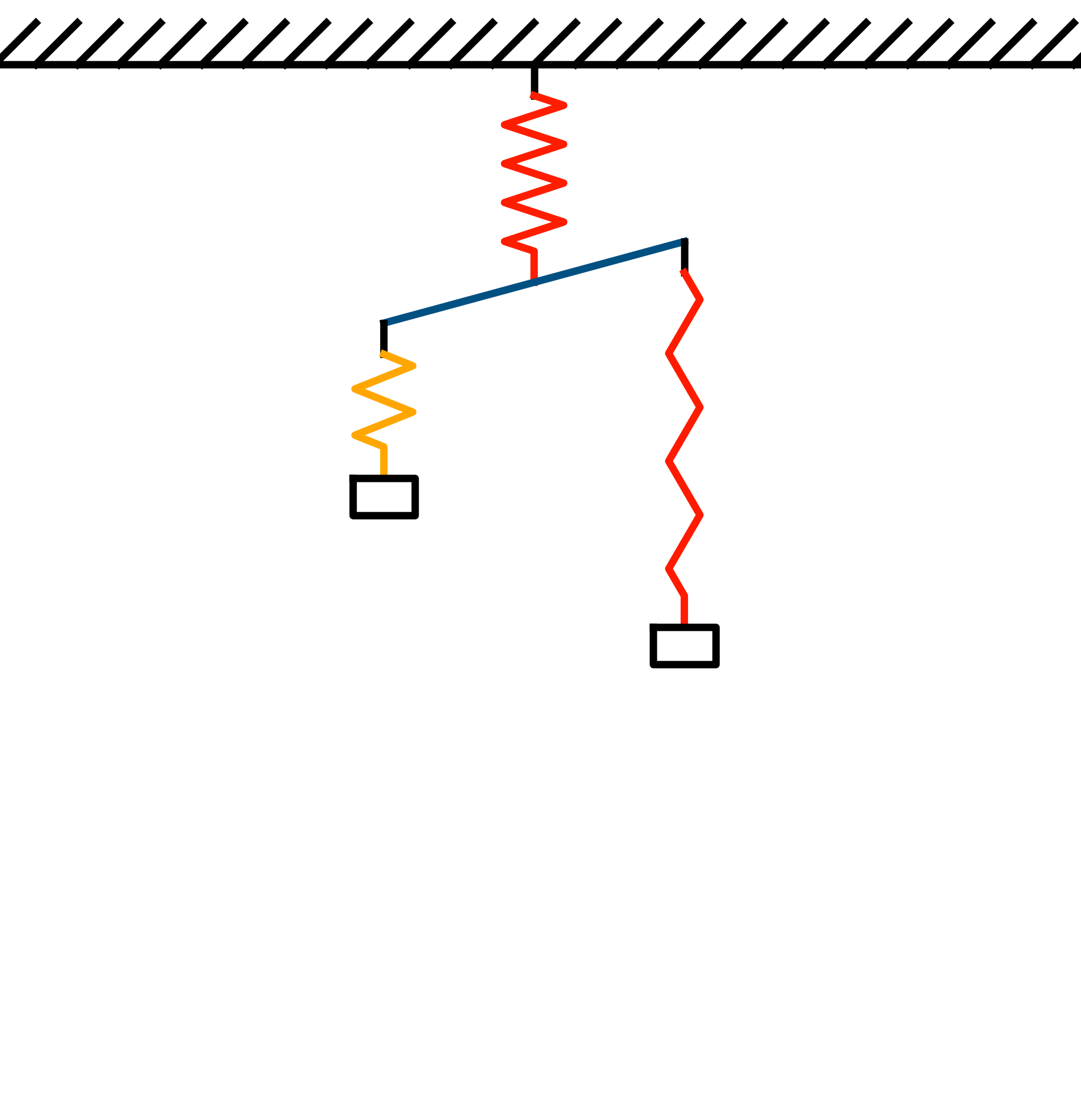}%
    \includegraphics[width=0.245\columnwidth]{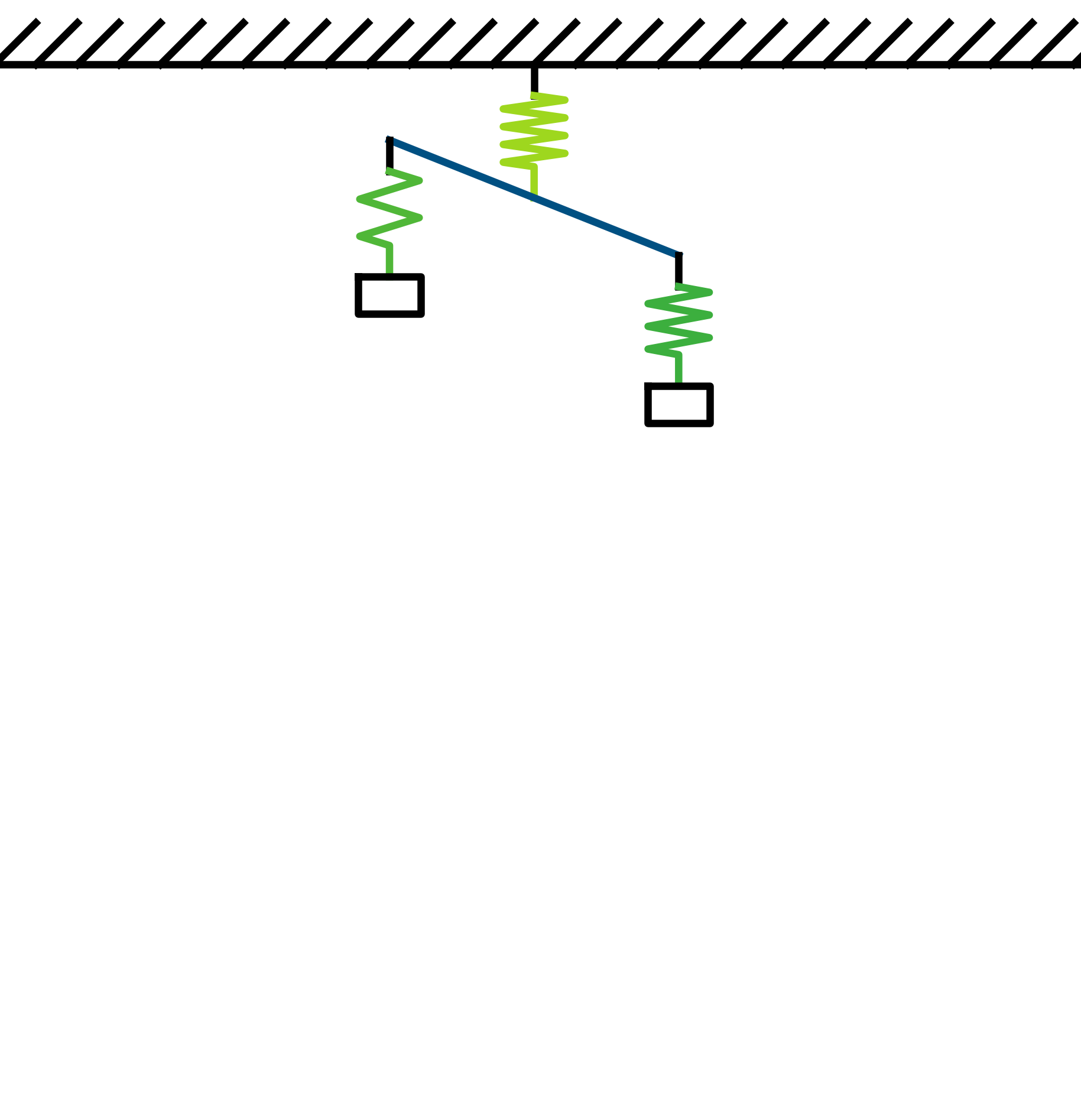}%
    \includegraphics[width=0.245\columnwidth]{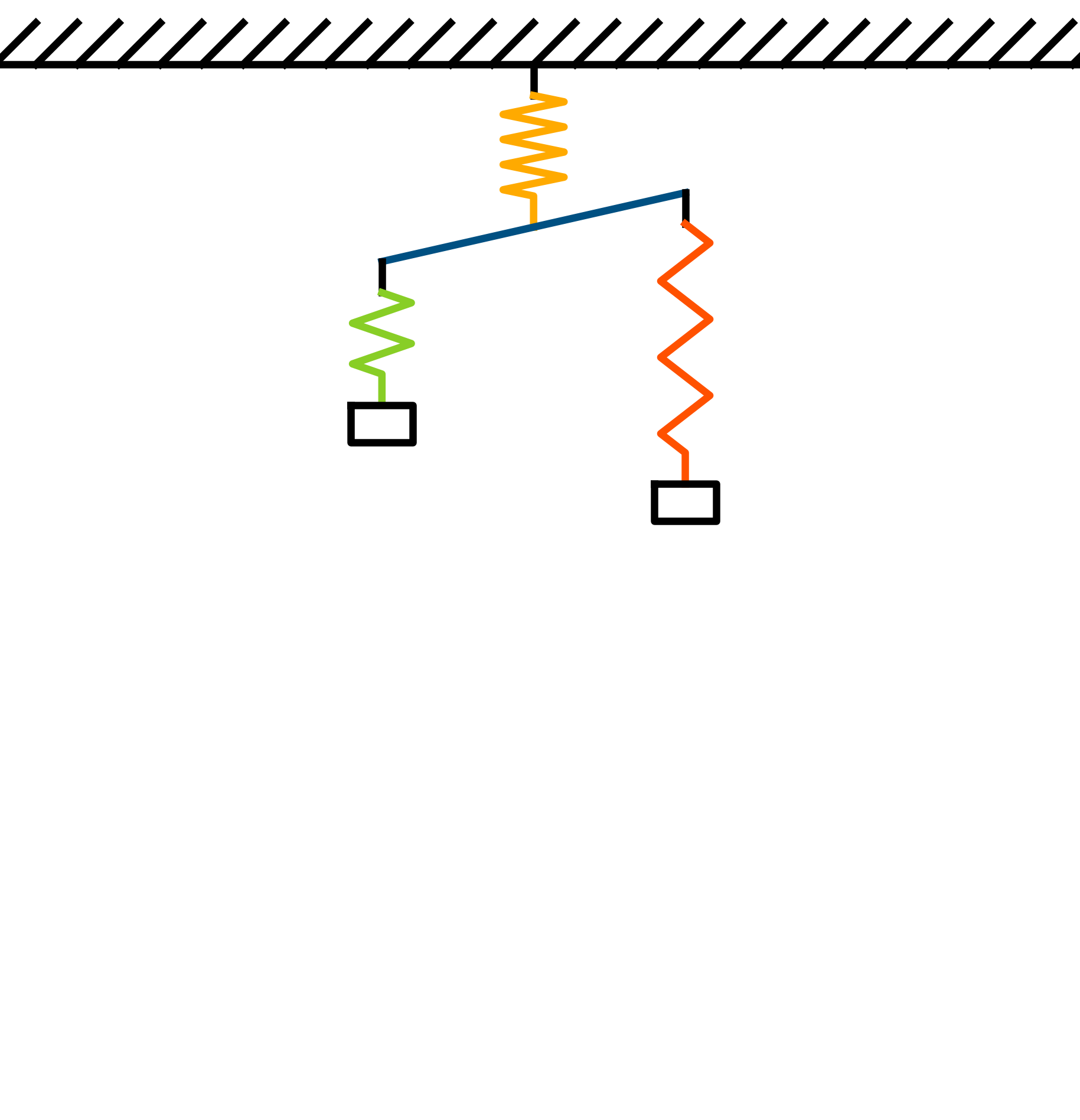}
    \vspace{-10pt}
    \caption{A textbook mass-spring system implemented in YASPS. Color indicates spring stress, with red corresponding to higher stress and green corresponding to lower stress.}
    \label{figure-spring}
\end{figure}

As an illustrative example, Fig.~\ref{figure-spring} shows a simple mass-spring system. Unlike the previous examples, where positions are parameterized linearly, this system is inherently non-linear. Each spring (zigzag segment) is parameterized by the angles between its segments, and the position of the spring endpoint is determined by these angles. The central lever is similarly controlled by an angular degree of freedom governing its tipping motion. As a result, the lower portion of the system is entirely driven by angular parameters, leading to a deeply nested non-linear dependency structure.


The energies used in this system are relatively simple: spring elasticity is defined by the deviation between the current angle and a rest angle, while inertia is applied only to the two mass blocks at the ends of the system, whose positions are defined by not only the springs at the end of the lever, but also by the rotation of the lever and the spring hanging from the ceiling.
Despite this simple setup, the Jacobian computation is more complex than the energy evaluation itself. Due to the nested dependency structure, the Jacobian contains repeated sub-expressions that can be naturally reused. This makes the example particularly well suited for demonstrating the benefits of modular code generation and compilation.
Using YASPS’ modular compilation strategy, the NVCC compilation time for inertia on the mass blocks is reduced from 6 seconds (without modular reuse) to 4 seconds.

\subsection{SpMV}\label{section-evaluation-spmv}
Here we compare YASPS’ SpMV routine against \textsc{cuSPARSE}-based SpMV implementations under different storage formats. All matrices are pre-compressed prior to multiplication so that there are no duplicate coordinates. 
\begin{figure}[h]
    \centering
    \includegraphics[width=1.0\columnwidth]{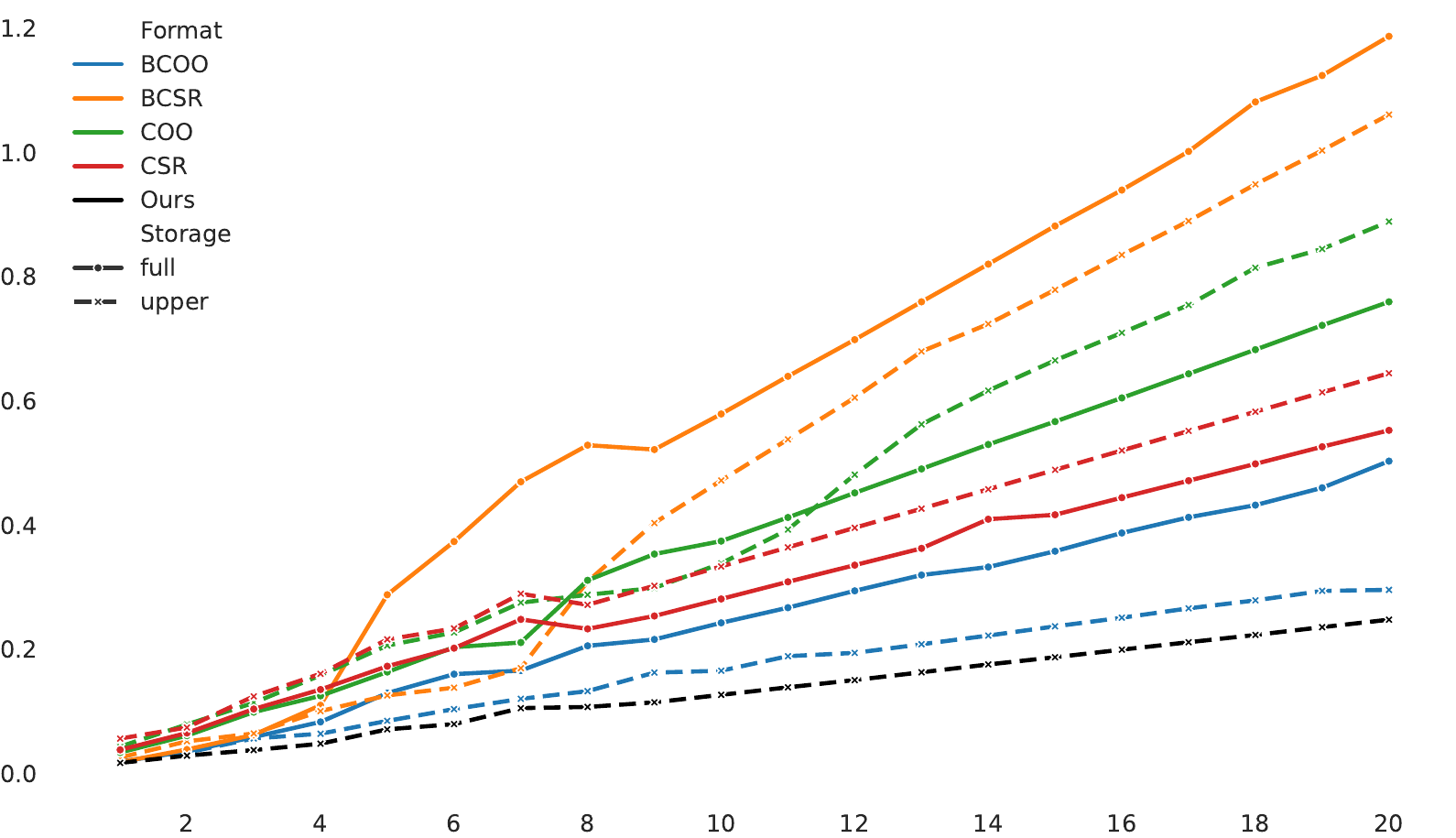}
    \caption{Time comparison for a single SpMV operation using different storage formats. The y-axis shows runtime in milliseconds, and the x-axis indicates the number of bunnies in the scene. Each bunny contains $57{,}579$ degrees of freedom and $238{,}279$ $3\times3$ blocks in the upper triangle of the system matrix.}
    \label{figure-spmv}
\end{figure}
To evaluate scalability, we construct scenes containing $N$ bunnies, each consisting of $19{,}193$ vertices. Stable Neo-Hookean energy is applied to the tetrahedral elements of each bunny, producing $238{,}279$ $3\times3$ blocks in the upper triangular portion of the Hessian matrix per bunny. Vertex indices are randomly permuted to avoid artificially favorable memory locality.

For \textsc{cuSPARSE}, since block sparse SpMV routines are deprecated, we manually implement block coordinate (BCOO) and block compressed sparse row (BCSR) SpMV kernels. These kernels explicitly unroll computations for fixed $3\times3$ blocks. In the BCSR implementation, each row is assigned to a single thread, while in the BCOO implementation, each block is processed independently by one thread. Additionally, we also expand the matrix to see the performance difference of full matrix SpMV against only upper-triangular symmetric matrix SpMV. 

As shown in Fig.~\ref{figure-spmv}, YASPS’ storage format outperforms all other tested formats. This improvement is attributable not only to reduced memory storage (Appendix \ref{section-hessian-compression}), but also to the reduction technique introduced in Appendix \ref{section-spmv}. Among the baselines, the closest competitor is the BCOO format that stores only the upper triangular portion of the matrix.

Note that this benchmark is run outside of the entire simulation pipeline. In the full simulation, using identical SpMV code on the same data can lead up to a $15\%$ slower execution. Additionally, if we instead do not generate kernels for each different block sizes at compile time, a $2-3\times$ slowdown is observed because loop unrolling will be absent for the dense block-vector multiplication for each thread.

\subsection{Separation of Hessian and Jacobian}\label{section-separation-evaluation}

At the beginning of Sec.~\ref{section-hessian-code} we mentioned how we give the user the choice to generate inner Hessian $\nabla^2_g f(g(x))$ and Jacobian matrix $J_g(x)$ instead of the entire multiplied Hessian matrix $\nabla_x^2 f(g(x))$ in Eq.~\eqref{equation:chain-rule}. In many cases this optimization will not affect the performance. However, GPU is memory sensitive, and a slightly larger matrix will make the Hessian computation run out of memory even though the theoretical memory usage is below the threshold. (We discuss this further in Appendix \ref{section-memory-appendix}.)

Take again the caged bunny example in Sec.~\ref{section-cage}. Each surface vertex is controlled by exactly 8 cage points. This means that in the worst scenario, a collision energy that involves 4 vertices on the caged bunny, will have $8 \times 4 \times 3 = 96$ DoF. The resulting Hessian then always has a theoretical maximum size of $96\times96$. Running the computation code that involves a matrix this large can trigger an out-of-memory from GPU.

However, a closer observation tells us that this $96\times96$ matrix can be nicely separated into the Jacobian matrix of size $12\times96$ and the inner Hessian matrix of size $12\times12$. Generating those two matrices instead leads to a $6\times$ size reduction compared to the $96\times96$ matrix. As a result, when this optimization is turned on, we can successfully run the simulation without any out-of-memory issue.
\section{Memory consumption}
Although YASPS executes the majority of its kernels on the GPU, we do not explicitly optimize for memory footprint. As a result, the observed GPU usage can exceed the theoretical amount required by the simulation.

This behavior arises from CUDA's management of thread-local stack and local memory. CUDA may automatically increase the per-thread stack size for kernels with large stack frames, and this setting is sticky for the rest of the execution.

Consequently, a kernel with large per-thread temporaries can increase the apparent memory footprint of subsequent kernels launched in the same context, even when those later kernels require much less local storage. 

In YASPS, this effect is amplified by the \texttt{UNION} operator, whose generated code may materialize the largest possible per-thread local Hessian matrix at compile time. This increases local-memory pressure and can therefore raise the persistent context-wide memory reservation.

To illustrate this effect, we report memory measurements in Sec.~\ref{section-memory-appendix} for two cases: mat-twist, where the cloth resolution is increased, and a collision example with multiple soft bunnies and affine-bodied bunnies. We will also discuss possible ways to help mitigate the memory usage in that same section of the appendix.
\section{Conclusion and Future Work}
In this paper, we introduced YASPS, a symbolic framework built around two differentiable operators, \texttt{JOIN} and \texttt{UNION}. 
These operators enable users to define new primitive types and parameterizations in a declarative manner, while allowing differentiation to propagate directly through user-defined constructions. 
By treating shape composition and attribute aggregation as differentiable operations, YASPS can assemble gradients and Hessians with known sparsity patterns, and efficiently construct and solve the resulting linear systems.

YASPS demonstrates that exposing structural operators as part of the differentiable program provides both flexibility for rapid prototyping and strong performance guarantees during optimization. Our results show that this approach enables concise implementations of complex simulation models while maintaining competitive—or superior—performance compared to existing systems.

Despite these results, YASPS is still in an early stage of development, and several important limitations remain. Addressing these limitations presents immediate directions for future work.

A first area for improvement is code generation. Currently, the generated code does not explicitly reuse previously allocated registers, whose owner will not participate in any further computations, for new intermediate values. Although \textsc{nvcc} performs a certain level of register pruning, explicitly optimized code generation would further improve performance. In addition, the current common subexpression elimination (CSE) algorithm is relatively basic. For example, in the stable Neo-Hookean energy, the generated code performs nearly $4\times$ more multiplications than code produced by \textsc{SymPy}, indicating substantial room for optimization.

A second area for improvement is differentiation. While the reuse strategy allows YASPS to expose and reason about the structure of energy compositions, it also increases the number of matrix operations required to multiply Jacobians. Moreover, explicitly constructed Jacobian matrices are often sparse, leading to unnecessary computation. Similarly, the current Hessian kernels always reserve space for the largest possible Hessian blocks, which significantly increases GPU memory pressure. More efficient differentiation algorithms that avoid explicit Jacobian construction and excessive memory reservation would greatly improve scalability.

A third area is to expose even more backend to the users. For example, if the Hessian computation is known to the user, as well as the analytic eigenvalues and eigenvectors, it would be beneficial to directly use the user-provided formulation instead of a system generated solution.

Several larger future directions are also promising.

The first is support for dynamic arities.
This limitation prevents efficient representation of relationships such as vertex-to-neighboring-triangle connectivity, and is imposed by the need to allocate static memory when compiling GPU kernels. 

Another direction is support for dynamically sized data attributes. Currently, any change in attribute length requires recomputation of index mappings and Hessian structures, which precludes applications that rely on adaptive remeshing or resolution changes during simulation.

Finally, an especially promising avenue is inverse simulation. Supporting inverse problems would require YASPS to introduce and manipulate the global Jacobian matrix, enabling optimization over control parameters, material properties, or target states. Extension in this direction would further position YASPS as a unified system for both forward and inverse physics-based computation.
\begin{acks}
We appreciate the insightful comments and feedback from the anonymous reviewers and the shepherd. 
The project is supported by NSF Grant 2238839 and gifts from Adobe, Google, and Activision.
Minchen Li acknowledges partial support from a Junior Faculty Startup Fund from Carnegie Mellon University and gift funding from Genesis AI. 
\end{acks}

\bibliographystyle{ACM-Reference-Format}
\bibliography{sections/citations}
\appendix
\section{Index Generator}\label{section-appendex-index}
\subsection{Gradient Size Computation}
The first step of index generation is to determine the size of the global gradient vector. Given a set of minimization-target attributes, the global gradient length is the sum over targets of
\[
(\text{per-instance dimension}) \times (\text{number of instances}).
\]
Concretely, for a target attribute \(\alpha\) with \(n_\alpha\) instances and per-instance shape \(r_\alpha\times c_\alpha\), it contributes \(n_\alpha r_\alpha c_\alpha\) scalar degrees of freedom.

Consider the example we used in Section \ref{section-reuse} for the mixed-material mesh. Suppose \(N\) vertices are free, each contributing \(3\) degrees of freedom, and there is a single affine body with affine matrix \(A\in\mathbb{R}^{3\times 3}\) and translation \(t\in\mathbb{R}^{3}\). Differentiating with respect to these targets yields a global gradient of length
\[
s = 3N + 9 + 3,
\]
and the global Hessian is therefore an \(s\times s\) matrix.

\begin{figure}[h]
  \centering
  \includegraphics[width=\columnwidth]{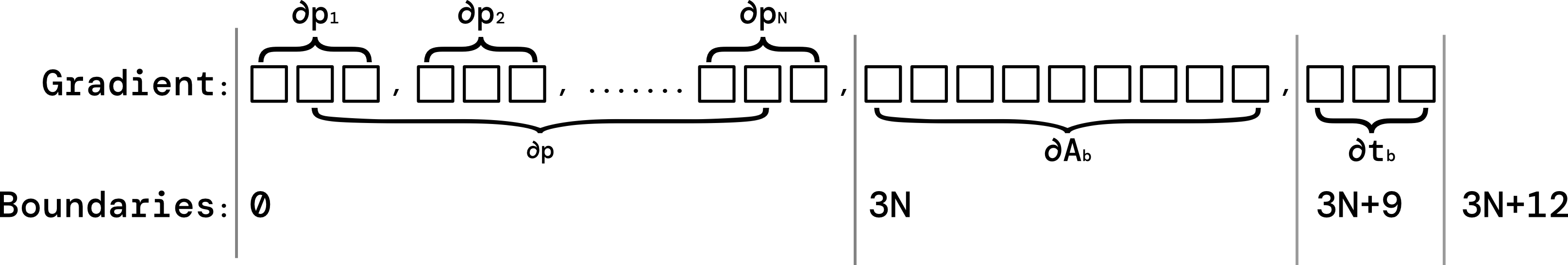}
  \caption{At the beginning of differentiation, YASPS fixes the layout of the global gradient and assigns each target attribute a contiguous segment. The resulting prefix-sum \texttt{Boundaries} array stores the segment boundaries; in this example, three target attributes induce four boundary values.}
  \label{figure-gradient-boundaries}
\end{figure}

During this initialization pass, YASPS assigns each target attribute a contiguous block in the global gradient vector and stores the corresponding prefix-sum offsets in the \texttt{Boundaries} array (Fig.~\ref{figure-gradient-boundaries}). These offsets allow the system to compute, for every attribute instance, the exact gradient range (start offset and block length) in the global layout.

\subsection{Index Size Computation}
With the global layout fixed, the next step is to determine—for each energy term, how many placement indices must be stored in order to scatter its local gradient into the global gradient vector. Importantly, this count is not necessarily equal to the number of scalar entries in the local gradient.

For example, consider a stable Neo-Hookean energy applied to four free vertices. Each vertex position has per-instance dimension \(3\). Although the local gradient has \(12\) scalar entries, it does not require twelve placement indices. Instead, it requires four indices, each indicating where to place a contiguous \(1\times 3\) segment of the local gradient in the global vector.

Because YASPS has access to the symbolic structure of each energy, it can traverse the computation graph to determine the \emph{maximum} number of index entries required at each node. This ensures that our computation kernel for index generation can allocate a static array even before the actual connectivity is set. 

Let \(\kappa\) denote the index-computation function (defined in the next subsection), and let \(|\kappa|(x)\) denote the number of index entries produced for node \(x\). YASPS computes \(|\kappa|\) using the following rules:
\begin{itemize}
\item For any data attribute \(d\), \(|\kappa|(d)=1\), since a single referenced instance of \(d\) corresponds to one contiguous block in the global gradient.

\item For any computation \(f\) that takes attributes \(x_1,\dots,x_n\) as inputs,
\[
|\kappa|\bigl(f(x_1,\dots,x_n)\bigr)
  = |\kappa|(x_1) + \dots + |\kappa|(x_n),
\]
because the placement indices are formed by concatenating the indices required by each input.

\item For a \texttt{JOIN} attribute of arity \(k\),
\[
|\kappa|\bigl(\mathrm{JOIN}_{C}(\alpha_B)\bigr)
  = k \cdot |\kappa|(\alpha_B),
\]
because \texttt{JOIN} gathers \(k\) referenced instances and therefore replicates the underlying index requirements \(k\) times.

\item For a \texttt{UNION} attribute,
\[
|\kappa|\bigl(\mathrm{UNION}(\alpha_{X_1},\dots,\alpha_{X_m})\bigr)
  = \max\bigl(|\kappa|(\alpha_{X_1}),\dots,|\kappa|(\alpha_{X_m})\bigr),
\]
since only one branch is active at runtime and the worst-case branch determines the required index capacity.
\end{itemize}

Using these rules, YASPS performs a symbolic pass over each energy’s computation graph to determine, for every node, the maximum number of indices it may need to store once the computation graph is fixed.

\subsection{Index Computation}\label{section-index-computation}
We now describe how YASPS computes the actual placement indices used to assemble local gradients into the global gradient vector. This procedure resembles the index-size computation above, but differs in one key aspect: whereas \(|\kappa|\) depends only on the symbolic graph structure, \(\kappa\) must be evaluated using runtime information (e.g., connectivity values and the active branch of a \texttt{UNION}).

For clarity, we describe index computation for a single instance. Computing indices for all instances is trivially parallelizable.

\paragraph{Data attributes}
Let \(\alpha\) be a target data attribute with per-instance shape \(r\times c\), and let \(i\) denote the instance index. Let \(\mathrm{start}(\alpha)\) be the (0-based) starting offset of \(\alpha\)’s block in the global gradient array as determined by \texttt{Boundaries}. Then the starting coordinate of the block corresponding to \(\alpha(i)\) is
\[
\mathrm{start}(\alpha) + (r c)\, i.
\]
In practice, YASPS stores placement indices using a 1-based convention so that the value \(0\) can be reserved to denote ``no contribution'' (used by \texttt{UNION} padding). We therefore define the stored index as
\[
\kappa(\alpha(i)) = \bigl(\mathrm{start}(\alpha) + (r c)\, i\bigr) + 1.
\]
(The assembly kernel subtracts \(1\) before indexing into 0-based arrays.)

\paragraph{\texttt{JOIN}}
For a \texttt{JOIN} attribute, indices are gathered according to the connectivity. Let \(C:I_A\to I_B^k\) be the arity-\(k\) connectivity used by \(\mathrm{JOIN}_C(\alpha_B)\). For instance \(i\in I_A\) with \(C(i)=(j_1(i),\dots,j_k(i))\), the placement indices are obtained by concatenating the indices of the referenced base instances:
\[
\kappa\!\left(\mathrm{JOIN}_{C}(\alpha_B)(i)\right)
  = \bigl[
      \kappa(\alpha_B(j_1(i))),\;
      \ldots,\;
      \kappa(\alpha_B(j_k(i)))
    \bigr].
\]

\paragraph{\texttt{UNION}}
Index computation for a \texttt{UNION} attribute is branch-dependent. Although all unioned attributes share the same value shape, they may require different numbers of placement indices. Suppose the precomputed capacity for the union node is \(|\kappa|=p\). For an instance identified by \((j,i)\) (meaning ``instance \(i\) from branch \(j\)''), we compute
\[
\kappa\bigl(\mathrm{UNION}(\alpha_{X_1},\dots,\alpha_{X_m})(j,i)\bigr)
  = \operatorname{pad}\!\left(\kappa(\alpha_{X_j}(i)),\, p\right)
  \in \mathbb{Z}^p,
\]
where \(\operatorname{pad}(\cdot,p)\) appends zeros so that the result has length \(p\). These zeros correspond to inactive index slots and are ignored during assembly.

\paragraph{General computations}
For any computation \(f\) that takes inputs \(x_1,\dots,x_n\), the placement indices are formed by concatenation:
\[
\kappa\bigl(f(x_1,\dots,x_n)\bigr)
  = \bigl[\kappa(x_1), \ldots, \kappa(x_n)\bigr].
\]

\section{Global Hessian Compression}\label{section-hessian-compression}

In Section~\ref{section-index-computation}, we computed, for each energy instance, how its local gradient and Hessian contributions map into the global system. After these per-instance indices (and block coordinates) are collected, YASPS constructs a compressed global Hessian layout.

\paragraph{Block-sparse representation.}
YASPS stores Hessian coordinates per \emph{block} (rather than per scalar entry), and represents the global Hessian in a compressed block-sparse format. 

The first step is to obtain the set of distinct block shapes. This is straightforward, as the possible shapes are given by the Cartesian product of all attributes' dimensions (we only consider the attributes we are minimizing against), which provides a superset of the block shapes that occur in practice. 

For each distinct pair of row/column sizes \((r,c)\), YASPS then gathers all block coordinates contributed by all energy instances (restricted to the upper-triangular part due to symmetry), sorts them lexicographically by \((\text{row},\text{col})\), and removes duplicates. This produces a unique list of global blocks to be stored for each block shape.

The resulting global layout is represented by the following arrays:
\begin{itemize}
  \item \emph{BlockRowSize} and \emph{BlockColSize}: arrays describing the set of unique block shapes \((r,c)\), ordered from smaller to larger (according to a fixed ordering).
  \item \emph{BlockCoordinateStart}: for each block shape, an offset into the global coordinate arrays indicating where the blocks of that shape begin.
  \item \emph{RowCoordinate} and \emph{ColCoordinate}: the row/column indices of each \emph{unique} block in the global Hessian (stored in upper-triangular form).
  \item \emph{HessianBlocks}: the numerical storage for all unique Hessian blocks, laid out contiguously by block shape.
  \item \emph{PositionInData}: a per-instance lookup table that maps each local sub-block produced by an energy instance to the corresponding destination location in \emph{HessianBlocks}.
\end{itemize}

\paragraph{Assembly.}
After evaluating and (locally) compressing a Hessian instance as described in Section~\ref{section-hessian-code}, the kernel first uses the per-instance index information from Section~\ref{section-index-computation} to determine the global block coordinates of all sub-blocks contributed by that instance. Using the precomputed lookup (\emph{PositionInData}), YASPS then identifies the destination block in \emph{HessianBlocks} and accumulates the local contribution into the global storage (typically via atomic additions).

This global compression reduces the number of stored blocks and eliminates duplicates, which in turn substantially lowers the cost of downstream operations such as sparse matrix-vector products (SpMV), since fewer blocks need to be stored and multiplied.

\section{SpMV}\label{section-spmv}

As the conjugate gradient (CG) solver is largely dominated by sparse matrix–vector multiplication (SpMV), optimizing this kernel is critical. While block compression significantly reduces the total number of nonzero blocks, further performance gains can be achieved by improving the SpMV kernel itself.

A naïve SpMV strategy for the block-sparse format is to launch a generic kernel in which each thread processes a single block: loading its coordinates, multiplying the dense block with the right-hand-side vector, and additionally accumulating the transpose contribution.

However, in our storage layout, blocks are first grouped by block dimension and then sorted by row index. This structure allows us to generate a specialized SpMV kernel for each block dimension. By doing so, the dense block–vector multiplication can be fully unrolled at compile time, eliminating loop overhead and improving instruction-level efficiency.

Furthermore, because blocks of the same dimension are sorted by row, it is common for multiple blocks processed within the same CUDA thread block to contribute to the same contiguous segment of the output vector. We exploit this property by performing an intra-block reduction, as shown in Algorithm~\ref{alg:spmv}. Unlike classical warp-level reductions, this reduction is linear rather than logarithmic, since blocks within a CUDA block are not guaranteed to all belong to the same row. Instead, threads perform a forward scan over contiguous row segments and accumulate partial results locally before atomically updating the output vector.

Although this reduction does not have logarithmic complexity, it still reduces the number of atomic operations and improves memory locality. We demonstrate that this strategy yields substantial performance improvements in Section ~\ref{section-evaluation-spmv}.

\section{Preconditioner}\label{section-preconditioner}
To effectively reduce the number of iterations required by the PCG method, YASPS uses a block Jacobi preconditioner.

Instead of solving the system
\[
H x = g,
\]
we solve
\[
M^{-\frac{1}{2}} H M^{-\frac{1}{2}} y = M^{-\frac{1}{2}} g, \quad x = M^{-\frac{1}{2}} y,
\]
where $M$ is a block-diagonal approximation of $H$, and $M^{-\frac{1}{2}}$ is the Cholesky factorization of $M^{-1}$. With proper variable substitutions, the system can be solved without factorizing $M^{-1}$ \cite{solomon2015numerical}. In YASPS, every time a local Hessian contribution is inserted into the global system, its diagonal block is simultaneously accumulated into a dedicated global array. 
Once the global Hessian assembly is complete, YASPS inverts each diagonal block independently. These blocks are typically small, and the inversions are trivially parallelizable on the GPU, yielding an explicit representation of $M^{-1}$.

As a reference, in the example shown in Section~\ref{section-cage}, computing the block-diagonal inverse takes $0.53$~ms, whereas the Hessian and gradient computation for the static component alone takes $20.63$~ms. The cost of forming the preconditioner therefore constitutes only a small fraction of the overall execution time.

While the reduction in CG iterations varies across scenarios, the block Jacobi preconditioner consistently improves convergence. In particular, for scenes composed entirely of affine body dynamics (ABD) meshes, inverting the diagonal blocks alone is sufficient to directly invert the system. More generally, we observe a $20$--$30\%$ reduction in CG iterations compared to using a scalar diagonal (Jacobi) preconditioner in many practical cases.

\section{GPU Memory}\label{section-memory-appendix}
\subsection{Mat Twist}
\begin{figure}
    \centering
    \includegraphics[width=1.0\columnwidth]{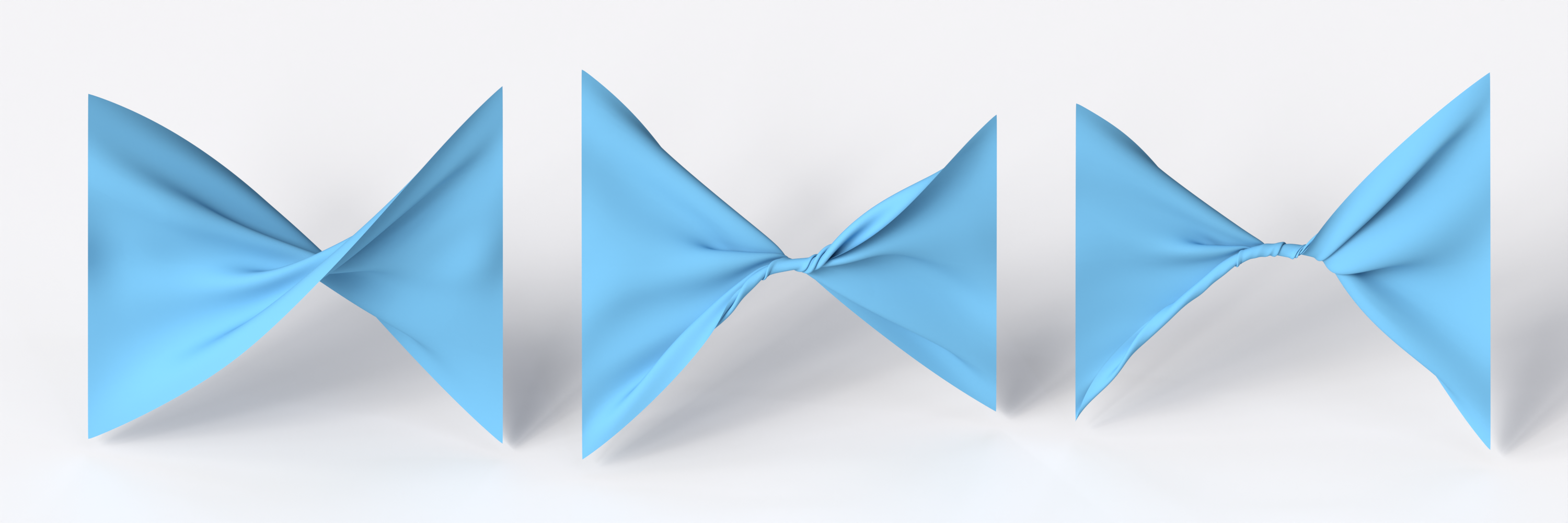}
    \vspace{-10pt}
    \caption{We continuously twist a piece of cloth for 2 seconds, with a time step of 0.001s, and a rotation of each side at 5 rad per second. We report the memory consumption for this same setting at different cloth resolution in Table~\ref{table-mat-twist}.}
    \label{figure-mat-twist}
\end{figure}

\begin{table*}[htbp] 
\small
\centering 
\noindent
\caption{Timing and memory consumption for the twisting mat example shown in Fig.~\ref{figure-mat-twist}. For all resolutions, we use the same CCD query capacity, with maximum limits of $10^8$ collision pairs and $10^8$ continuous collision pairs. For each case, we report the mesh resolution (number of vertices, faces, and edges), the maximum number of collision pairs, the total runtime of each major component, the GPU memory allocated for the CCD module, and the minimum as well as the maximum GPU memory used by YASPS over the 2{,}000 frames. YASPS memory is computed by subtracting the memory initialization for CCD from the total memory consumed by the program. This means that the minimum YASPS memory will always record the memory allocated by YASPS itself (excluding CCD module) when no collision happens.
However, if the CCD module allocates any additional memory during the computation, it will be captured in the max YASPS memory entry.}

\begin{tabularx}{\textwidth}{|l|l|l|X|X|X|X|r|r|r|}
    \hline
    \thead{\# Vertices} & \thead{\# Faces} & \thead{\# Edges} & \thead{\# Collision\\Pairs(Max)} & \thead{CCD \& CD\\Total (s)} & \thead{Diff\\Total (s)} & \thead{CG\\Total (s)} & \thead{CCD\\Memory (MB)} & \thead{Min YASPS\\Memory (MB)} & \thead{Max YASPS\\Memory (MB)} \\ \hline \hline
    10,000 & 19,602 & 29,601 & 109,898 & 172.69 & 109.93 & 30.58 & 9216.98 & 2027.75 & 2086.87\\ \hline
    40,000 & 79,202 & 119,201 & 467,006 & 224.77 & 356.51 & 54.85 & 9250.54 & 2137.00 & 2248.47\\ \hline
    90,000 & 178,802 & 268,801 & 1,097,477 & 305.79 & 705.38 & 120.90 & 9317.65 & 2361.39 & 2766.01\\ \hline
    160,000 & 318,402 & 478,401 & 2,014,214 & 435.10 & 1190.16 & 248.65 & 9380.56 & 2562.66 & 3290.43\\ \hline
    250,000 & 498,002 & 748,001 & 3,301,877 & 609.22 & 1839.86 & 439.90 & 9477.03 & 2858.42 & 4053.27\\ \hline
    360,000 & 717,602 & 1,077,601 & 4,545,714 & 682.14 & 2277.02 & 733.69 & 9598.66 & 3290.37 & 4920.97\\ \hline
    490,000 & 977,202 & 1,467,201 & 6,204,614 & 933.33 & 3147.01 & 1161.76 & 9724.49 & 3694.46 & 5866.13\\ \hline
    640,000 & 1,276,802 & 1,916,801 & 8,393,014 & 1170.94 & 4111.04 & 1780.36 & 9892.27 & 4209.83 & 6529.87\\ \hline
    810,000 & 1,616,402 & 2,426,401 & 10,798,135 & 1516.92 & 5381.39 & 2524.11 & 10070.52 & 4806.68 & 8428.46\\ \hline
    1,000,000 & 1,996,002 & 2,996,001 & 12,806,211 & 2054.95 & 6828.68 & 3819.17 & 10261.36 & 5427.57 & 9718.21\\ \hline
\end{tabularx}
    
\label{table-mat-twist}
\end{table*}
In this first example, we stress test the system by continuously twisting a piece of cloth shown in Fig.~\ref{figure-mat-twist}, and report the timing as well as the memory usage in Table~\ref{table-mat-twist} under different resolutions.

Under this setting, the largest per-thread local Hessian among all energies is of size $12 \times 12$, produced by the point-triangle collision energy and the bending energy (which involves 4 vertices in a hinge stencil, with 2 adjacent triangles).

To demonstrate that the memory increase scales steadily with respect to both mesh resolution and the number of collision pairs, we also report the following two metrics:
\begin{itemize}
    \item \textbf{Minimum YASPS memory increase per 10k vertices}, defined as
    \begin{equation}\label{equation:memory-increase-vertices}
    \frac{
        M_{\text{min}}^{(i)} - M_{\text{min}}^{(i-1)}
    }{
        |V^{(i)}| - |V^{(i-1)}|
    }
    \times 10{,}000,
    \end{equation}
    where $M_{\text{min}}^{(i)}$ denotes the minimum observed YASPS memory at resolution level $i$, and $|V^{(i)}|$ is the corresponding number of vertices. This checks if YASPS' memory allocation for the simulation without any collision indeed scales linearly with the number of primitive instances (vertices, triangle, edges) in the scene. Note that the number of triangles and edges also scales linearly with the number of vertices, as shown in Table~\ref{table-mat-twist}.

    \item \textbf{Memory increase per 100k collision pairs}, defined as
    \begin{equation}\label{equation:memory-increase-cp}
    \frac{ M_{\text{max}}^{(i)} - M_{\text{min}}^{(i)}}{C^{(i)}} \times 100{,}000,
    \end{equation}
    where $M_{\text{max}}^{(i)}$ denotes the maximum observed YASPS memory at resolution level $i$, and $C^{(i)}$ is the maximum number of collision pairs at resolution $i$. This evaluates how memory behaves when we increase the number of collision pairs, which in turn increase the number of non-zero entries in the global Hessian matrix, as well as the number of local Hessian blocks generated for collision energies.
\end{itemize}

\begin{figure}
    \centering
    \includegraphics[width=1.0\columnwidth]{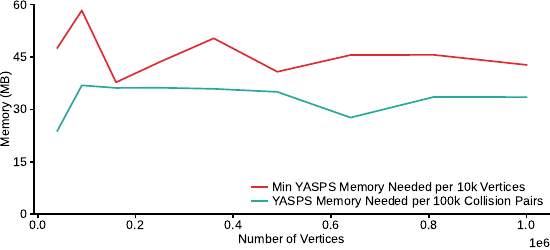}
    \vspace{-10pt}
    \caption{The two metrics defined in Eq.~\eqref{equation:memory-increase-vertices} and Eq.~\eqref{equation:memory-increase-cp} for the simulation shown in Fig.~\ref{figure-mat-twist}. All numbers are computed from Table~\ref{table-mat-twist}.}
    \label{figure-mat-twist-memory}
\end{figure}

As shown in Fig.~\ref{figure-mat-twist-memory}, both metrics stabilize as the mesh resolution increases, indicating that the amortized memory growth approaches a constant. 

\subsection{Soft and Affine Body Bunnies}
\begin{figure}
    \centering
    \includegraphics[width=1.0\columnwidth]{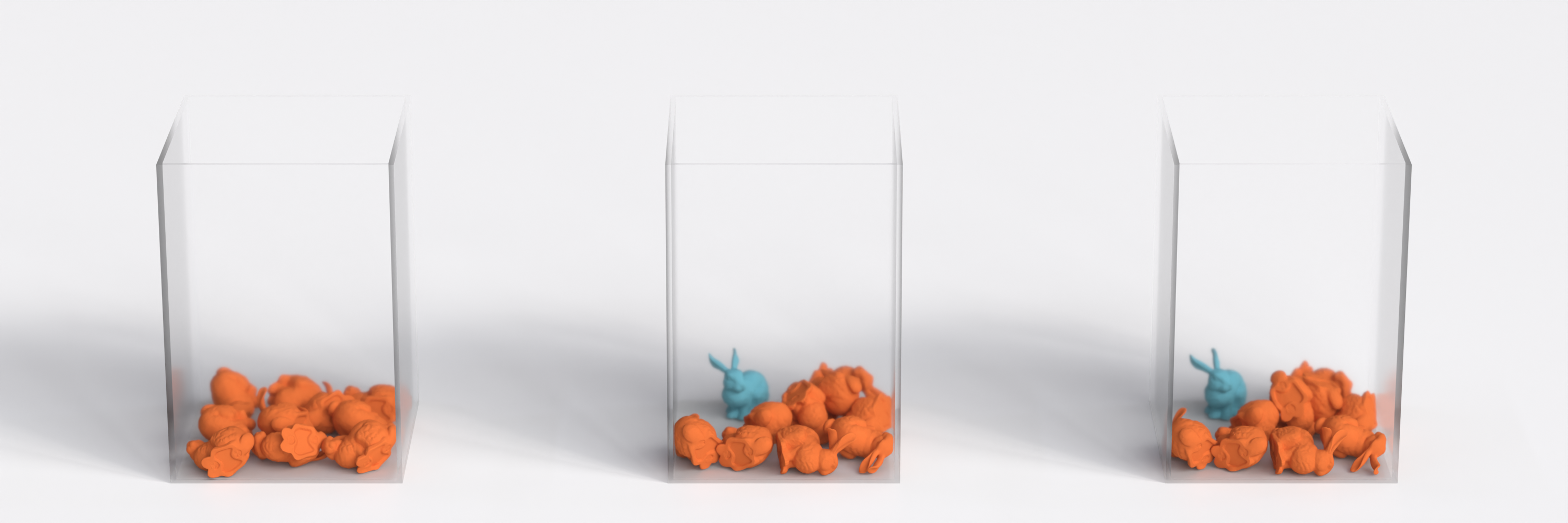}
    \vspace{-10pt}
    \caption{We drop 10 bunnies in a container with different settings to show how different matrix sizes generated in Hessian computation affect the total memory consumption. \textbf{Left}: all 10 bunnies are soft. \textbf{Middle}: 9 bunnies are soft and 1 bunny is controlled by affine body (colored light blue). \textbf{Right}: same as middle, but we turn on the optimization where we separate the generation of inner Hessian and Jacobian matrices, which reduces the memory size.}
    \label{figure-dropping-in-container-memory}
\end{figure}

\begin{table*}[htbp] 
\small
\centering 
\noindent
\caption{Performance and memory statistics for the simulation shown in Fig.~\ref{figure-dropping-in-container-memory}. For every group of three rows, the number of bunnies is increased by one (each bunny contains 19{,}193 vertices), and we evaluate three settings: fully soft, mixed soft and affine, and mixed with turning on the optimization introduced in Sec.~\ref{section-separation-evaluation}.
For each configuration, we report runtime breakdowns (collision detection, differentiation, and conjugate gradient solve), along with the maximum YASPS GPU memory consumption and the maximum CUDA per-thread stack size limit during the simulation.}

\begin{tabularx}{\textwidth}{|l|l|l|X|X|X|X|r|r|}
    \hline
     \thead{\# Vertices} & \thead{\# Soft\\Bunnies} & \thead{\# Affine\\Bunnies} &  \thead{Separate\\Jacobian}  & \thead{CCD \& CD\\Total (s)} & \thead{Diff\\Total (s)} & \thead{CG\\Total (s)} & \thead{Max YASPS\\Memory (MB)} & \thead{Thread Stack\\Limit (KB)} \\ \hline \hline
    115,158 & 6 & 0 & False & 71.48 & 137.36 & 53.14 & 2438.92 & 8.72\\ \hline
    115,158 & 5 & 1 & False & 71.46 & 188.78 & 46.44 & 10661.92 & 43.53\\ \hline
    115,158 & 5 & 1 & True & 73.73 & 192.48 & 46.61 & 5389.02 & 21.28\\ \hline
    134,351 & 7 & 0 & False & 99.97 & 192.31 & 77.93 & 2503.93 & 8.72\\ \hline
    134,351 & 6 & 1 & False & 100.43 & 257.92 & 66.55 & 10710.15 & 43.53\\ \hline
    134,351 & 6 & 1 & True & 105.45 & 277.91 & 72.67 & 5438.63 & 21.28\\ \hline
    153,544 & 8 & 0 & False & 126.19 & 246.42 & 98.10 & 5274.27 & 8.72\\ \hline
    153,544 & 7 & 1 & False & 121.27 & 331.57 & 88.41 & 10785.78 & 43.53\\ \hline
    153,544 & 7 & 1 & True & 129.92 & 329.18 & 92.73 & 5515.64 & 21.28\\ \hline
    172,737 & 9 & 0 & False & 159.13 & 313.58 & 149.46 & 11541.55 & 8.72\\ \hline
    172,737 & 8 & 1 & False & 144.59 & 387.55 & 108.66 & 10848.56 & 43.53\\ \hline
    172,737 & 8 & 1 & True & 155.20 & 392.04 & 113.75 & 5582.61 & 21.28\\ \hline
    191,930 & 10 & 0 & False & 181.48 & 358.41 & 156.24 & 2696.42 & 8.72\\ \hline
    191,930 & 9 & 1 & False & 166.88 & 434.80 & 132.92 & 10905.32 & 43.53\\ \hline
    191,930 & 9 & 1 & True & 183.93 & 462.10 & 138.83 & 5635.04 & 21.28\\ \hline
\end{tabularx}
    
\label{table-dropping-in-container}
\end{table*}

For the second example, we re-run the dropping-bunnies-in-a-container simulation shown in Fig.~\ref{figure-dropping-in-container} with $N$ bunnies, where $6 \leq N \leq 10$. For each value of $N$, we evaluate three settings (the rendered result is shown in Fig.~\ref{figure-dropping-in-container-memory}):
\begin{itemize}
    \item All bunnies are soft, where each vertex is controlled by its own 3 DoF.
    \item One bunny is parameterized as an affine body. The collision kernel is constructed over a \texttt{UNION} of soft vertices and affine-body vertices.
    \item Same as the previous setting, but with the optimization in Sec.~\ref{section-separation-evaluation} enabled.
\end{itemize}

Since each affine-body vertex is associated with 12 degrees of freedom, when the collision energy involves the \texttt{UNION} of soft vertices and affine-body vertices, the largest possible per-thread Hessian for collision energies in configurations involving affine bodies increases to $48 \times 48 = 2304$ entries after local matrix expansion.

When the optimization that separates the inner Hessian and the Jacobian is enabled, the required storage reduces to $12 \times 12 + 12 \times 48 = 720$ entries. Although this optimization introduces an additional temporary $12 \times 48$ matrix to materialize partial multiplication results, the total number of temporary variables remains significantly smaller than $2304$.

As shown in Table~\ref{table-dropping-in-container}, the maximum YASPS memory is positively correlated with the per-thread stack limit. In fact, the ratio of maximum YASPS memory between different settings closely matches the ratio of their per-thread stack limits. 

Notably, although not explicitly shown in the table, the number of collision pairs is similar across settings with the same number of soft and affine bunnies. This indicates that the observed reduction in memory consumption when enabling the optimization is not driven by differences in collision workload, but rather by reduced per-thread temporary storage requirements.

\subsection{Mitigation}
Since the GPU memory overhead primarily stems from the persistent per-thread stack size within a single CUDA execution context, two seemingly straightforward solutions arise:
\begin{itemize}
    \item Resetting the stack size after each kernel execution.
    \item Separating execution into multiple CUDA contexts.
\end{itemize}

However, neither approach is well-suited for our simulation setting. For the first approach, the kernels are invoked repeatedly within each simulation loop. Even if the stack size is reset, subsequent kernel launches will increase it again to meet their requirements. This leads CUDA to repeatedly resize the per-thread stack allocation, introducing additional overhead without reducing the overall memory footprint.

For the second approach, although separating kernels into different CUDA contexts can in principle isolate stack growth, it is not effective in practice. Memory allocated within one context is not shared with others. As a result, splitting execution across multiple contexts can lead to duplicated memory usage rather than reduction. Without a mechanism to dynamically manage contexts based on stack requirements, blindly assigning kernels to different contexts may increase the total memory footprint. Furthermore, frequent context switching and independent allocation patterns across contexts can exacerbate memory fragmentation.

That being said, there are several ways to reduce memory usage.

\subsubsection{Front-End Code Optimization}
From the user's perspective, memory usage can be reduced by reformulating the computation. For example, in point--triangle collision energy, instead of expressing the computation as a \texttt{JOIN} over four vertices, each from a possibly independent affine body, one can formulate it as a computation over one vertex and one triangle, where the triangle is represented by an affine transformation (e.g., an affine matrix and a translation). In this formulation, the maximum Hessian size can be reduced to $24 \times 24$ (corresponding to collision of two affine bodies), significantly lowering per-thread memory pressure.

\subsubsection{Code Generation}
Another approach to alleviating memory pressure is to optimize the code generation routine. Two optimizations are particularly relevant.

The first is to offload intermediate computations to global memory. This reduces per-thread stack usage at the cost of additional memory accesses, resulting in a modest performance trade-off.

The second is specific to YASPS' Hessian code generation routine. Currently, YASPS materializes many matrices, either to store the uncompressed final Hessian in memory or to compress the Hessian matrix locally. However, these matrices do not necessarily need to be materialized if accumulation and compression are performed directly in global memory.

\subsubsection{Locally Sparse Representation}
Probably the most effective approach is to exploit sparsity in local Jacobian matrices. Consider the position of a vertex in an affine body mesh, computed as
\[
p = Ar + t,
\]
where $p$ is the current position, $A$ is the affine body matrix, $r$ is the rest position, and $t$ is the affine body translation.

The derivative of $p$ with respect to $A$ (with column-major flattening) is
\[
\frac{\partial p}{\partial A} =
\begin{bmatrix}
r^T & 0   & 0 \\
0   & r^T & 0 \\
0   & 0   & r^T
\end{bmatrix},
\]
which is highly sparse. 

In YASPS, if we directly multiply $\frac{\partial p}{\partial A}$ with another variable, YASPS will automatically exclude the zeros from computation, which is also reflected in the generated code by never performing the zero multiplications. However, if this Jacobian is \texttt{UNION}ed or \texttt{JOIN}ed with other variables, YASPS will materialize this matrix as an \textsc{Eigen} dense matrix, and offload the multiplication to the \textsc{Eigen} library. This wastes both computation on redundant operations and memory on storing zeros. Thus, designing a sparse local representation that integrates naturally with \texttt{JOIN} and \texttt{UNION} operations and works on GPU could therefore significantly reduce memory usage while preserving computational structure.
\section{Core Syntax}\label{section-core-syntax}

\begin{figure*}[t!]
\centering
\small

\begin{tabular}{@{}p{0.59\textwidth}
p{0.40\textwidth}@{}}
\multicolumn{2}{@{}c@{}}
{\rule{\textwidth}{0pt}}\\

$
\begin{array}{rcl@{\qquad}l}
\multicolumn{4}{@{}c}{\sigma ::= (n_r, n_c), \qquad n_r, n_c \in \mathbb{N}_{>0}} \\[5pt]
\multicolumn{4}{@{}c}{sid, mid, pid, uid, cid, aid \in \mathsf{String}, \qquad i, k \in \mathbb{N}_{\geq 0}} \\[5pt]
\multicolumn{4}{@{}l}{\rule{\linewidth}{1pt}} \\
\multicolumn{4}{@{}c}{\textbf{Scene Definition}} \\[-0.5em]
\multicolumn{4}{@{}l}{\rule{\linewidth}{1pt}}\\[5pt]

\texttt{scene} & ::= & \mathsf{scene}(sid)
& \begin{tabular}[t]{@{}l@{}}
\text{initialize a \texttt{scene} with a name} $sid$
\end{tabular} \\[10pt]

\texttt{mesh} & ::= & 
\mathsf{mesh}(mid, \texttt{scene})
& \begin{tabular}[t]{@{}l@{}}
create a \texttt{mesh} with a name $mid$\\[-0.2em]
under a \texttt{scene}
\end{tabular} \\[10pt]

& \mid & \texttt{scene}.mid
& \begin{tabular}[t]{@{}l@{}}
access a \texttt{mesh} with a name $mid$\\ [-0.2em]
from a \texttt{scene}
\end{tabular} \\[15pt]

\texttt{prim} & ::= & \mathsf{primitive}(pid, \texttt{mesh}, i)
& \begin{tabular}[t]{@{}l@{}}
create a \texttt{primitive type}  with\\ [-0.2em]
a name $pid$, number of instances $i$, \\[-0.2em]
under a \texttt{mesh} 
\end{tabular} \\[10pt]

& \mid & \texttt{mesh}.pid
& \begin{tabular}[t]{@{}l@{}}
access a \texttt{primitive type} with\\ [-0.2em]
a name $pid$ from a \texttt{mesh}
\end{tabular} \\[15pt]

\texttt{pUnion} & ::= & 
\begin{aligned}[t]
\mathsf{primitiveUnion}(&uid,\texttt{mesh}, \\
&[\delta_1, \ldots, \delta_n])
\end{aligned}
& \begin{tabular}[t]{@{}l@{}}
create a \texttt{primitive union} with\\ [-0.2em]
a name $uid$, a list of $\delta$s,\\ [-0.2em]
under a \texttt{mesh} \\ 
\end{tabular} \\[10pt]

& \mid & \texttt{mesh}.uid
& \begin{tabular}[t]{@{}l@{}}
access a \texttt{primitive union} with\\ [-0.2em]
a name $uid$ from a \texttt{mesh}
\end{tabular} \\[15pt]

\delta & :: = & \texttt{prim} \mid \texttt{pUnion} 
& \begin{tabular}[t]{@{}l@{}}
a $\delta$ type is either a \texttt{primitive type} \\ [-0.2em]
or a \texttt{primitive union}
\end{tabular} \\[10pt]

& \mid & c.\texttt{from}
& \begin{tabular}[t]{@{}l@{}}
accessing $\delta_1$\\ [-0.2em]
from a \texttt{connectivity} $c$
\end{tabular} \\[10pt]

& \mid & c.\texttt{to}
& \begin{tabular}[t]{@{}l@{}}
accessing $\delta_2$\\ [-0.2em]
from a \texttt{connectivity} $c$
\end{tabular} \\[15pt]

c & :: = & \mathsf{connectivity}(cid, \delta_1, \delta_2, k)
& \begin{tabular}[t]{@{}l@{}}
create a connectivity with\\[-0.2em]
a name $cid$, from $\delta_1$ to $\delta_2$,\\ [-0.2em]
with arity $k$, \\[-0.2em]
under $\delta_1$ \\
\end{tabular} \\[10pt]

& \mid & \delta.cid
& \begin{tabular}[t]{@{}l@{}}
access a connectivity with\\ [-0.2em]
a name $cid$ from a $\delta$ type
\end{tabular} \\[15pt]

h & ::= &\texttt{scene} \mid \texttt{mesh} \mid \delta & \text{any host that can have attributes} \\[15pt]
\end{array}
$

&
$
\begin{array}{rcl@{\qquad}l}
\multicolumn{4}{@{}l}{\rule{\linewidth}{1pt}} \\
\multicolumn{4}{@{}c}{\textbf{Attribute Definition}} \\[-0.5em]
\multicolumn{4}{@{}l}{\rule{\linewidth}{1pt}}\\[5pt]

\alpha & ::= & \mathsf{data}(aid, \sigma, h)
& \begin{tabular}[t]{@{}l@{}}
create a data attribute \\[-0.2em]
with a name $aid$,\\ [-0.2em]
a dimension $\sigma$, \\[-0.2em]
under a host $h$
\end{tabular} \\[10pt]
& \mid & \mathsf{constant}(aid, \sigma, h)
& \begin{tabular}[t]{@{}l@{}}
create a constant attribute \\[-0.2em]
with a name $aid$,\\ [-0.2em]
a dimension $\sigma$, \\[-0.2em]
under a host $h$
\end{tabular} \\[10pt]

& \mid & \mathsf{attr}(aid, e, h)
& \begin{tabular}[t]{@{}l@{}}
create a compute attribute \\[-0.2em]
with a name $aid$,\\ [-0.2em]
an expression $e$, \\[-0.2em]
under a host $h$
\end{tabular} \\[10pt]

\multicolumn{4}{@{}l}{\rule{\linewidth}{1pt}} \\
\multicolumn{4}{@{}c}{\textbf{Attribute Expressions}} \\[-0.5em]
\multicolumn{4}{@{}l}{\rule{\linewidth}{1pt}}\\[5pt]

e & ::= & v \in \mathbb{R}
  & \text{float values} \\[0.1em]

  & \mid & h.aid
  & \text{attribute access} \\[0.1em]

  & \mid & [e_1,\ldots,e_n]
  & \text{array construction} \\[0.1em]

  &\mid& e[i]
  & \text{array access} \\[0.1em]

  & \mid & \mathsf{op}(e_1,\ldots,e_n)
  & \text{operator application} \\[0.1em]

  & \mid & \mathsf{JOIN}_c(e)
  & \text{\texttt{JOIN} attribute through \(c\)} \\[0.1em]

  & \mid & \mathsf{UNION}(e_1,\ldots,e_n)
  & \text{\texttt{UNION} attribute} \\[0.4em]

\mathsf{op}
  & ::= & + \mid - \mid \times \mid /
  & \text{} \\[0.1em]

  & \mid & \sin \mid \cos \mid \exp \mid \log
  & \text{transcendental} \\[0.1em]

  & \mid & \mathsf{select} \mid \geq \mid \leq \mid \cdots
  & \text{logical operations} \\[0.1em]

  & \mid & \mathsf{reshape}
  & \text{reshaping the dimension} \\[0.1em]

  & \mid & \mathsf{row} \mid \mathsf{col}
  & \text{row, column access} \\[0.1em]

  & \mid & \mathsf{cross} \mid \mathsf{dot} \mid \mathsf{norm}
  & \text{vector operations} \\[0.1em]
  
  & \mid & \mathsf{inv} \mid \mathsf{det} \mid \mathsf{transpose}
  & \text{matrix operations} \\[0.1em]

\multicolumn{4}{@{}l}{\rule{\linewidth}{1pt}} \\
\multicolumn{4}{@{}c}{\textbf{Declarations}} \\[-0.5em]
\multicolumn{4}{@{}l}{\rule{\linewidth}{1pt}}\\[5pt]

d & ::= & scene \mid mesh \mid \delta \mid c \\
 & \mid & \alpha \\[5pt]

D & ::= & d
  & \text{declaration} \\
  & \mid & D \, ; \, d
  & \text{declaration list}\\

\end{array}
$
\\[1.2em]




\end{tabular}

\vspace{-2pt}
\caption{
Core YASPS syntax. The left grammar defines declarations for scenes, meshes, primitives, connectivity relations. The right grammars define attribute initialization, symbolic attribute expressions and declarations in YASPS. 
Here $d$ denotes a declaration and $D$ a sequence of declarations, while $e$ denotes an expression. 
The tuple $\sigma = (n_r,n_c)$ denotes the per-instance shape of an attribute. The symbol $\delta$ denotes a primitive domain, which is either a primitive type $p$ or a primitive union $u$, and $h$ denotes an attribute host (a scene, mesh, primitive, or primitive union). Thus $h.aid$ refers to an attribute stored on $h$ with the name $aid$. 
}
\label{figure-syntax}
\end{figure*}
\begin{table}
\centering
\caption{A one to one correspondence for some entries in our core syntax in Fig.~\ref{figure-syntax} to the \textsc{Python} API. On the right column, we use 3 different typography for 3 different purposes. A \texttt{mono-spaced} font means function call. A \textbf{bold} font indicates an object (as opposed to a class). $Math$ $mode$ for string values, numeric values or an attribute expression.}
\begin{tabular*}{\columnwidth}{@{\extracolsep{\fill}}ll}

\midrule[1pt]
$\mathsf{scene}(sid)$        & \texttt{addScene($sid$)} \\

$\mathsf{mesh}(mid, \texttt{scene})$ & \textbf{scene}.\texttt{addMesh($mid$)}\\

$\mathsf{primitive}(pid, \texttt{mesh}, i)$ & \textbf{mesh}.\texttt{addPrimitive($pid$, $i$)} \\

$
\begin{aligned}[t]
\mathsf{primitiveUnion}(&uid,\texttt{mesh}, \\
&[\delta_1, \ldots, \delta_n])
\end{aligned}
$
&
$
\begin{aligned}[t]
\textbf{me}&\textbf{sh}.\texttt{addPrimitiveUnion(}\\
&uid,\\
&\texttt{[}\textbf{prim\_1}\texttt{,}\textbf{$\mathbf{\ldots}$}\texttt{,}\textbf{prim\_n}\texttt{]}\\
\texttt{)}
\end{aligned}
$
\\
$\mathsf{connectivity}(cid, \delta_1, \delta_2, k)$ 
& 
$
\begin{aligned}[t]
\textbf{pr}&\textbf{im\_1}.\texttt{addConnectivity(}\\
&uid\texttt{,}\textbf{prim\_2}\texttt{,}\\
&[\ldots]\texttt{, }k\\
\texttt{)}
\end{aligned}
$
\\

$\mathsf{data}(aid, (n_r, n_c), h)$
&
$
\begin{aligned}[t]
\textbf{h}.&\texttt{addAttribute($aid$}\\
&\text{rows=$n_r$\texttt{,} cols=$n_c$}\\
\texttt{)}
\end{aligned}
$
\\

$\mathsf{constant}(aid, (n_r, n_c), h)$
&
$
\begin{aligned}[t]
\textbf{h}.&\texttt{addConstant($aid$}\\
&\text{rows=$n_r$\texttt{,} cols=$n_c$}\\
\texttt{)}
\end{aligned}
$
\\

$\mathsf{attr}(aid, e, h)$
&
$
\begin{aligned}[t]
\textbf{h}.&\texttt{addAttribute($aid$}\\
&\text{computed\_attribute=$e$}\\
\texttt{)}
\end{aligned}
$
\\



$\mathsf{attr}(aid, \mathsf{JOIN}_c(e), h)$
&
$
\begin{aligned}[t]
\textbf{h}.&\texttt{addAttribute($aid$}\\
&\text{through=$c$,}\\
&\text{source=$e$}\\
\texttt{)}
\end{aligned}
$
\\

$\mathsf{attr}(aid, \mathsf{UNION}(e_1\ldots e_n), \texttt{pUnion})$
&
\textbf{pUnion}.\texttt{addAttribute($aid$})
\\

\midrule[1pt]
\end{tabular*}
\label{syntax-reference}
\end{table}

We present a compact abstract syntax for how symbolic attributes can be declared and how the scene, mesh and primitive can be declared in YASPS in Fig.~\ref{figure-syntax}. This excludes syntax for declaring energies, obtaining the differentiation and minimization, whose API calls are shown in Sec.~\ref{section-energy} and Sec.~\ref{section-solution-extraction}. 

In addition, we also add Table.~\ref{syntax-reference} for one-to-one core syntax to the \textsc{Python} API reference, whose semantics are presented in Sec.~\ref{section-frontend}.
\section{Comparison of Tet4 and Hex8}\label{section-hex8}

\begin{table}
\centering
\caption{Statistics for the simulation scene, meshes and their material properties shown in Fig.~\ref{figure-cage}.}
\begin{tabular*}{\columnwidth}{@{\extracolsep{\fill}}lr}

\midrule[1pt]
\multicolumn{2}{c}{Soft and ABD Bunnies} \\
\midrule[1pt]
Vertices        & 19,193 \\
Tetrahedra      & 79,935 \\
Surface triangles & 20,832 \\
Edges           & 31,248 \\
Surface vertices & 10,418 \\
Young's modulus (Soft) & 3.17 $\times 10^1$ kPa \\
Poisson ratio (Soft) & 0.226 \\
Mass (soft) & $1.0$ kg \\
Young's modulus (ABD) & 9.88 $\times 10^3$ kPa \\
Poisson ratio (ABD) & 0.485 \\
Mass (ABD) & $0.5$ kg \\

\midrule[1pt]
\multicolumn{2}{c}{Cloth} \\
\midrule[1pt]
Vertices  & 10,201 \\
Triangles & 20,000 \\
Edges     & 30,200 \\
Stretch stiffness & 3.55 $\times 10^2$ kPa \\
Shear stiffness & 1.00 $\times 10^2$ kPa \\
Bending stiffness & 0.25 \\
Thickness & $0.001$ m \\
Mass & $1.0$ kg \\

\midrule[1pt]
\multicolumn{2}{c}{Caged Bunny} \\
\midrule[1pt]
Vertices & 6,172 \\
Triangles    & 12,340 \\
Edges    & 18,510 \\
Mass & $6.2$ kg \\

\midrule[1pt]
\multicolumn{2}{c}{Cages} \\
\midrule[1pt]
Vertices     & 736 \\
Number of cages & 451 \\
Number of tetrahedra per cage & 6 \\
Number of Gaussian quadrature points per cage & 8 \\
Young's modulus (Soft) & $0.1$ kPa \\
Poisson ratio (Soft) & 0.25 \\

\midrule[1pt]
\multicolumn{2}{c}{Scene} \\
\midrule[1pt]
Frames & 200 \\
Time step size & 0.01s \\

\bottomrule
\end{tabular*}
\label{table-hex8-stats}
\end{table}
\begin{table}[t]
\centering
\caption{Performance stats for the simulation shown in Fig.~\ref{figure-cage} under two settings. For this example we also include total compilation time for the solver kernel, computation kernels, and index kernels, as well as the total number of kernels generated by YASPS during the simulation.}
\label{table-hex-8-performance}

\begin{tabular*}{\columnwidth}{@{\extracolsep{\fill}}lr}
\midrule[1pt]

\multicolumn{2}{c}{\textbf{Tet4}} \\
\midrule[1pt]
Differentiation total time & $143.35\ \mathrm{s}$ \\
Differentiation average time & $59.04\ \mathrm{ms}$ \\
CG total time & $127.21\ \mathrm{s}$ \\
CG average time & $0.075\ \mathrm{ms}$ \\
\# Newton iterations & 2428 \\
\# CG iterations & 1695130 \\
Max YASPS memory & $15302.13\ \mathrm{MB}$ \\
Generated files & 595 \\
Compile time & $346.95\ \mathrm{s}$ \\

\midrule[1pt]
\multicolumn{2}{c}{\textbf{Hex8}} \\
\midrule[1pt]
Differentiation total time & $197.76\ \mathrm{s}$ \\
Differentiation average time & $87.66\ \mathrm{ms}$ \\
CG total time & $114.55\ \mathrm{s}$ \\
CG average time & $0.075\ \mathrm{ms}$ \\
\# Newton iterations & 2256 \\
\# CG iterations & 1535411 \\
Max YASPS memory & $15820.91\ \mathrm{MB}$ \\
Generated files & 705 \\
Compile time & $458.60\ \mathrm{s}$ \\

\bottomrule
\end{tabular*}

\end{table}

\begin{figure}
    \centering
    \includegraphics[width=1.0\columnwidth]{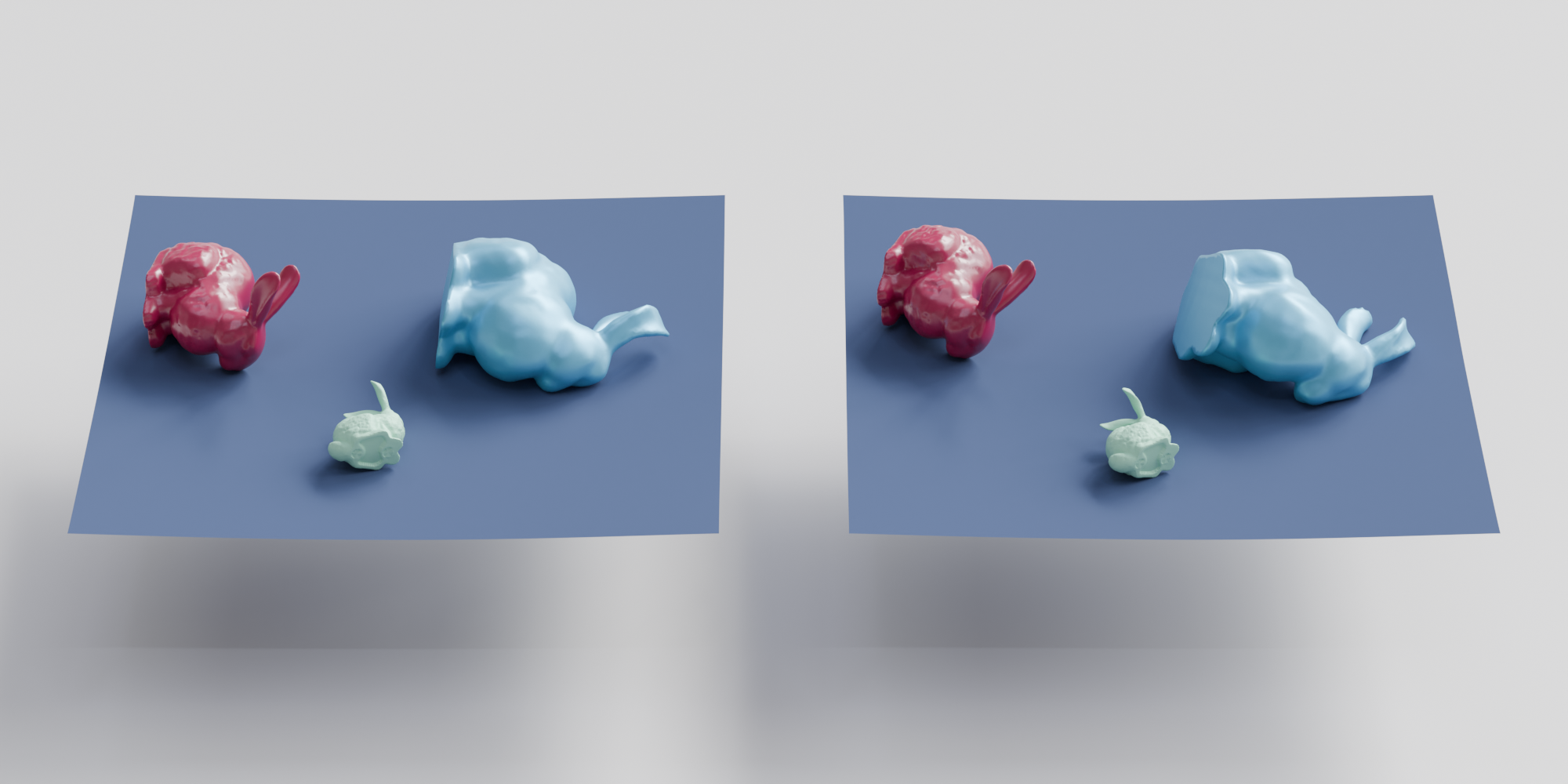}
    \vspace{-10pt}
    \caption{Results after 1.5 seconds of simulation. Left: each cage is treated as a hexahedral element and its energy is evaluated using 8-point Gaussian quadrature (Hex8). Right: each cage is decomposed into 6 tetrahedra (Tet4).}
    \label{figure-hex8}
\end{figure}

Here we show all the statistics for the simulation shown in Fig.~\ref{figure-cage}. The simulation settings including the material properties, stats for meshes are shown in Table \ref{table-hex8-stats}. The performance stats, including the compilation time for all the kernels, and the number of kernels are shown in Table \ref{table-hex-8-performance}. The simulation result after 1.5 seconds (150 time steps) is shown in Fig.~\ref{figure-hex8}.

We report the statistics for this example because it represents an extreme case. Each vertex on the caged bunny is connected to up to eight cage vertices. As a result, a point–triangle collision pair can, in the worst case, be influenced by $8\times 4 = 32$ Hex8 mesh vertices, leading to a maximum Hessian size of $96\times96$. In turn, this puts a lot of pressure on the per-thread stack memory, which is reflected through the maximum YASPS memory entry in Table \ref{table-hex-8-performance}.

In addition, YASPS generates specialized kernels for different compressed Hessian sizes, as illustrated in Fig.~\ref{figure-hessian-code-generation}. Consider, for example, a point–point collision pair where both vertices belong to the caged bunny. Depending on how many cage vertices the two points share (0, 1, 2, 4, or 8), the resulting compressed Hessian sizes can be $48\times 48$, $45\times 45$, $42\times 42$, $36\times 36$ or $24\times 24$, respectively.

In fact, out of the 705 files generated (this includes solver kernels, block diagonal inverse kernels, index kernels, and all the computation kernels) for the Hex8 discretization, 495 of them are for variations of different compressed Hessian size. By comparison, for the simulation shown in Fig.~\ref{figure-dropping-in-container-memory}, which involves an affine bunny and multiple soft bunnies (and in addition, the container itself whose vertices are in another primitive type), YASPS generates 151 kernels in total, of which only 38 correspond to different Hessian size variations.


\section{Algorithms}
Here we list all the algorithms used in this paper.

\begin{algorithm}
\caption{Find Local Boundary Pairs}
\label{alg:boundary-dfs}
\begin{algorithmic}[1]

\State \textbf{Input:} root node $r$
\State \textbf{Output:} mapping $\mathsf{Succ}$ from boundary nodes to boundary neighbors

\State $\mathsf{Succ} \gets$ empty mapping
\State $\mathsf{visited} \gets \emptyset$

\Procedure{DFS}{$v, t$}
  \If{$v \in \mathsf{visited}$}
    \State \Return
  \EndIf
  \State $\mathsf{visited} \gets \mathsf{visited} \cup \{v\}$

  \For{each child $w$ of $v$}
    \If{$\mathrm{type}(w) = \texttt{JOIN}$ \textbf{or} $\mathrm{type}(w) = \texttt{UNION}$}
      \State $\mathsf{Succ}[t] \gets \mathsf{Succ}[t] \cup \{w\}$
      \For{each child $u$ of $w$}
        \State \Call{DFS}{$u, w$}
      \EndFor

    \ElsIf{$\mathrm{type}(w) = \texttt{DATA}$}
      \State $\mathsf{Succ}[t] \gets \mathsf{Succ}[t] \cup \{w\}$

    \Else
      \State \Call{DFS}{$w, t$}

    \EndIf
  \EndFor
\EndProcedure

\State $\mathsf{Succ}[r] \gets \emptyset$
\State \Call{DFS}{$r, r$}

\State \Return $\mathsf{Succ}$

\end{algorithmic}
\end{algorithm}

\begin{algorithm}
\caption{Differentiate Local Pairs}
\label{alg:neighbor-differentiation}
\begin{algorithmic}[1]

\State \textbf{Input:} neighbor mapping $\mathsf{Succ}$ from Algorithm~\ref{alg:boundary-dfs}

\Function{GetSelfChildren}{$v$}
  \State $s \gets \emptyset$
  \For{each child $u$ of $v$}
    \If{$\mathrm{type}(u) \in \{\texttt{JOIN}, \texttt{UNION}, \texttt{DATA}\}$}
      \State $s \gets s \cup \{u\}$
    \Else
      \State $s \gets s \cup \Call{GetSelfChildren}{u}$
    \EndIf
  \EndFor
  \State \Return $s$
\EndFunction

\For{each boundary node $v$ in $\mathrm{dom}(\mathsf{Succ})$}
  \If{$\mathrm{type}(v) = \texttt{JOIN}$}
    \State $u \gets v.\mathsf{children}[0]$ \Comment{\texttt{JOIN} has exactly one child attribute}
    \State $w \gets \mathsf{Succ}[v]$
    \If{$\dfrac{\partial u}{\partial w}$ not yet computed}
      \State $J \gets \dfrac{\partial u}{\partial w}$
      \State $H \gets \dfrac{\partial^2 u}{\partial w^2}$ \Comment{Obtained by performing $\dfrac{\partial J}{\partial w}$}
      \State store $J, H$ on $u$'s primitive
    \EndIf

  \ElsIf{$\mathrm{type}(v) = \texttt{UNION}$}
    \For{each child $u$ of $v$}
      \State $p \gets \Call{GetSelfChildren}{u} \cap \mathsf{Succ}[v]$
      \If{$\dfrac{\partial u}{\partial p}$ not yet computed}
        \State $J \gets \dfrac{\partial u}{\partial p}$
        \State $H \gets \dfrac{\partial^2 u}{\partial p^2}$
        \State store $J, H$ on $u$'s primitive
      \EndIf
    \EndFor

  \Else
    \State $w \gets \mathsf{Succ}[v]$
    \If{$\dfrac{\partial v}{\partial w}$ not yet computed}
      \State $J \gets \dfrac{\partial v}{\partial w}$
      \State $H \gets \dfrac{\partial^2 v}{\partial w^2}$
      \State store $J, H$ on $v$'s primitive
    \EndIf
  \EndIf
\EndFor

\end{algorithmic}
\end{algorithm}
\clearpage
\begin{algorithm}
\caption{Compute Final Hessian}
\label{alg:final-hessian}
\begin{algorithmic}[1]

\State \textbf{Input:} neighbor mapping $\mathsf{Succ}$ from Algorithm~\ref{alg:boundary-dfs}, root node $r$
\State \textbf{Output:} global Hessian $H$ and gradient $g$

\State initialize caches $J_v, H_v$ as undefined for all boundary nodes $v$

\Function{ComputeHJ}{$v$}
  \If{$J_v$ is already computed}
    \State \Return $(J_v, H_v)$
  \EndIf

  \If{$\mathrm{type}(v) = \texttt{JOIN}$}
    \State $u \gets v.\mathsf{children}[0]$
    \State $(J_u, H_u) \gets \Call{ComputeHJ}{u}$
    \State $J_\text{flat} \gets \texttt{JOIN}(J_u, v.\texttt{connectivity})$ \Comment{Derivative of \texttt{JOIN} is a \texttt{JOIN} of derivatives plus reordering}
    \State $H_\text{flat} \gets \texttt{JOIN}(H_u, v.\texttt{connectivity})$
    \State $J_v \gets \texttt{reorder}(J_\text{flat})$
    \State $H_v \gets \texttt{reorder}(H_\text{flat})$

  \ElsIf{$\mathrm{type}(v) = \texttt{UNION}$}
    \State $J_\text{flat} \gets \emptyset$
    \State $H_\text{flat} \gets \emptyset$
    \For{each child $u$ of $v$}
      \State $(J_u, H_u) \gets \Call{ComputeHJ}{u}$
      \State $J_\text{flat} \gets J_\text{flat} \cup J_u$
      \State $H_\text{flat} \gets H_\text{flat} \cup H_u$
    \EndFor
    \State $J_v \gets \texttt{reorder}(\texttt{UNION}(J_\text{flat}))$ \Comment{Derivative of \texttt{UNION} is a \texttt{UNION} of derivatives plus reordering; padding omitted here for clarity}
    \State $H_v \gets \texttt{reorder}(\texttt{UNION}(H_\text{flat}))$

  \Else
    \Comment{Generic boundary node (typically a child of \texttt{JOIN}/\texttt{UNION}, or the root $r$)}
    \State $J_{g,\text{flat}} \gets \emptyset$
    \State $H_{g,\text{flat}} \gets \emptyset$
    \For{each neighbor $u \in \mathsf{Succ}[v]$}
      \State $(J_u, H_u) \gets \Call{ComputeHJ}{u}$
      \State $J_{g,\text{flat}} \gets J_{g,\text{flat}} \cup J_u$
      \State $H_{g,\text{flat}} \gets H_{g,\text{flat}} \cup H_u$
    \EndFor
    \State $J_g \gets \texttt{reorder}(J_{g,\text{flat}})$ \Comment{Construct global Jacobian of $g$}
    \State $H_g \gets \texttt{reorder}(H_{g,\text{flat}})$
    \State $J_f \gets \dfrac{\partial v}{\partial u}$ \Comment{Local Jacobian of $f\circ g$ cached from Algorithm~\ref{alg:neighbor-differentiation}}
    \State $H_f \gets \dfrac{\partial^2 v}{\partial u^2}$ \Comment{Local Hessian of $f\circ g$  cached from Algorithm~\ref{alg:neighbor-differentiation}}
    \State $(J_v, H_v) \gets \texttt{ApplyChainRule}(J_f, H_f, J_g, H_g)$
  \EndIf

  \State \Return $(J_v, H_v)$
\EndFunction

\State $(J_r, H_r) \gets \Call{ComputeHJ}{r}$
\State $g \gets J_r$
\State $H \gets H_r$
\State \Return $(H, g)$

\end{algorithmic}
\end{algorithm}

\begin{algorithm}
\caption{Generate Code Order}
\label{alg:gen-code-order}
\begin{algorithmic}[1]

\State \textbf{Input:} root node $r$
\State \textbf{Output:} code order stack $S$, important nodes $N$

\Procedure{DFS}{$v, S, N, \mathsf{visited}$}
  \If{$v \in \mathsf{visited}$}
    \State \Return
  \EndIf
  \State $\mathsf{visited} \gets \mathsf{visited} \cup \{v\}$
  \State $S \gets S \cup \{v\}$

  \If{$\mathrm{type}(v) \in \{\texttt{JOIN}, \texttt{UNION}\}$ \textbf{or} $v.\mathsf{hasName}$}
    \State \Call{GenObj}{$v$} \Comment{Defined in Algorithm~\ref{alg:gen-code-and-compile}}
    \State $N \gets N \cup \{v\}$
  \Else
    \For{each child $w$ of $v$}
      \State \Call{DFS}{$w, S, N, \mathsf{visited}$}
    \EndFor
  \EndIf
\EndProcedure

\Function{GenerateCodeOrder}{$r$}
  \State $S \gets \emptyset$
  \State $N \gets \emptyset$
  \State $\mathsf{visited} \gets \emptyset$
  \State \Call{DFS}{$r, S, N, \mathsf{visited}$}
  \State \Return $S, N$
\EndFunction

\end{algorithmic}
\end{algorithm}
\clearpage
\begin{algorithm}
\caption{Generate Code}
\label{alg:codegen-dfs}
\begin{algorithmic}[1]

\State \textbf{Input:} code order stack $S$ from Algorithm~\ref{alg:gen-code-order}
\State \textbf{Output:} generated code as a string $\mathsf{Code}$

\Function{AttributeToCode}{$v, t, \mathsf{visited}, \mathsf{intermediates}$}
\If{$v\in \mathsf{visited}$}
\State \Return $\mathsf{intermediates}[v]$
\EndIf
\State $t \gets t + 1$
\State $\mathsf{tmp} \gets \texttt{"x\_"} + \texttt{toString}(t)$ \Comment{fresh temporary name as a string}
\State $\mathsf{intermediates}[v] \gets \mathsf{tmp}$
\State $\mathsf{visited} \gets \mathsf{visited} \cup v$
\State $\mathsf{inputs}\gets\texttt{""}$
\For{each child $u$ of $v$}
\State $\mathsf{inputs} \gets \mathsf{inputs} + \texttt{", "} +$ \Call{AttributeToCode}{$u$}
\EndFor
\If{$v.\mathsf{hasKernel}$}
\State \Return $\mathsf{tmp} + \texttt{" = "} \ + v.\mathsf{kernelName} +\texttt{"("}  + \mathsf{inputs} + \texttt{")"}$
\Else
\State \Return  $\mathsf{tmp} + \texttt{" = "}  + $ \Call{ToCodeString}{$\mathsf{type}(v), \mathsf{inputs}$}
\EndIf
\EndFunction

\Function{CodeGen}{$S$}
    \State $\mathsf{Code} \gets \texttt{""}$ \Comment{The final code}
    \State $\mathsf{visited} \gets \emptyset$ \Comment{Visited nodes}
    \State $\mathsf{intermediates} \gets$ empty mapping \Comment{For any visited nodes, we replace the code call with a reference to an intermediate value which should already be computed}
    \State $t \gets 0$ \Comment{Records the number of intermediates}
    \While{$S \neq \emptyset$}
        \State $v \gets S.\texttt{pop}()$
        \State $\mathsf{Code} \gets \mathsf{Code}\, + \, $\Call{AttributeToCode}{$v, t, \mathsf{visited}, \mathsf{intermediates}$}
    \EndWhile
    \State \Return{$\mathsf{Code}$}
\EndFunction

\end{algorithmic}
\end{algorithm}

\begin{algorithm}
\caption{Generate And Compile Code}
\label{alg:gen-code-and-compile}
\begin{algorithmic}[1]

\State \textbf{Input:} root node $r$
\Function{GenObj}{$v$}
\State $(S, N) \gets $ \Call{GenerateCodeOrder}{$v$} 
\State $\mathsf{Code} \gets$ \Call{CodeGen}{$S$}
\State $O \gets $ \Call{Compile}{Code}
\State $v.\mathsf{obj} \gets O$
\State \Return{$O, N$}
\EndFunction

\State{$\mathsf{Objs} \gets \emptyset$}
\State $(O, N) \gets $ \Call{GenObj}{$r$}
\State $\mathsf{Objs} \gets \mathsf{Objs}  \cup O$
\For{\textbf{each} $v \in N$}
\State $\mathsf{Objs} \gets \mathsf{Objs}  \cup v.\mathsf{obj}$
\EndFor

\State $\mathsf{Kernel} \gets $ \Call{CompileFinalKernel}{$\mathsf{Objs}$}
\State \Return{$\mathsf{Kernel}$}
\end{algorithmic}
\end{algorithm}
\begin{algorithm}
\caption{CUDA Block Level SpMV}
\label{alg:spmv}
\begin{algorithmic}[1]
\State \textbf{Input:}
Block values $\mathbf{B}$,
positions array $\mathbf{P}$,
input vector $\mathbf{x}$,
output vector $\mathbf{y}$,
$r = \texttt{BLOCK\_ROW\_SIZE}$,
$c = \texttt{BLOCK\_COL\_SIZE}$

\State $id \gets \texttt{blockIdx.x} \cdot \texttt{blockDim.x} + \texttt{threadIdx.x}$
\State $t \gets \texttt{threadIdx.x}$

\State Allocate shared memory:
\[
\texttt{allResults}[32 \cdot r],\;
\texttt{rows}[32],\;
\texttt{cols}[32]
\]

\For{$i \gets t$ to $32r$ step $32$}
  \State $\texttt{allResults}[i] \gets 0$
\EndFor
\State \textbf{Synchronize threads}

\If{$id < N$} \Comment{$N = \texttt{POSITIONS\_END} - \texttt{POSITIONS\_START}$}
  \State $(\texttt{rows}[t], \texttt{cols}[t]) \gets \mathbf{P}[\texttt{POSITIONS\_START} + id]$
  \State
  $\texttt{allResults}[t \cdot r : (t+1)r]
   \gets \mathbf{B}[id] \cdot \mathbf{x}[\texttt{cols}[t]]$
\Else
  \State $\texttt{rows}[t] \gets \bot$ \Comment{invalid segment}
\EndIf
\State \textbf{Synchronize threads}

\If{$id < N$}
  \If{$t = 0 \;\lor\; \texttt{rows}[t] \neq \texttt{rows}[t-1]$}
    \Comment{Start of a new row segment}
    \State $\mathbf{s} \gets \mathbf{0} \in \mathbb{R}^r$
    \For{$i \gets t$ while $i < 32$ and $\texttt{rows}[i] = \texttt{rows}[t]$}
      \For{$j \gets 0$ to $r-1$}
        \State $s_j \gets s_j + \texttt{allResults}[i \cdot r + j]$
      \EndFor
    \EndFor
    \For{$j \gets 0$ to $r-1$}
      \State $\mathbf{y}[\texttt{rows}[t] + j] \mathrel{+}= s_j$ \Comment{atomic add to global result}
    \EndFor
  \EndIf
\EndIf

\If{$id < N$ and $\texttt{rows}[t] \neq \texttt{cols}[t]$}
  \State
  $\mathbf{y}[\texttt{cols}[t]]
  \mathrel{+}= \mathbf{B}[id]^T \cdot \mathbf{x}[\texttt{rows}[t]]$ \Comment{Add the transpose result}
\EndIf

\end{algorithmic}
\end{algorithm}

\end{document}